\documentclass[sigconf,nonacm]{acmart} 
\AtBeginDocument{%
  }

\acmISBN{978-1-4503-XXXX-X/2018/06}

\usepackage{dsfont}
\usepackage{multicol}
\usepackage{multirow}
\usepackage{mathtools}
\usepackage{microtype}
\usepackage[normalem]{ulem}

\usepackage{amsmath}
\usepackage{subcaption}
\usepackage{float}
\usepackage{url}

\usepackage{booktabs}
\usepackage{pifont}
\usepackage{makecell}
\usepackage{graphicx}

\usepackage[linesnumbered, ruled, noline, noend]{algorithm2e}
\SetKwProg{Procedure}{Procedure}{:}{}

\usepackage{xcolor}
\usepackage{comment}

\definecolor{mycolor}{RGB}{153, 0, 0}
\SetKwFor{ForEach}{\textcolor{mycolor}{foreach}}{\textcolor{mycolor}{do}}{\textcolor{mycolor}{end}}%
\SetKwFor{For}{\textcolor{mycolor}{for}}{\textcolor{mycolor}{do}}{\textcolor{mycolor}{end}}%
\SetKwFor{While}{\textcolor{mycolor}{while}}{\textcolor{mycolor}{do}}{\textcolor{mycolor}{end}}%

\SetCommentSty{mycommfont}

\SetKwIF{If}{ElseIf}{Else}{\textcolor{teal}{if}}{\textcolor{teal}{then}}{\textcolor{teal}{else if}}{\textcolor{teal}{else}}{\textcolor{teal}{end if}}%

\newcommand{\splitsplat}{\textbf{Split\&Splat}}
\acmSubmissionID{652}

\setcopyright{acmlicensed}
\acmYear{}
\copyrightyear{}
\acmDOI{}
\acmISBN{}
\acmPrice{}

\newcommand{\ie}{\textit{i.e.}}
\newcommand{\eg}{\textit{e.g.}}

\begin{document}

\title{{Split\&Splat: Zero-Shot Panoptic Segmentation via Explicit} {Instance Modeling and 3D Gaussian Splatting}}

\author{Leonardo Monchieri, Elena Camuffo, Francesco Barbato, Pietro Zanuttigh, Simone Milani}

\authornote{%
Code: \textcolor{cyan}{\ttfamily \url{https://github.com/LTTM/Split_and_Splat}},\\
Contact: {\{leonardo.monchieri, elena.camuffo, francesco.barbato, pietro.zanuttigh, simone.milani\}@unipd.it}
}

\affiliation{  
\vspace{0.5em}
  \institution{University of Padova}
  \city{Padova}
  \country{Italy}
}

\renewcommand{\shortauthors}{Monchieri et al.} %

\begin{abstract}
3D Gaussian Splatting (GS) enables fast and high-quality scene reconstruction, but it lacks an object-consistent and semantically aware structure.
We propose \textbf{Split\&Splat}, a framework for panoptic scene reconstruction using 3DGS. Our approach explicitly models object instances. It first propagates instance masks across views using depth, thus producing view-consistent 2D masks. Each object is then reconstructed independently and merged back into the scene while refining its boundaries. Finally, instance-level semantic descriptors are embedded in the reconstructed objects, supporting various applications, including panoptic segmentation, object retrieval, and 3D editing.
Unlike existing methods, \textbf{Split\&Splat} tackles the problem by first segmenting the scene and then reconstructing each object individually. This design naturally supports downstream tasks and allows \textbf{Split\&Splat} to achieve state-of-the-art performance on the ScanNetv2 segmentation benchmark.
\end{abstract}


\begin{teaserfigure}
  \centering\includegraphics[width=\textwidth]{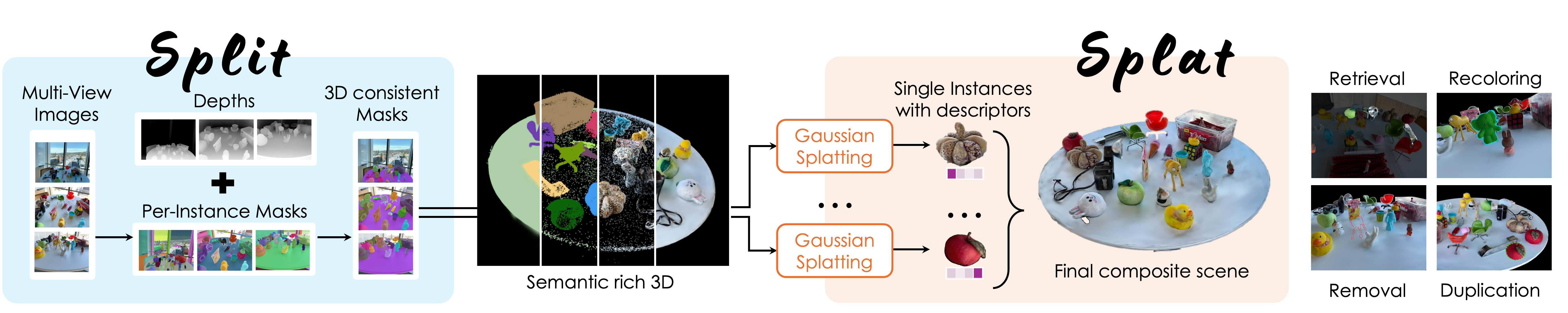}
  \caption{
  Split\&Splat pipeline.
Starting from multi-view images, segmentation masks are generated, enabling the split of a scene into single objects with explicit geometric consistency. These subsets are then independently reconstructed via 3D Gaussian splatting, merged, and augmented with object-level descriptors, enabling object-level reasoning, retrieval and editing, within a coherent global scene.
  }
  \Description{Split and Splat graphical abstract.}
  \label{fig:teaser}
  \vspace{2.5em}
\end{teaserfigure}

\maketitle

\newpage
\section{Introduction} \label{sec:intro}
3D Gaussian splatting \cite{kerbl3Dgaussians} represents a milestone advancement in the field of volume rendering, enabling ultra-fast novel-view synthesis with a photorealistic appearance.
The growing popularity of this approach has paved the way for its application in various tasks \cite{gaussianGrouping,sage,qin2024langsplat,wang2025gaussiangraph}. 
Traditionally, the technique focuses solely on pixel rendering and does not capture the semantics of scene content, which is crucial in applications such as object detection, open vocabulary semantic segmentation, and panoptic segmentation.

Nowadays, this limitation can be effectively addressed by leveraging Vision Foundation Models (VFMs) \cite{sam2,oquab2023dinov2,simeoni2025dinov3,clip}. 
These systems are capable of producing dense, high-level feature embeddings that support robust mask generation and region-level classification. 
Leveraging such capabilities, recent works \cite{gaussianGrouping,langsplatv2_2025,wu2024opengaussian,occams_lgs_2024,li2025instancegaussian,vala} have begun integrating GS-based 3D reconstructions with VFM-derived segmentation cues, thereby enabling the propagation of object masks and category labels within the reconstructed scene.

However, achieving view-consistent segmentation in a GS representation remains a challenging problem. Inconsistencies in mask assignment or feature aggregation across different viewpoints often lead to disjointed object representations and degraded segmentation quality. Such issues propagate to downstream tasks, ultimately limiting the effectiveness of GS-based approaches for reliable 3D-aware segmentation and scene understanding.

Compared to approaches that simply fit the semantic features from the 2D images into an already existing 3DGS, our 2-stage approach (\textbf{Split\&Splat}) integrates reconstruction and segmentation tasks into a single, coherent, and highly-performing framework. Indeed, explicit modeling of object instances enables a sharper and more reliable separation of objects in the 3D scene, facilitating the integration of semantic information as instance-level descriptors.

More specifically, our pipeline reconstructs the scene as follows: (1) we extract 2D instance masks from the multi-view images of the scene; (2) the 2D object masks are propagated and refined by exploiting depth cues, ensuring multi-view and geometric consistency; (3) each object is independently reconstructed using Gaussian splatting; (4) a semantic descriptor is embedded in each segment.
Overall, our pipeline achieves a segmented and semantically structured Gaussian splatting representation of the scene that is well-suited for downstream applications such as panoptic segmentation, object retrieval, and 3D editing.

\section{Related Work} \label{sec:related}
Recent works have extended Gaussian splatting beyond photorealistic rendering toward semantic and open vocabulary 3D scene understanding.
Early efforts incorporated per-Gaussian feature embeddings to associate each Gaussian with semantic or instance-level descriptors. 
\textit{Gaussian Grouping}~\cite{gaussianGrouping} introduced class-aware features to enable semantic segmentation directly within the explicit Gaussian representation. 
\textit{SAGE}~\cite{sage} uses semantic-aware Gaussian splatting for real-time optimization.
\textit{GAGS}~\cite{peng2024gags} introduces a granularity-aware 2D to 3D feature distillation scheme, while  \cite{10943563} integrates a 2D segmentation step in propagating semantic features to 3D reconstruction. \textit{GOI}~\cite{goi} learns an optimizable manifold in an open vocabulary (CLIP-like) semantic space to identify Gaussians of Interest, enabling text-guided selection and segmentation.
Overall, these methods confirm the 3DGS's capability to produce explicit and differentiable models able to generate real-time semantic masks without the heavy computation typically associated with implicit NeRF-style models.

A more recent line of research focuses on language-driven segmentation by aligning Gaussian features with multimodal embeddings. 
\textit{LangSplat}~\cite{qin2024langsplat,langsplatv2_2025} and \textit{LEGaussians}~\cite{shi2024language} encode CLIP-aligned features inside each Gaussian, enabling text-conditioned segmentation, retrieval, and editing in 3D space. 
\textit{VALA}~\cite{vala} proposes a visibility-aware aggregation of language features by integrating text embeddings across multiple views while weighting them according to Gaussian visibility, thereby improving open vocabulary segmentation under occlusions and partial observations.
\textit{InstanceGaussian}~\cite{li2025instancegaussian} jointly learns appearance and instance semantics within a unified Gaussian representation. 
Unlike language-driven or purely feature-distillation approaches, it focuses on balancing appearance and semantic attributes through a \textit{clustered} representation, where multiple appearance Gaussians share instance features. Additionally, it introduces a bottom-up, category-agnostic instance aggregation strategy based on over-segmentation and graph connectivity, enabling adaptive instance discovery without relying on predefined object counts or category priors.

While the contribution of these approaches cannot be understated, they are still limited by two major issues: (1) they distill 2D features into the 3DGS framework without ensuring that the per-view embeddings maintain 3D spatial consistency, and (2) the addition of dense descriptors and features drastically increases the memory required to store each Gaussian parameter, leading to a resource-intensive training process.
To tackle these issues, our approach addresses the instance segmentation task in a bottom-up fashion. Rather than directly learning a unified 3DGS representation that must be split at test time, we model all object instances independently before aggregating them into a global representation. 
This allows us to achieve higher quality segments and reduce the memory burden by embedding only sparse (one per segment) descriptors in the Gaussians, if necessary.

\begin{figure*}[t]
    \centering
    \includegraphics[width=\linewidth]{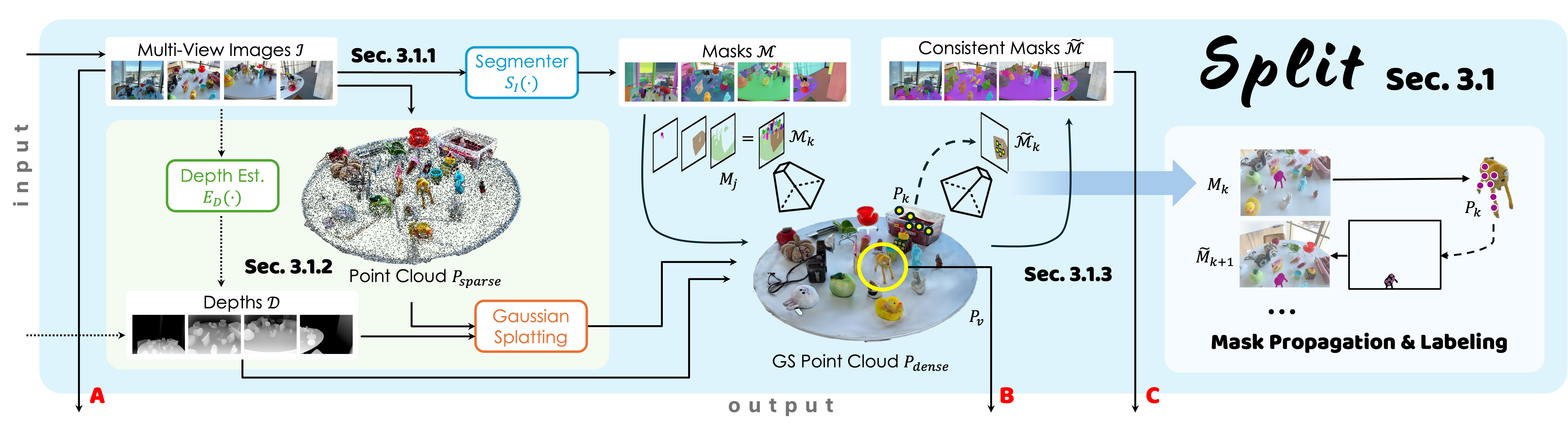}
    \Description[Split\&Splat Architecture: Detail on Split]{Split\&Splat Architecture: Detail on Split}
    \caption{
    During \textbf{Split}, multi-view images are processed to estimate depth and instance masks, which are propagated in 3D to produce refined, view-consistent segmentations. \textbf{\textcolor{red}{A}}, \textbf{\textcolor{red}{B}}, and \textbf{\textcolor{red}{C}} denote the output of Split, which serves as an input to Splat.} %
\label{fig:pipeline_split}
\end{figure*}

\section{Method}
\label{sec:method}
We introduce \splitsplat, a novel 3D scene reconstruction framework based on Gaussian splatting that directly embeds semantic descriptors into 3DGS objects, supporting multiple downstream tasks such as panoptic segmentation. 
Our method consists of two main stages: instance segmentation (\textbf{Split}) and reconstruction (\textbf{Splat}).

In the \textbf{Split} stage (Sec.~\ref{subsec:sceneseg}), a Structure-from-Motion (SfM) algorithm processes the initial multi-view data to create a 3D point cloud model of the scene and to estimate the camera parameters. At the same time, a monocular depth estimator associates each view with its corresponding depth map. Depth and geometry are used together to create an initial 3D Gaussian splatting representation.
Note that this step is only necessary when depth, pose estimation, and point cloud are missing; when they are available, it can be skipped.
In parallel, 2D instance segmentation masks for each view are generated and subsequently refined by exploiting the coarse reconstruction to achieve multi-view and geometric consistency. 

In the \textbf{Splat} stage (Sec.~\ref{subsec:instancerec}), the refined masks are used to identify object instances and reconstruct each of them via 3DGS. 
The separate objects are then merged sequentially into the global scene while embedding semantic information.
Finally, visual descriptors are computed on a set of masked views cropped around the instance. These descriptors are then applied to the instance-level reconstructions, producing semantically informed 3D representations.
The final output, composed of Gaussian objects associated with semantic visual descriptors, supports a variety of downstream tasks in addition to panoptic segmentation, such as object retrieval and scene editing.

\subsection{Split: 3D Instance Segmentation} \label{subsec:sceneseg}
In this section, we describe the \textbf{Split} stage, where we obtain a set of segmentation maps from multi-view images. Using the estimated depth maps along with an initial dense point cloud, we refine the masks and enforce consistency across the 2D segmentations and the 3D scene. A schematic overview is shown in Fig.~\ref{fig:pipeline_split}.

\subsubsection{Mask generation}\label{subsec:mask_prop}
Given a set of $K$ multi-view representations $\mathcal{I}\!=\!\{I_{k}\}_{k=1}^K$ with resolution $H\times W$, capturing a 3D scene, we compute per-view instance segmentation masks using a segmenter $S_I(\cdot)$. 
This yields $K$ segmentation mask sets $\mathcal{M}_k=\{M_{k,j} \}$, each containing $J_k=|\mathcal{J}_k|$ binary masks $M_{k,j}$ (with ${j \in \mathcal{J}_k}$) associated with the object instances $j$ detected in view $k$.
The cardinality $J_k$ depends on the number of visible instances in each view $k$. 
The masks $M_{k,j}$ are automatically initialized using \textit{ping} points from SAM2's fine-grid policy~\cite{sam2}. The masks are then merged using a coarse-to-fine approach \cite{gaussianGrouping} to promote larger masks (see  Fig.~\ref{fig:quali_grid} for some visual examples). 
Note that standard 2D instance segmenters struggle to recover objects that are partially occluded or missing in a given view; therefore, at this step, the mask labels are arbitrary and require further processing to ensure cross-view consistency. These issues are reflected in some of the failure cases of approaches like DEVA~\cite{cheng2023tracking} or those using foundation model descriptors \cite{oquab2023dinov2,simeoni2025dinov3,sam2,clip}, which directly embed semantic features into Gaussians without particular attention to the underlying 3D structures. As a result, they cannot ensure cross-view consistency before performing instance segmentation.
To tackle this issue, we introduce a mask propagation technique, detailed in Sec.~\ref{subsec:mask_cons}, to produce high-quality, multi-view consistent instance masks without any manual input. 

\subsubsection{Initial 3D reconstruction}
Concurrently with the mask generation, we process the images in $\mathcal{I}$ using the COLMAP SfM algorithm \cite{schonberger2016structure} to obtain a sparse point cloud reconstruction $P_{sparse} \in \mathbb{R}^{3\times n}$.
Similarly, we use a monocular depth estimator $E_D(\cdot)$  (refer to Sec.~\ref{sec:impl_details} for more details) to obtain a set of $K$ depth maps $D_k$.
The two are used as initialization for depth-regularized Gaussian splatting, which will generate a dense point cloud of the scene $P_{dense} \in \mathbb{R}^{3\times n}$. 

\begin{figure*}[t]
\centering
\includegraphics[width=\linewidth]{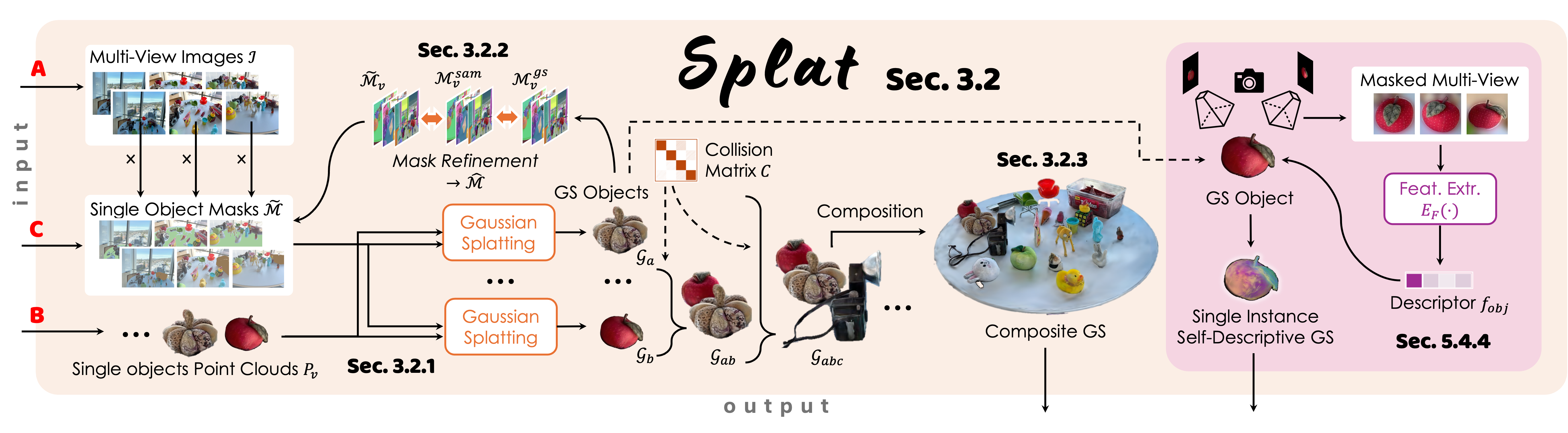}
    \Description[Split\&Splat Architecture: Detail on Splat]{Split\&Splat Architecture: Detail on Splat}
    \caption{
    During \textbf{Splat}, each instance is reconstructed independently using 3DGS, then merged into a global model enriched with per-instance descriptors. \textbf{\textcolor{red}{A}}, \textbf{\textcolor{red}{B}}, and \textbf{\textcolor{red}{C}} denote the output of Split, which serves as an input to Splat.}
\label{fig:pipeline_splat}
\end{figure*}

\subsubsection{Mask propagation}\label{subsec:mask_cons}
After the reconstruction, we begin iteratively assigning the instance labels in $\mathcal{J}$ to the points in $P_{dense}$, updating the labels of the points after each iteration to enforce 3D consistency. The result is a dense and view-consistent point cloud $P_{labeled} \subseteq P_{dense}$ (\textit{floaters} or inner-object points completely covered by surface points are removed), constructed as follows.

For each view $k$, the points $p \in P_{dense}$ are mapped onto the image plane using the camera projection $\Pi_k(\cdot)$, to obtain their pixel coordinates $(w,h)$ and projected depth $\hat{d}$. Points lying outside camera $k$'s field of view are discarded:
\begin{equation}
\hat{P}_{k} \!=\! \{p \!\mid\! {p \in P_{dense}}, {(w,h,\hat{d}) \!=\! \Pi_k(p)}, {0\!\leq\!w\!<\!W}, {0\!\leq\!h\!<\!H}\} \;\;.
\end{equation}
In the following, we omit the ranges of $h$ and $w$ for ease of notation. 
To ensure the visibility of the selected points in the current view, we preserve only the points satisfying \mbox{$|D_k(w,h) - \hat{d}| < \tau_{depth}$}; we empirically set $\tau_{depth} = 0.02$ (see Sec. \ref{sec:depth_th} for more details).
This constraint yields the set of surface-consistent points for view $k$ as:
\begin{equation}
P_{k}=\{p \mid {p \in \hat{P}_{k}},  {(w,h,\hat{d})=\Pi_k(p)}, {|D_k(w,h)-\hat{d}|<\tau_{depth}}\} \;\;.
\end{equation}
These sets $P_k \subseteq P_{dense}$ (one for each view $k$) are used to progressively update the masks $\mathcal{M}_k$ starting from frame $k=1$. More in detail, we extract the subsets $P_{k,j}$ of points $p \in P_k$ whose projections fall inside the eroded $M_{k,j}$ (avoiding border effects): 
\begin{equation}
P_{k,j}=\{ p \mid p \in P_k, \ M_{k,j}(w,h)=1\} \;\;.
\end{equation}
This allows us to compute the sets $P_k^{*} = 
\bigcup_{j \in \mathcal{J}_k}P_{k,j} \! \subset \! P_k$ of all labeled points (\ie, non-background) visible by camera $k$. To refine $P_k^{*}$, we apply DBSCAN~\cite{dbscan}, removing isolated points.
The sets $P_{k,j}$ are used to sequentially re-project the semantic labels $j \in \mathcal{J}_{k-1}$ from $P_{k-1,j}$ onto the next view $k$, thus obtaining virtual masks $M^{*}_{k,j}$. 
If there is no intersection between $M^{*}_{k,j}$ and any mask in $\mathcal{M}_{k}$, the warped region corresponds to a different object. 
Otherwise, labels $i$ in masks $M_{k,i}$ are remapped into the label $j$ that maximizes the intersection with $M^{*}_{k,j}$ (see Alg. 1 in the Appendix for details). These labels are then transferred to $P^*_k$ for 3D-consistent processing.

\begin{figure*}[t]
    \centering
    \scriptsize

    \begin{minipage}{\textwidth}
        \centering
        \begin{minipage}{.3\textwidth}
            \centering
            Ground truth mesh
        \end{minipage}%
        \begin{minipage}{.3\textwidth}
            \centering
            $P_{labeled}$ 
        \end{minipage}%
        \begin{minipage}{.3\textwidth}
            \centering
             Overlay
        \end{minipage}%
    \end{minipage}\\[2mm]

    \begin{minipage}{\textwidth}
       \makebox[0pt][r]{%
            \raisebox{-0.3\height}{\rotatebox{90}{ScanNetv2/{0062\_00}}}%
            \hspace{2mm}%
        }%
        \centering
        \begin{minipage}{.3\textwidth}
            \includegraphics[width=\textwidth]{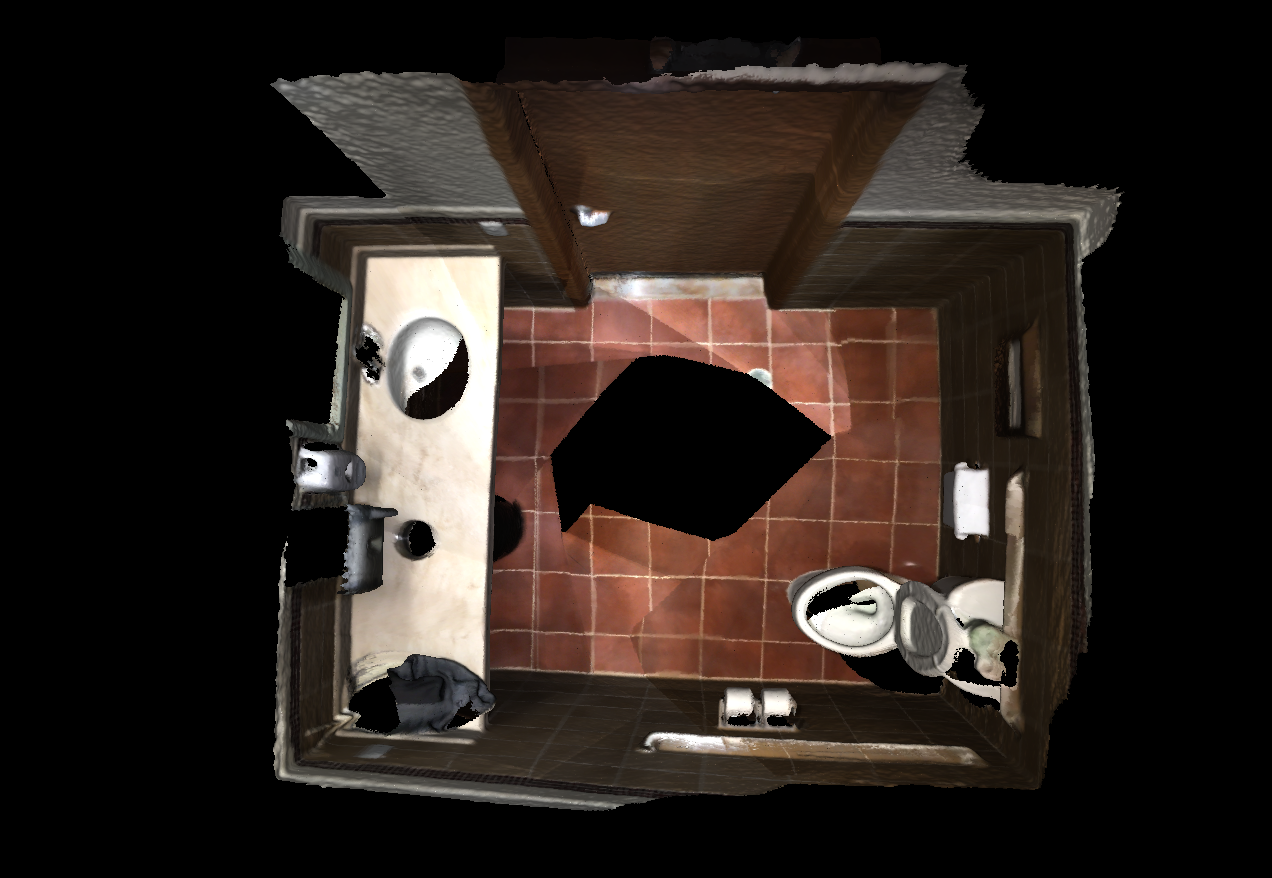}
        \end{minipage}%
        \hspace{0.03cm}
        \begin{minipage}{.3\textwidth}
            \includegraphics[width=\textwidth]{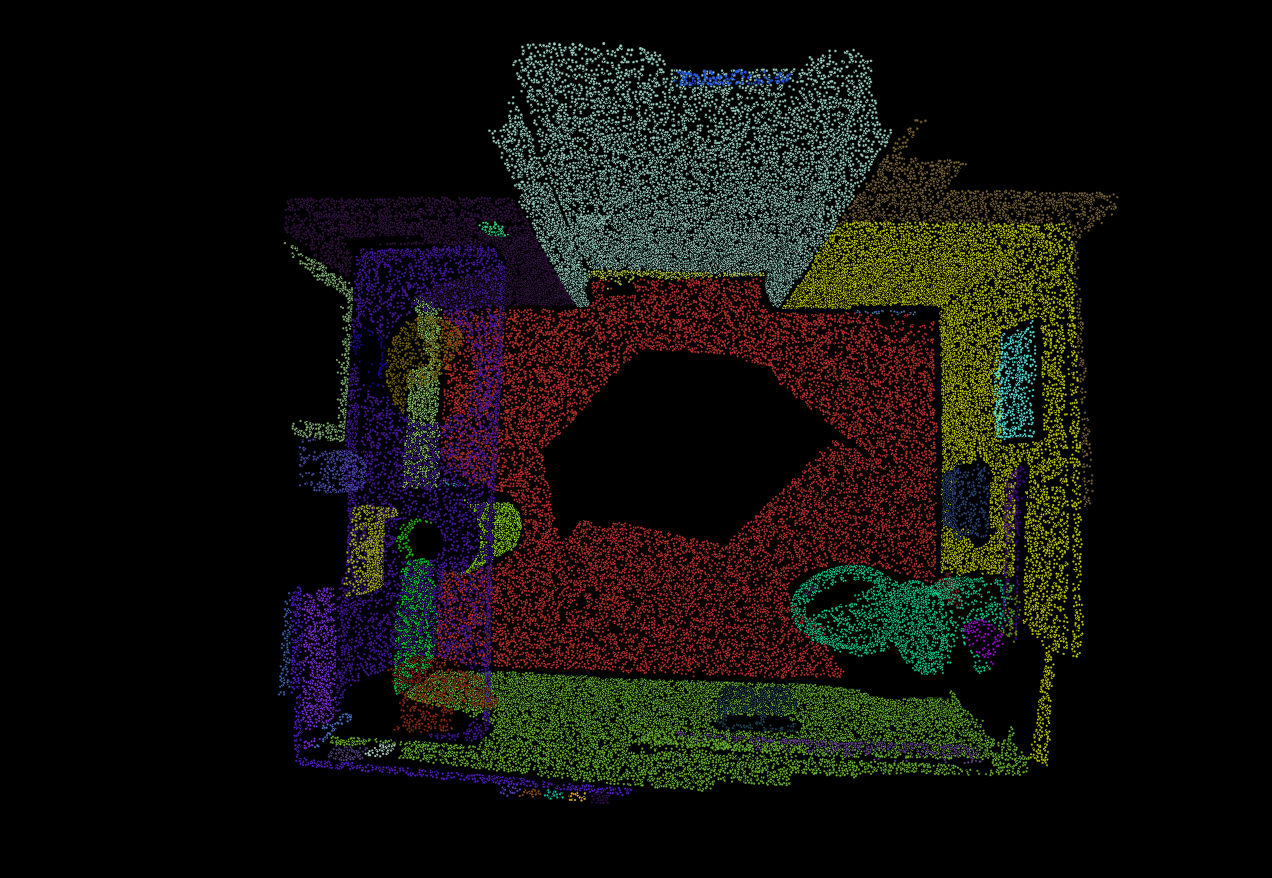}
        \end{minipage}%
        \hspace{0.03cm}
        \begin{minipage}{.3\textwidth}
            \includegraphics[width=\textwidth]{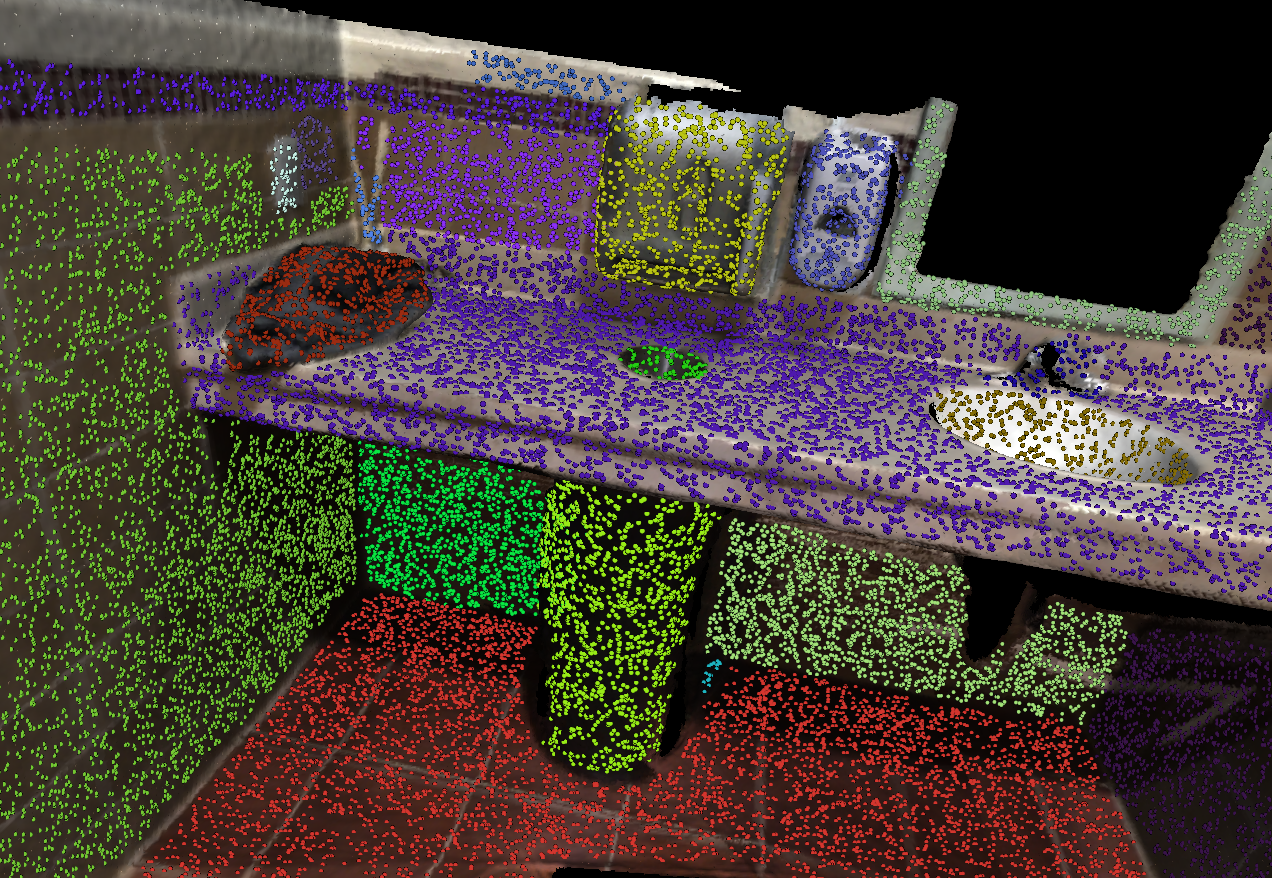}
        \end{minipage}%
    \end{minipage}\\[1mm]

    \begin{minipage}{\textwidth}
        \makebox[0pt][r]{%
            \raisebox{-0.3\height}{\rotatebox{90}{ScanNetv2/{0347\_00}}}%
            \hspace{2mm}%
        }%
        \centering
        \begin{minipage}{.3\textwidth}
            \includegraphics[width=\textwidth]{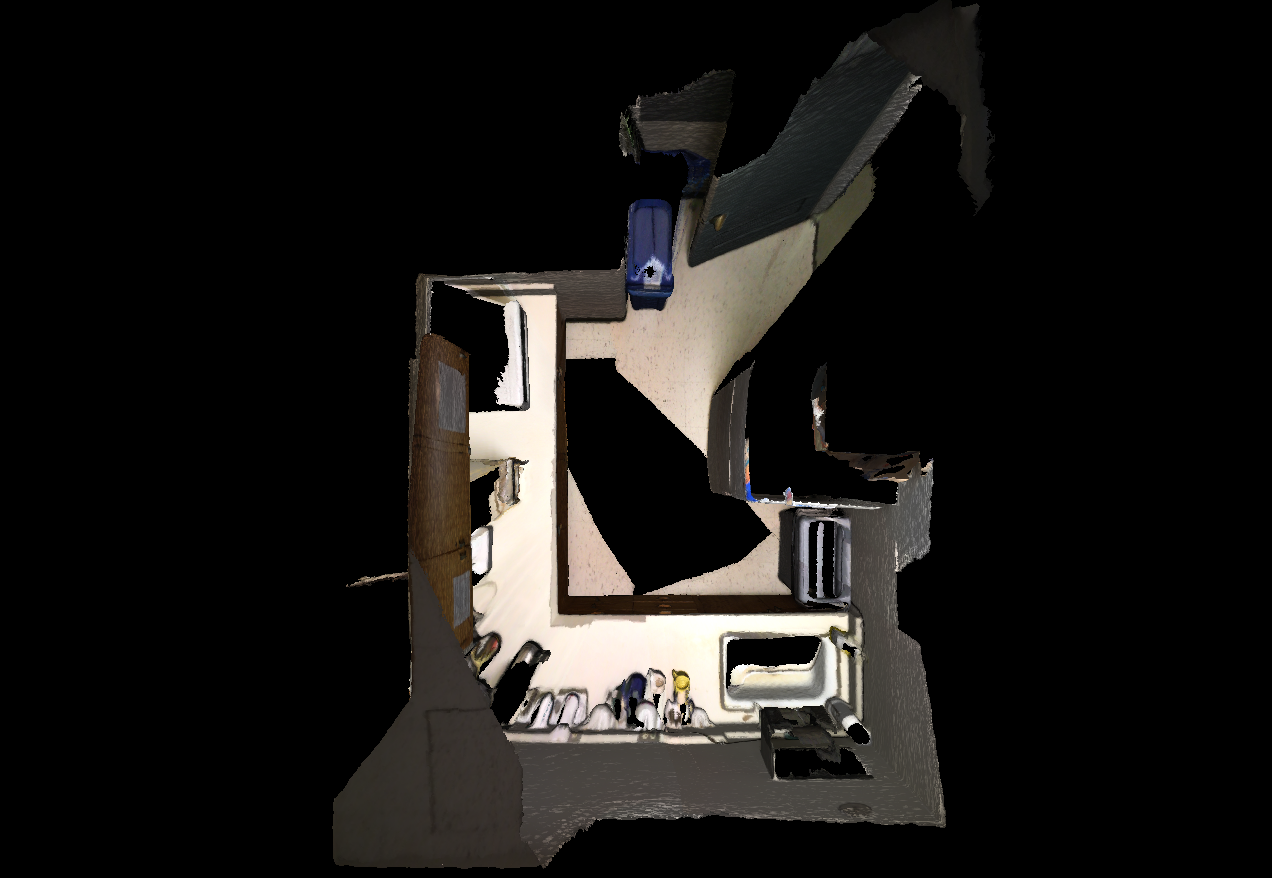}
        \end{minipage}%
        \hspace{0.03cm}
        \begin{minipage}{.3\textwidth}
            \includegraphics[width=\textwidth]{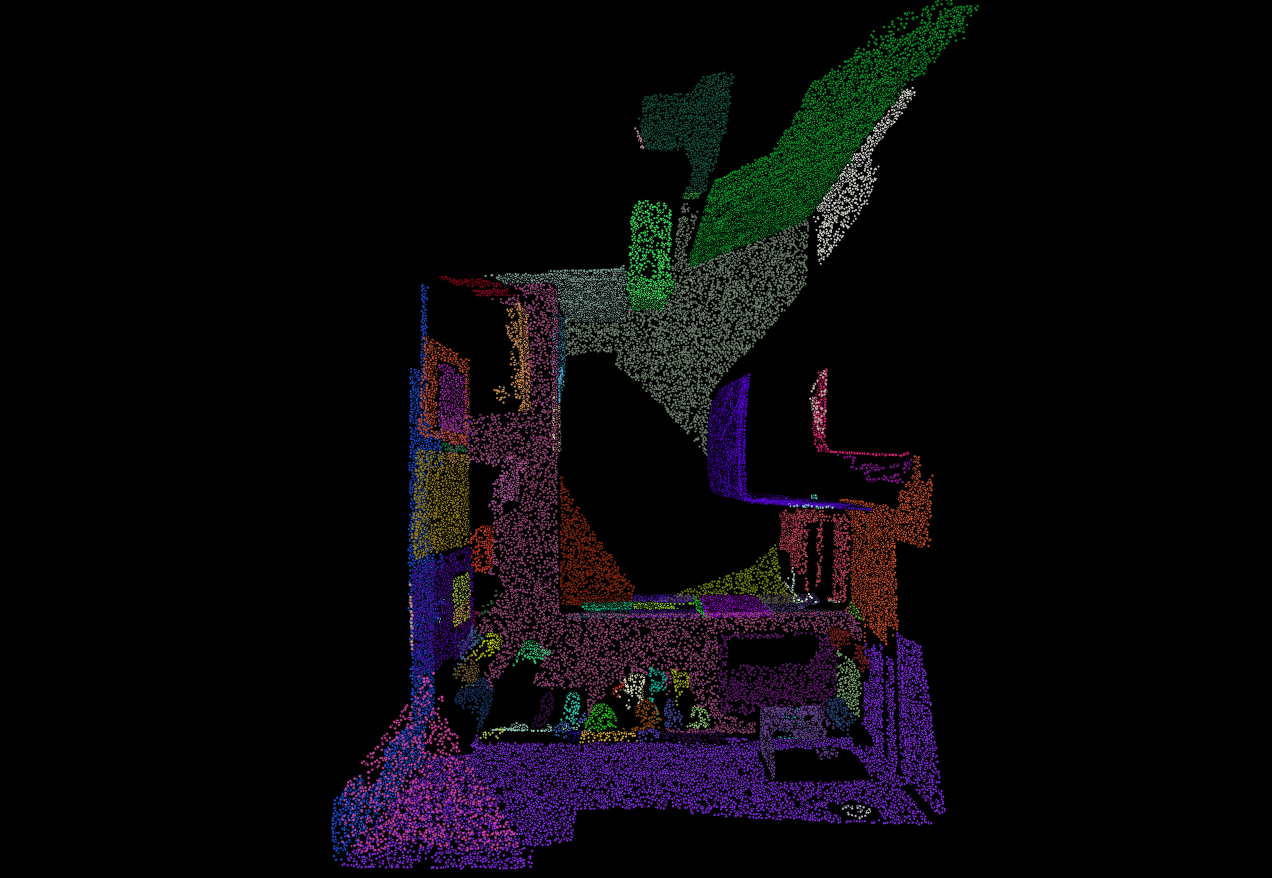}
        \end{minipage}%
        \hspace{0.03cm}
        \begin{minipage}{.3\textwidth}
            \includegraphics[width=\textwidth]{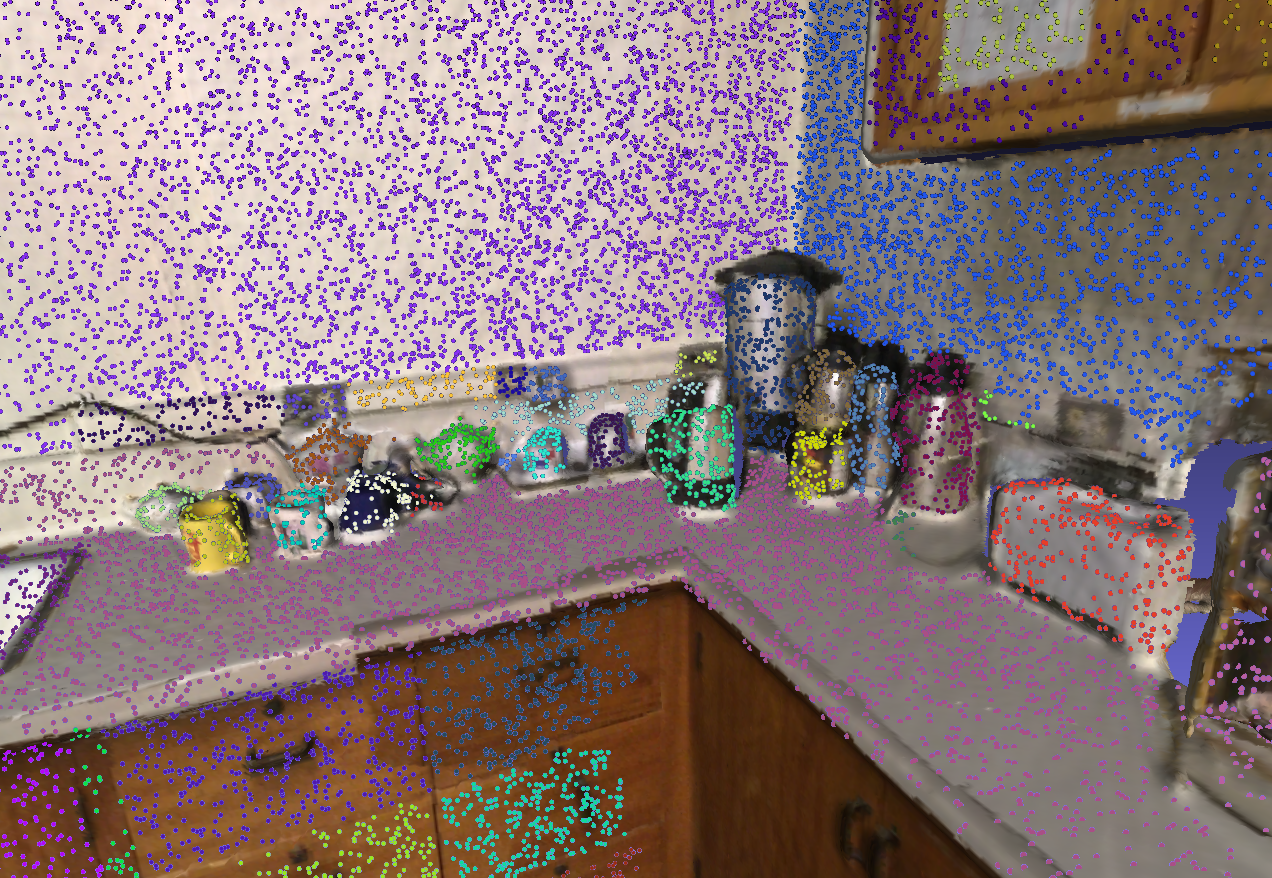}
        \end{minipage}%
    \end{minipage}

    \Description[Qualitative Results]{Qualitative results on LERF/ScanNet datasets}
    \caption{Labeled point cloud ($P_{labeled}$) produced by mask propagation stage, ensuring 3D consistency. }
    \label{fig:Propagation_qualitaives}
\end{figure*}

More specifically, to create a scene-level labeling, we assign a weight vector $w_p \in \mathbb{R}^{|\mathcal{J}|}$ to each 3D point $p$, where $w_p[j]$ represents the aggregated scores of label $j \in \mathcal{J}$ from different views.
Whenever $w_p$ is uninitialized, the view score is set to  $1+\lambda_{\text{init}}$, with $0<\lambda_{\text{init}}<1$, thus ensuring priority to the first view in case of ties. 
Otherwise, each view adds a score equal to $1$.
After all views have been processed, the array $w_p$ is normalized, and the final instance label $l_p$ for $p$ is assigned via majority voting, \ie, $l_p = \arg\!\max_{j \in \mathcal{J}} w_p[j]$.
We discard points whose normalized frequency lies below the threshold $\tau_{label}$. We empirically set $\tau_{label}=0.7$. 
This strategy ensures that the final labeling is consistent across views and robust to local segmentation errors.
Once all views have been processed, the resulting labeled point cloud $P_{labeled}$ is re-projected to update 2D masks by merging oversegmented instances, ensuring 3D-consistency across views (Fig.~\ref{fig:Propagation_qualitaives}).

The output of the \textbf{Split} stage is a set of view-consistent instance masks $\tilde{\mathcal{M}}$. The number of masks may differ across views compared to the original masks $\mathcal{M}$, since the object indices are now global. This stage resolves over-segmentation when objects share overlapping mask support across views, ensuring that multiple masks corresponding to the same object are merged. However, when objects do not overlap in any view, residual over-segmentation may persist and can be addressed with a refinement stage (see Sec.~\ref{subsec:mask_refine} for details). Any region of the dense point cloud that remains unlabeled after propagation is treated as background. 

\subsection{Splat: Object Reconstruction}\label{subsec:instancerec}
In the \textbf{Splat} stage (Fig.~\ref{fig:pipeline_splat}), the refined instance masks are used to reconstruct each object independently via Gaussian splatting. A separate 3DGS reconstruction is performed for each object using the masked multi-view images; then, splats are merged back into the full scene (as detailed in Section \ref{subsec:instancemerging}). 

\subsubsection{Per-instance 3D Gaussian splatting} \label{subsec:3dgs}
More in detail, given the refined instance masks $\tilde{\mathcal{M}} \!=\! \{\tilde{\mathcal{M}}_{l}\}_{l=1}^{L}$, grouped by instance label $l$ (independently of the source view $k$), we extract a multi-view image set $\mathcal{I}_l$ for each instance $l$.
Note that the images in $\mathcal{I}_l$ are extracted from $\mathcal{I}$ by masking each instance $l$, ensuring that only that instance remains visible. These images are used to perform instance-specific 3D Gaussian splatting, resulting in a set of $\mathcal{G}_l$ reconstructions. To initialize the splatting algorithm, we select a subset of points from $P_{labeled}$ that have $l$ as their label.

\subsubsection{Mask reprojection and refinement} 
\label{subsec:mask_refine}
After instance reconstruction, we refine the 2D masks to enhance 3D geometric consistency.
For each view $k$ and instance $l$, we extract the set of Gaussians $\mathcal{G}_{k,l}$ visible from view $k$. These sets are used to produce new masks $M^{gs}_{k,l}$ by rendering the scene with full-opacity (\ie, $\alpha=1$).
Then, we apply a greedy 2D sampling procedure (similar to KMeans++) to the projections of $\mathcal{G}_{k,l}$: starting from the point closest to the instance centroid (the average of the Gaussians' positions), we iteratively select the farthest point in the current sample set. This yields a compact and uniformly distributed set of 2D locations that cover the spatial extent of the visible instance. These sampled points are used to prompt the segmenter $S_I(\cdot)$, producing a refined mask ${M}^{sam}_{k,l}$.

\begin{figure*}[htbp]
\centering
\scriptsize

\begin{minipage}{\textwidth}
\centering

\begin{minipage}{\textwidth}
\centering
\hspace{0.2cm} %
\begin{minipage}{0.13\textwidth}\centering Image $I$\end{minipage}%
\begin{minipage}{0.13\textwidth}\centering Rendered mask $M^{gs}$\end{minipage}%
\begin{minipage}{0.13\textwidth}\centering SAM2 mask $M^{sam}$\end{minipage}%
\begin{minipage}{0.13\textwidth}\centering Original mask  $\tilde{M}$\end{minipage}%
\hspace{0.9cm}
\begin{minipage}{0.13\textwidth}\centering Image $I$\end{minipage}%
\begin{minipage}{0.13\textwidth}\centering Rendered mask $M^{gs}$\end{minipage}%
\begin{minipage}{0.13\textwidth}\centering SAM2 mask $M^{sam}$\end{minipage}%
\end{minipage}\\

\begin{minipage}{0.52\textwidth}
\centering
\begin{tabular}{@{}p{0cm}@{}p{\textwidth}@{}}

\makebox[0pt][r]{%
\rotatebox{90}{\shortstack{\textbf{LERF/teatime}\\``old camera"}}%
\hspace{0.4em}} &
\includegraphics[trim=0cm 9.2cm 0cm 0cm,
    clip,
    width=\linewidth
]{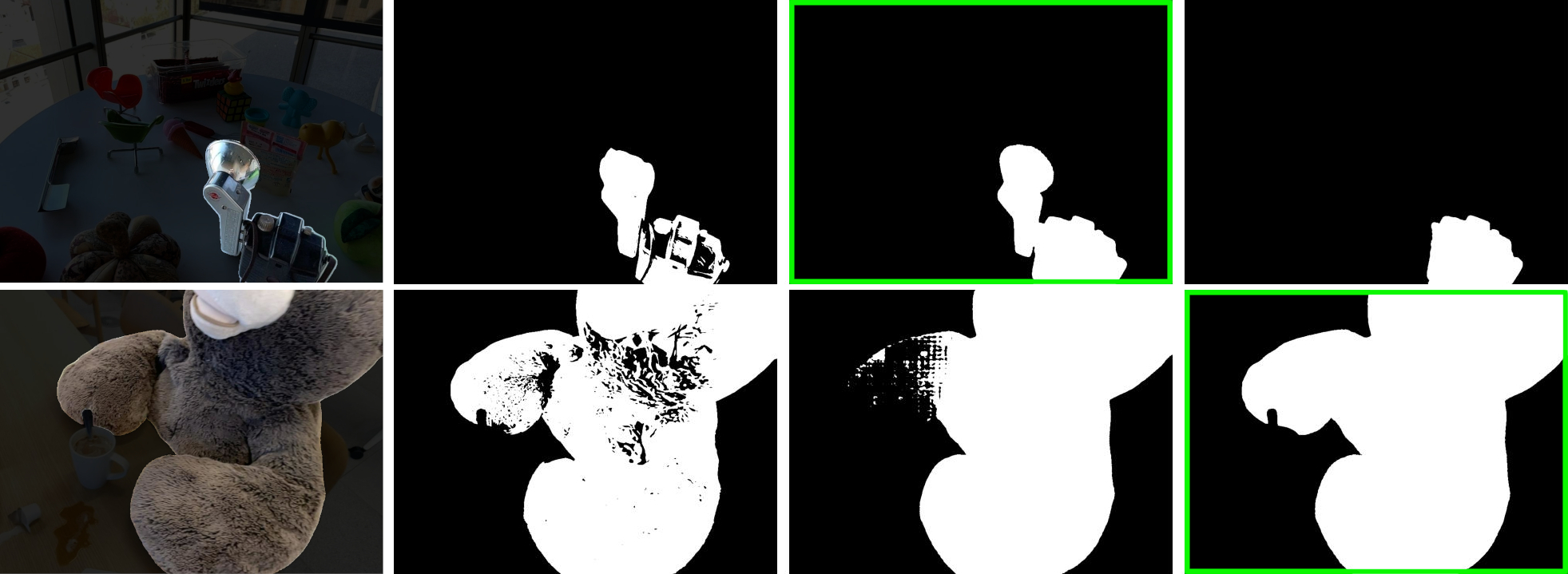} \\[0.1em]

\makebox[0pt][r]{%
\rotatebox{90}{\shortstack{\textbf{LERF/figurines}\\``stuffed bear"}}%
\hspace{0.4em}} &
\includegraphics[
    trim=0cm 0cm 0cm 9.2cm,
    clip,
    width=\linewidth
]{figs/qual/qual_3/ref_1.jpg}

\end{tabular}
\end{minipage}%
\hspace{1cm}%
\begin{minipage}{0.39\textwidth}
\centering
\begin{tabular}{@{}p{0cm}@{}p{\textwidth}@{}}

\makebox[0pt][r]{%
\rotatebox{90}{\shortstack{\textbf{LERF/ramen}\\``sake bottle"}}%
\hspace{0.4em}} &
\includegraphics[
    trim=0cm 10.5cm 0cm 0cm,
    clip,
    width=\linewidth
]{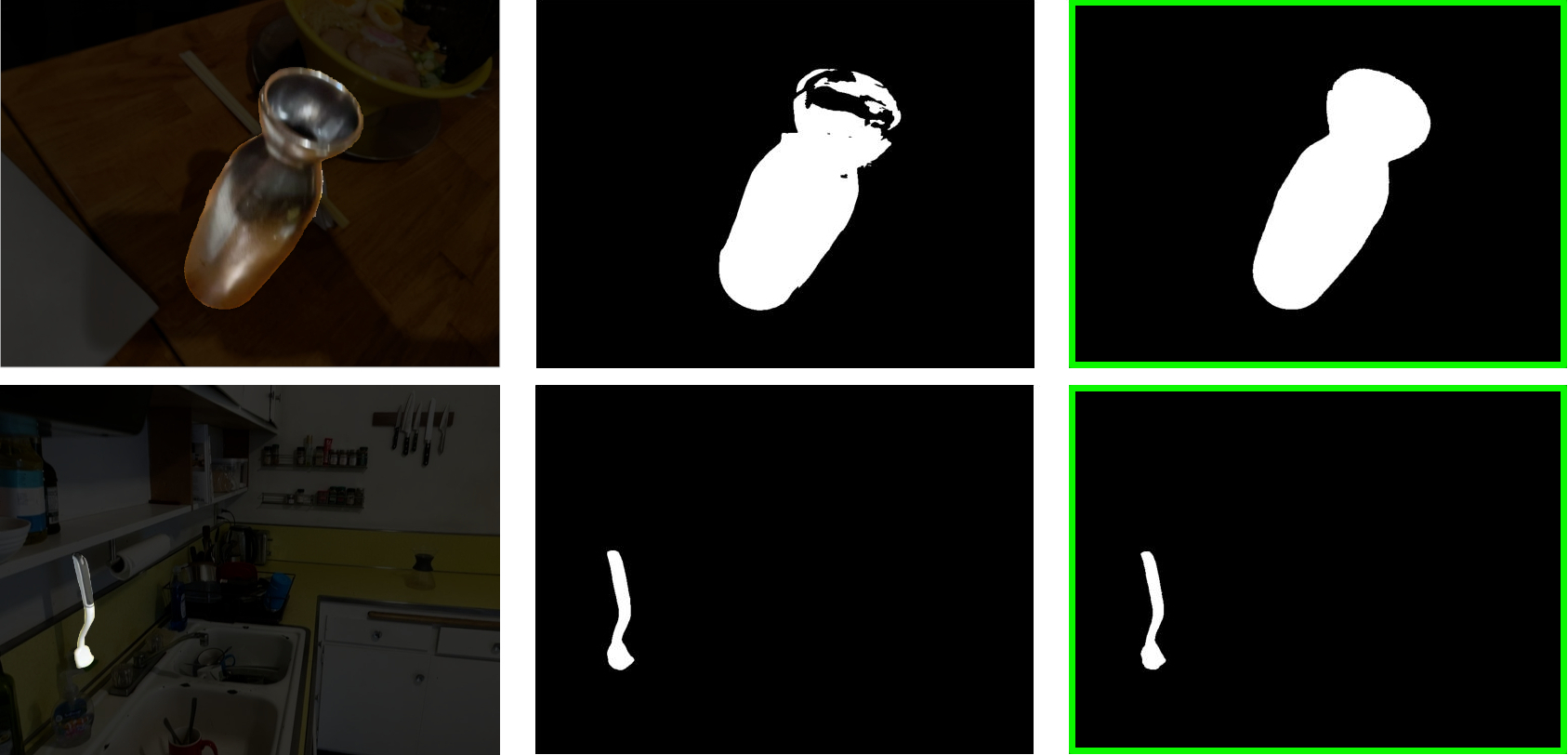} \\[0.1em]

\makebox[0pt][r]{%
\rotatebox{90}{\shortstack{\textbf{LERF/waldo}\\ \textbf{kitchen}``spatula"}}%
\hspace{0.4em}} &
\includegraphics[
    trim=0cm 0cm 0cm 10.5cm,
    clip,
    width=\linewidth
]{figs/qual/qual_3/ref_2_inverted.jpg}

\end{tabular}
\end{minipage}

\end{minipage}

\Description{Qualitative Results on LERF/ScanNet datasets}
\caption{Examples of mask refinement process. 
For ``old camera'' the refined mask is $M^{sam}$ while for ``stuffed bear'' $\tilde{M}$ present a lower IoU with $M^{gs}$. For ``sake bottle'' and ``spatula'', $\tilde{M}$ is not present and is discovered throughout the refinement process. Selected masks for each object are highlighted with \textcolor{green}{\textbf{green}} borders.}
\label{fig:mask_ref_qualitatives}
\end{figure*}

At this stage, we have two candidate segmentation masks: the propagated mask from the initial segmentation stage $\tilde{M}_{k,l}$ and the refined mask ${M}^{sam}_{k,l}$. We compare these masks by computing the Intersection-over-Union (IoU) with mask ${M}^{gs}_{k,l}$, rendered from the reconstructed Gaussians. If $\tilde{M}_{k,l}$ exists, we prefer it over $M^{sam}_{k,l}$ whenever its IoU is higher.
If it is not available (\eg, due to missed detections or because it has been removed in the geometry consistency check of Sec.~\ref{subsec:mask_cons}), we keep ${M}^{sam}_{k,l}$ only if its IoU with ${M}^{gs}_{k,l}$ exceeds a threshold $\tau_{iou}$ (empirically set to $0.95$).
This process yields a final refined mask $\hat{M}_{k,l}$ that is view-consistent and coherent with the scene geometry, improving segmentation quality, especially in cases where objects are small, occluded, or visually ambiguous (visual examples are provided in Fig.~\ref{fig:mask_ref_qualitatives}).

\subsubsection{Instance merging} \label{subsec:instancemerging}
To assemble a complete scene reconstruction from the independent instances $\mathcal{G}_{l}$ while preserving object boundaries, we merge them based on spatial overlap. 
For each instance $l$, we compute an axis-aligned (in the world reference system) 3D bounding box $B_l$ around $\mathcal{G}_{l}$ and  define a collision matrix $C \in \mathbb{R}^{L \times L}$, where each entry $C_{ab}$  measures the amount of overlap between the two instances $\mathcal{G}_{a}$ and $\mathcal{G}_{b}$: %
\begin{equation}
C_{ab} = \frac{1}{|\mathcal{G}_{a}|}\sum_{g \in \mathcal{G}_{a}} \mathds{1}_{B_{b}}(g) \;\;.
\end{equation}
Note that we normalize the entries of $C$ to allow comparison between instances of different sizes.

We iteratively select the pair of instances $(a, b)$ with the highest normalized overlap and merge them until a GS representation of the complete scene is achieved.
The merging is performed directly at the Gaussian level by merging the sets $\mathcal{G}_{a}$ and $\mathcal{G}_{b}$.
Since each instance was reconstructed from masked views, boundary regions may contain occlusion artifacts, such as dark or oversized Gaussians. To suppress them, we reinitialize the opacity of all Gaussians in the merged object to $\alpha = 0$ (following \cite{kerbl3Dgaussians}), and run a short Gaussian splatting refinement with densification disabled. 
This stage allows for the smoothing of collision borders and the removal of Gaussians that correspond to occlusions.

To refine object boundaries during merging, we include a mask consistency term in the reconstruction objective:
\begin{equation}
\ell = \ell_{RGB} + w_{mask} \cdot \ell_{mask},
\end{equation}
where $\ell_{RGB}$ is the standard GS optimization objective, while $\ell_{mask}$ encourages alignment between the rendered instance mask $M^{gs}_{k,l}$ and the chosen per-view refined mask $\hat{M}_{k,l}$ (Sec.~\ref{subsec:mask_refine}):
\begin{equation}
    \ell_{mask} = \frac{1}{KL} \sum^{K}_{k=1} \sum^{L}_{l=1} \left\lVert M^{gs}_{k,l} - \hat{M}_{k,l} \right\rVert_1.
\end{equation}

The weight $w_{mask}$ is gradually increased after each merge, as detailed in Sec.~\ref{sec:impl_details}, progressively reinforcing boundary sharpness and suppressing residual occlusions.
We merge instance pairs in parallel whenever they do not share collisions, update the collision matrix, and repeat until no further merges are required. This produces the final full-scene Gaussian representation in which all instances are jointly reconstructed, and boundary transitions are geometrically and semantically consistent.
The result of the \textbf{Splat} stage is a fully assembled scene in which object boundaries are geometrically consistent, and each Gaussian carries its corresponding instance label. While our approach is not explicitly designed for open vocabulary segmentation, Sec~\ref{sec:results-openvoc} explains how we adapted it to this task by assigning a semantic descriptor $f_{obj}$ to each $\mathcal{G}_{l}$ computed from a multi-view feature extractor $E_F(\cdot)$ (implemented using CLIP~\cite{clip}, see Fig.~\ref{fig:pipeline_splat}).

\section{Implementation Details} \label{sec:impl_details}
\begin{figure*}[t]
    \centering
    \scriptsize

    \begin{minipage}{\textwidth}
        \centering
        \begin{minipage}{.18\textwidth}
            \centering      
            Coarser \textit{ping} grid
        \end{minipage}%
        \begin{minipage}{.18\textwidth}
            \centering
             Coarse \textit{ping} grid
        \end{minipage}%
        \begin{minipage}{.18\textwidth}
            \centering
           Medium \textit{ping} grid
        \end{minipage}%
        \begin{minipage}{.18\textwidth}
            \centering
            Fine \textit{ping} grid
        \end{minipage}%
        \begin{minipage}{.18\textwidth}
            \centering
             \textbf{Final masks}
        \end{minipage}%
    \end{minipage}\\[2mm]

    \begin{minipage}{\textwidth}
        \makebox[0pt][r]{%
            \raisebox{-0.5\height}{\rotatebox{90}{LERF/Figurines}}%
            \hspace{2mm}%
        }%
        \centering
        \begin{minipage}{.18\textwidth}
            \includegraphics[width=\textwidth]{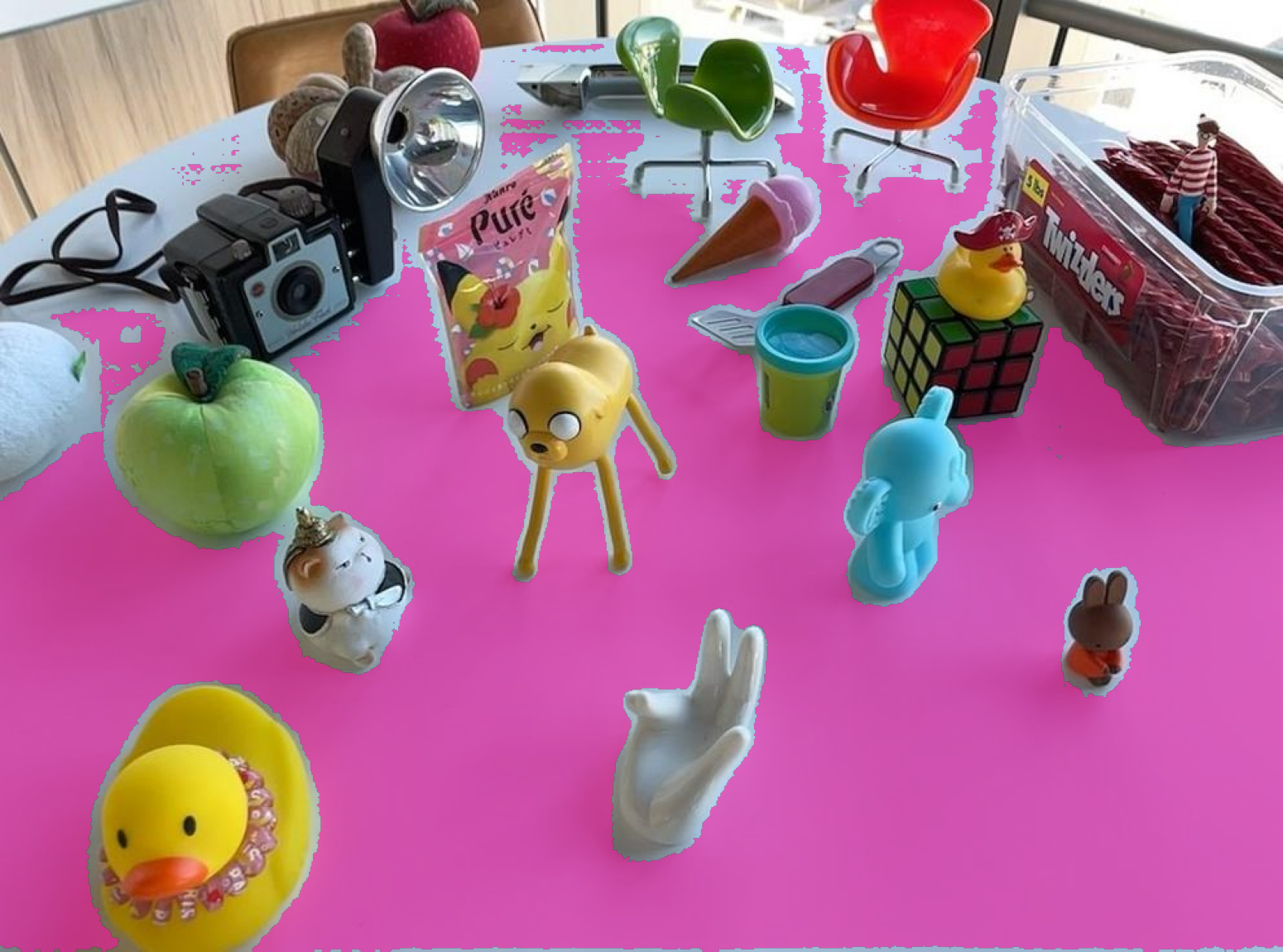} %
        \end{minipage}
        \begin{minipage}{.18\textwidth}
            \includegraphics[width=\textwidth]{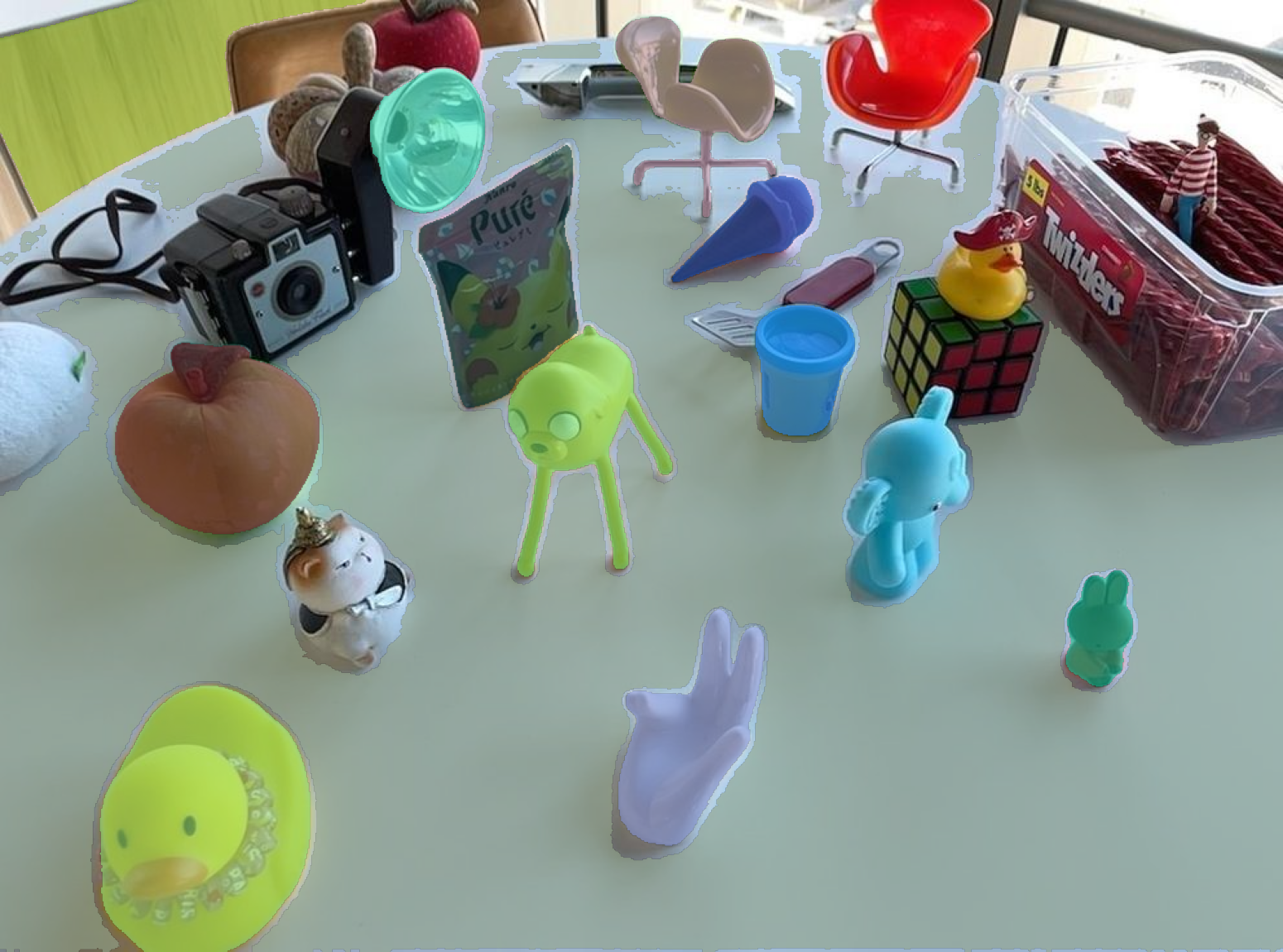} %
        \end{minipage}
        \begin{minipage}{.18\textwidth}
            \includegraphics[width=\textwidth]{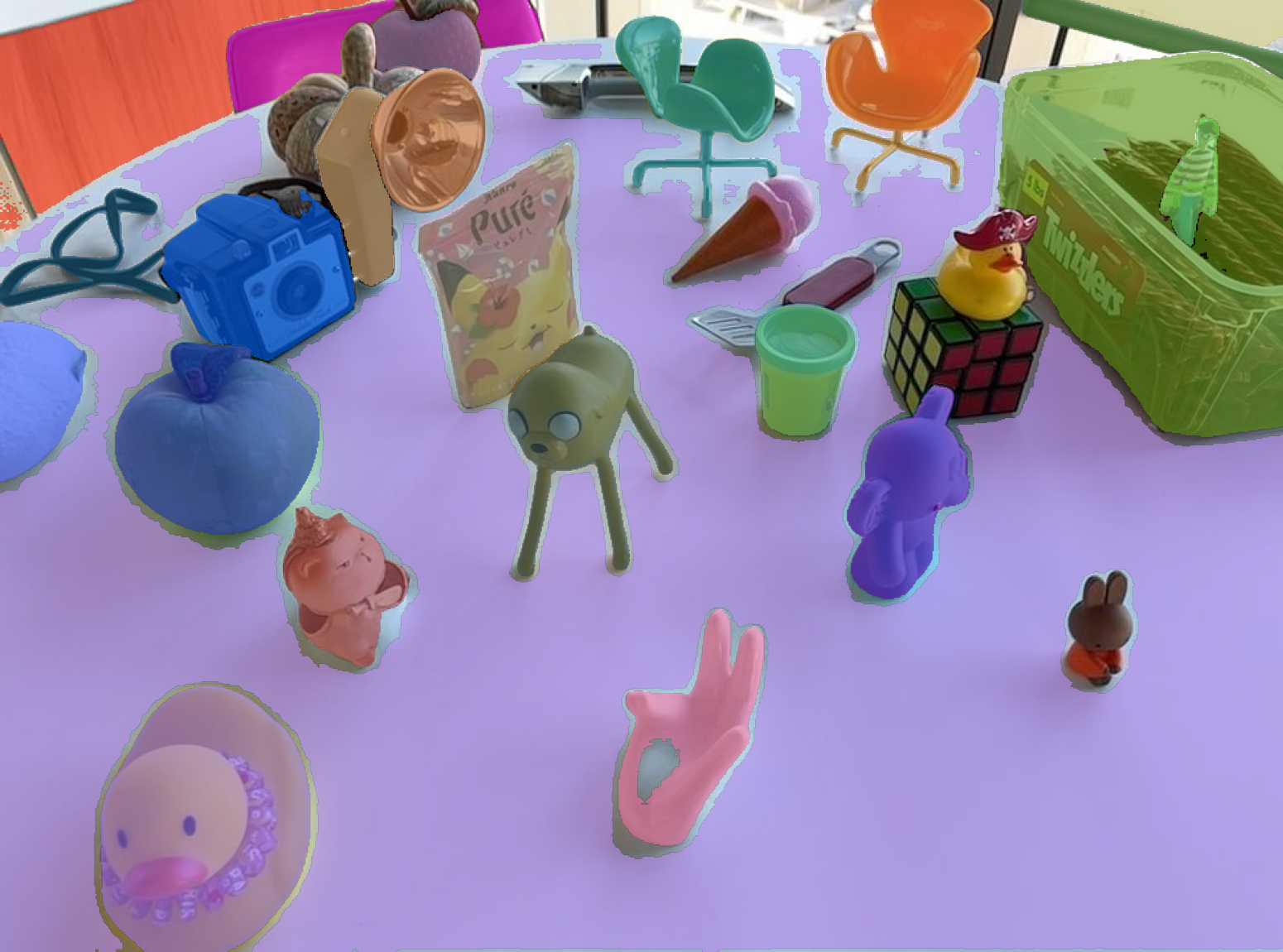} %
        \end{minipage}
        \begin{minipage}{.18\textwidth}
            \includegraphics[width=\textwidth]{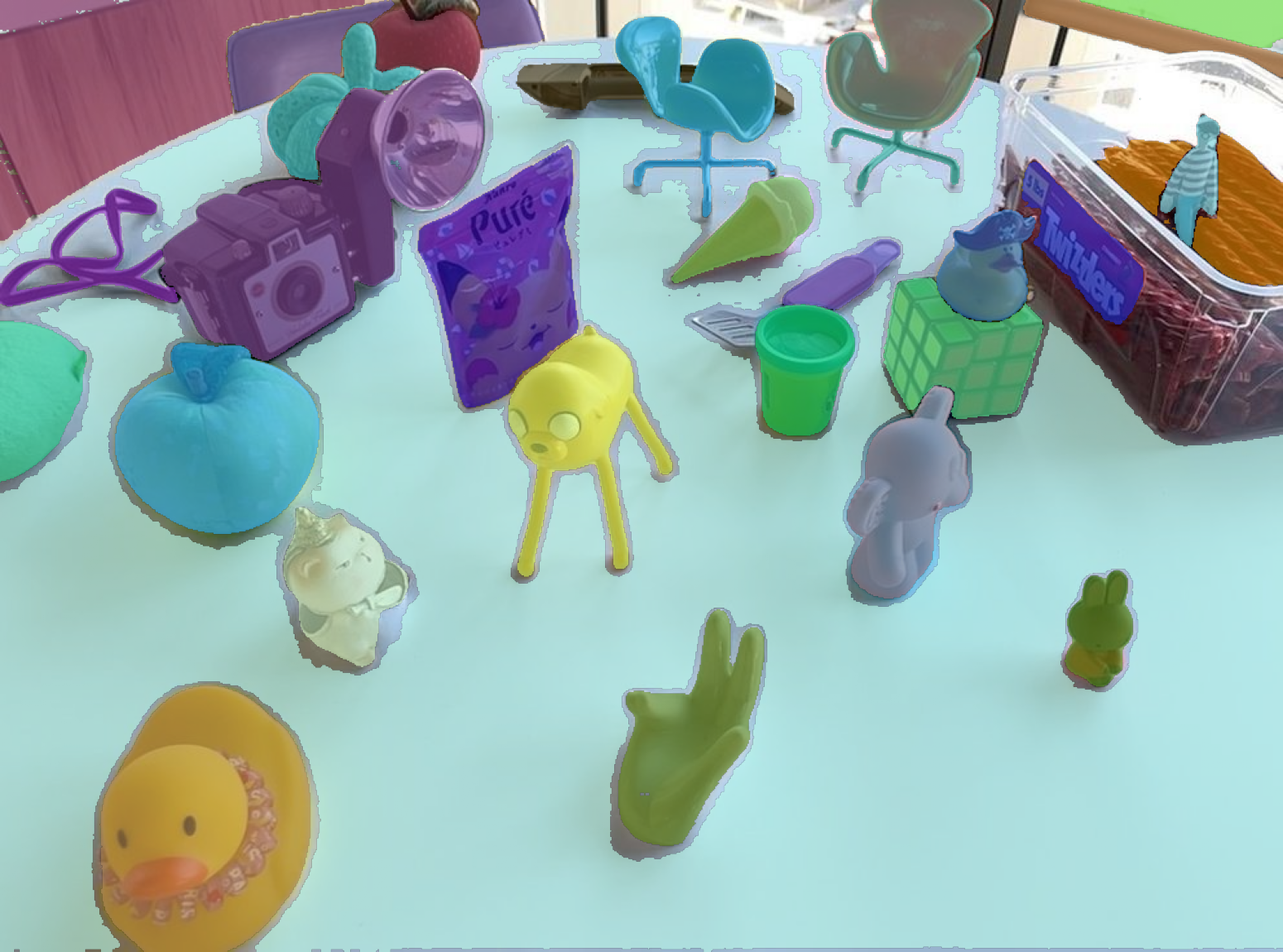} %
        \end{minipage}
        \begin{minipage}{.18\textwidth}
            \includegraphics[width=\textwidth]{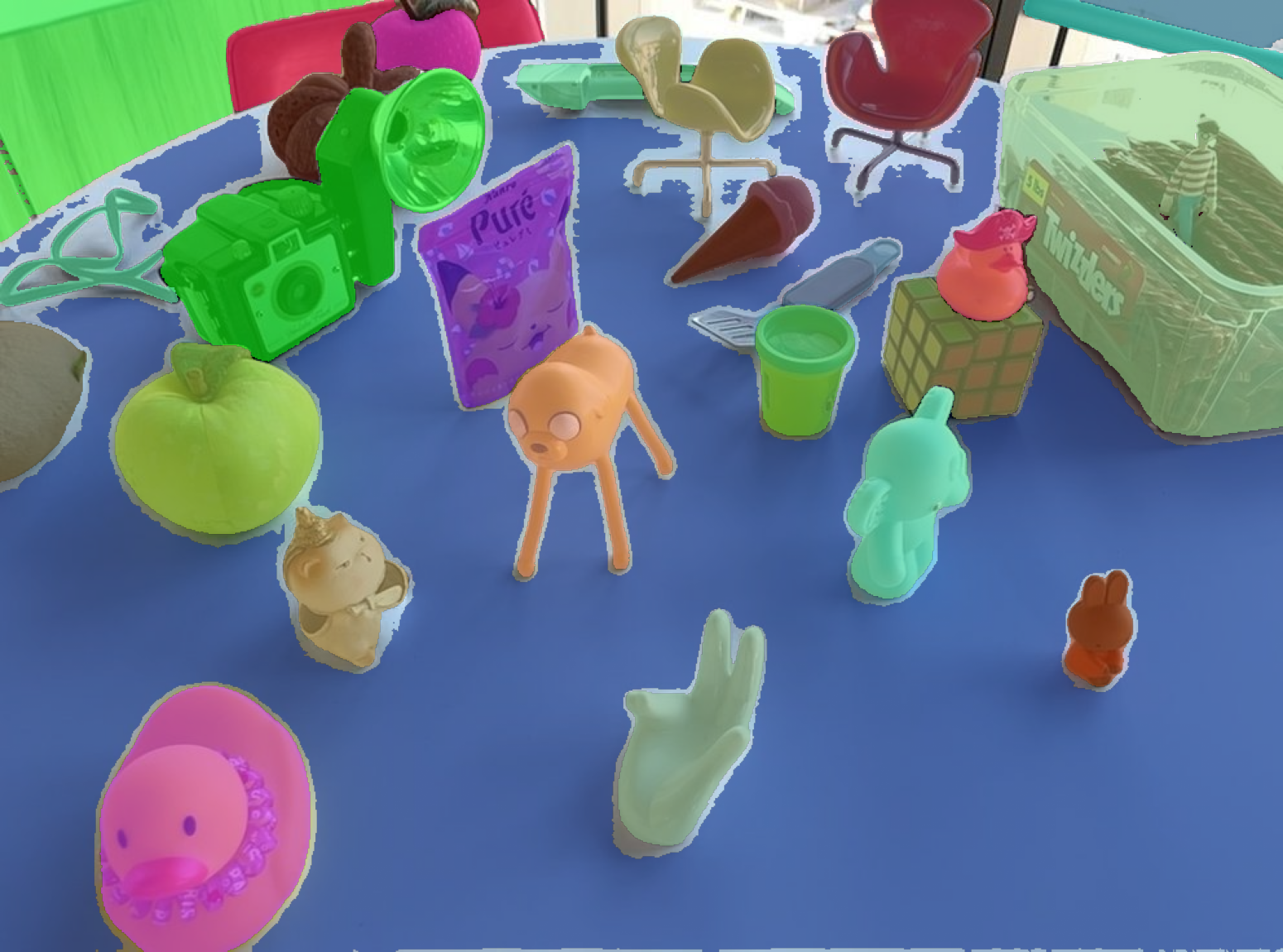} %
        \end{minipage}
    \end{minipage}\\[1mm]

    \begin{minipage}{\textwidth}
        \makebox[0pt][r]{%
            \raisebox{-0.3\height}{\rotatebox{90}{LERF/Teatime}}%
            \hspace{2mm}%
        }%
        \centering
        \begin{minipage}{.18\textwidth}
            \includegraphics[width=\textwidth]{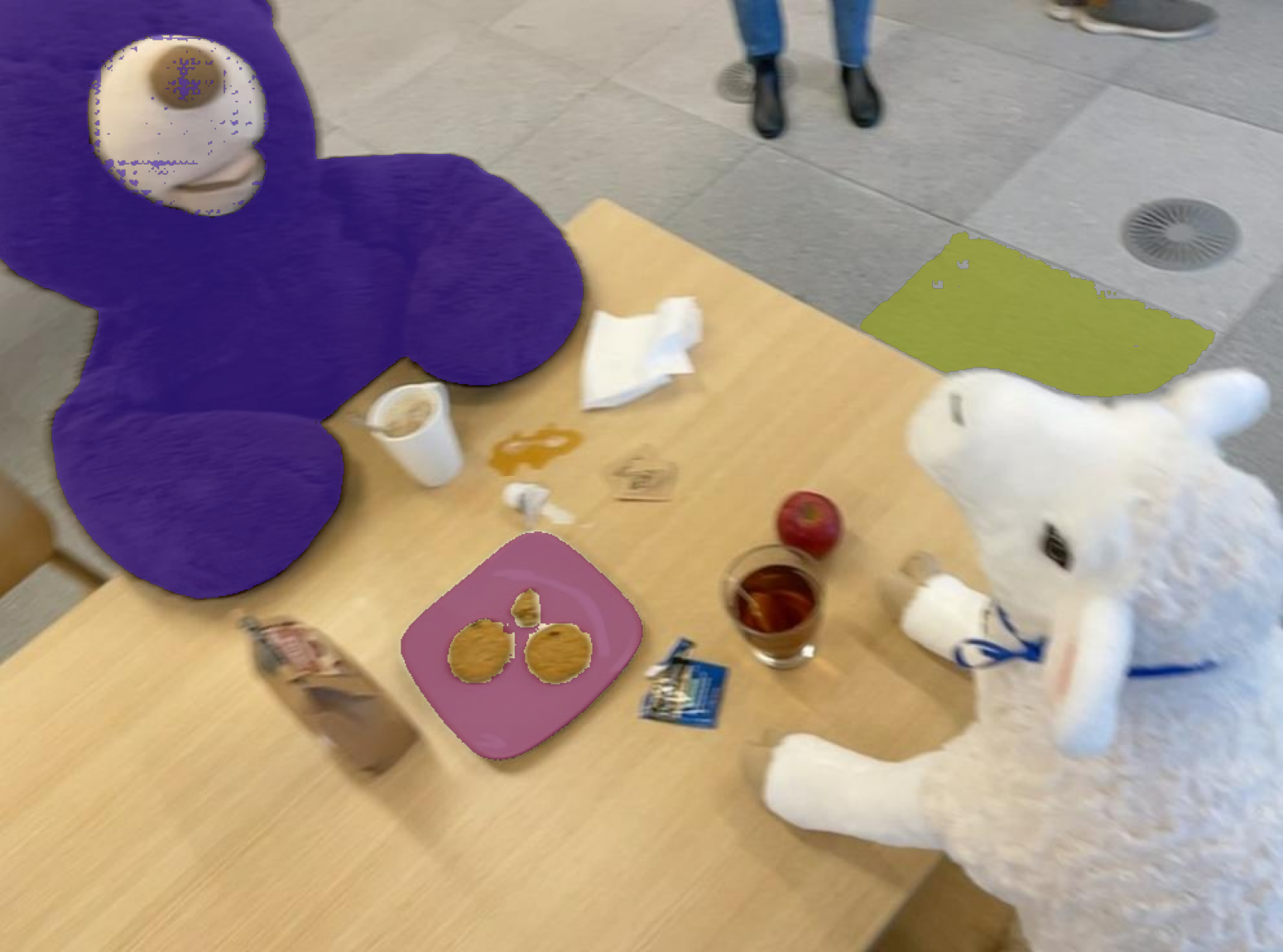} %
        \end{minipage}
        \begin{minipage}{.18\textwidth}
            \includegraphics[width=\textwidth]{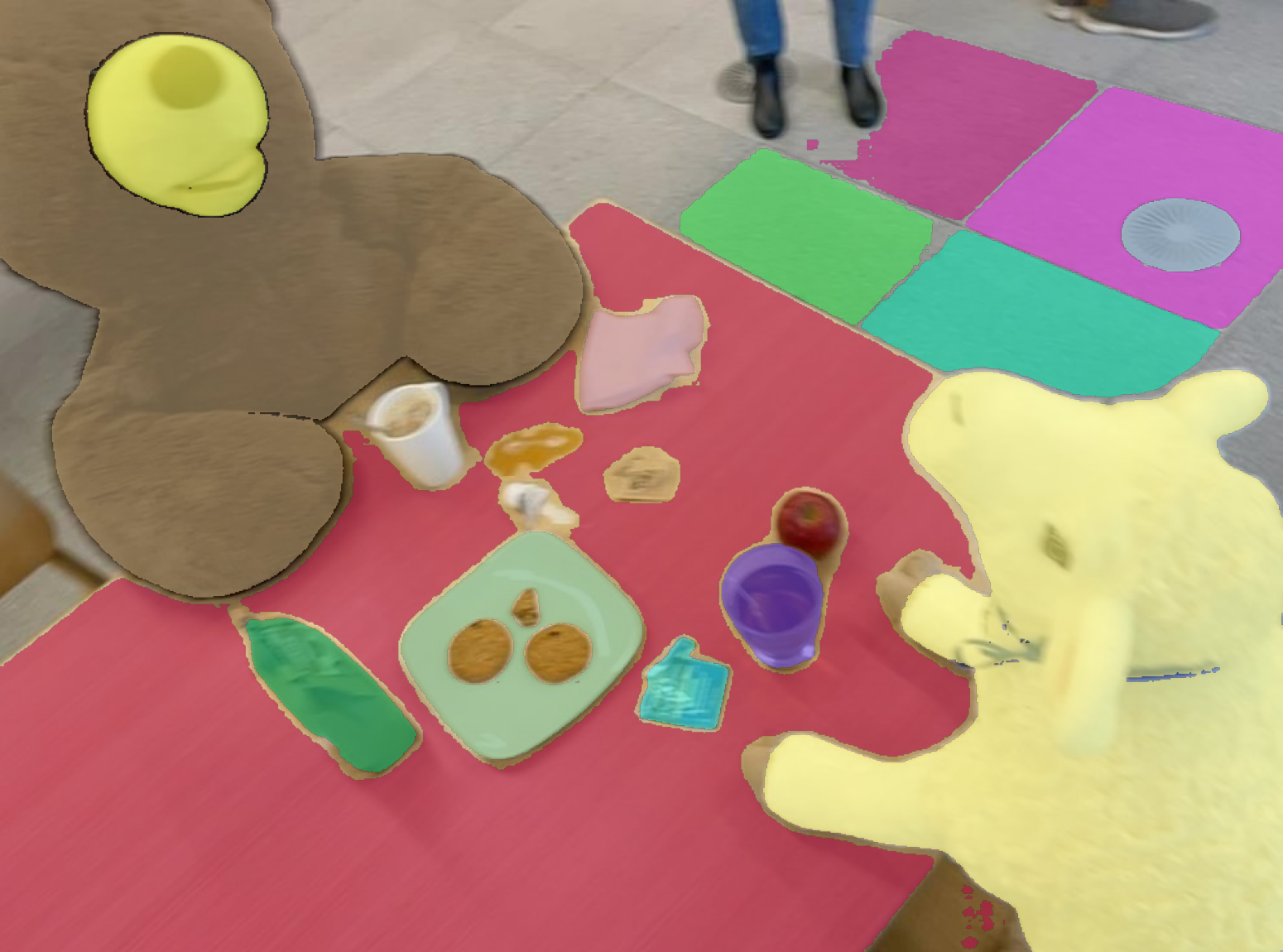} %
        \end{minipage}
        \begin{minipage}{.18\textwidth}
            \includegraphics[width=\textwidth]{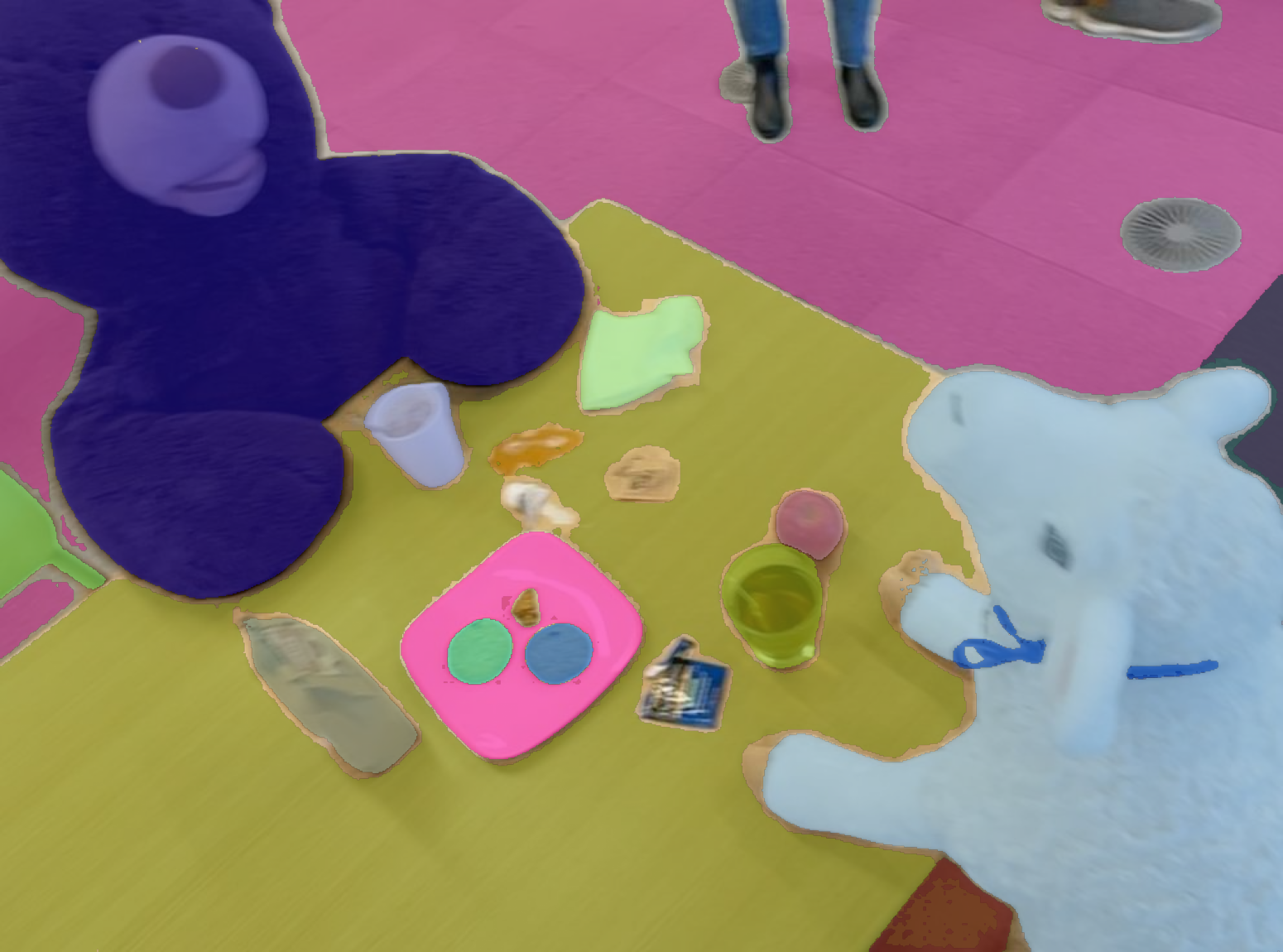} %
        \end{minipage}
        \begin{minipage}{.18\textwidth}
            \includegraphics[width=\textwidth]{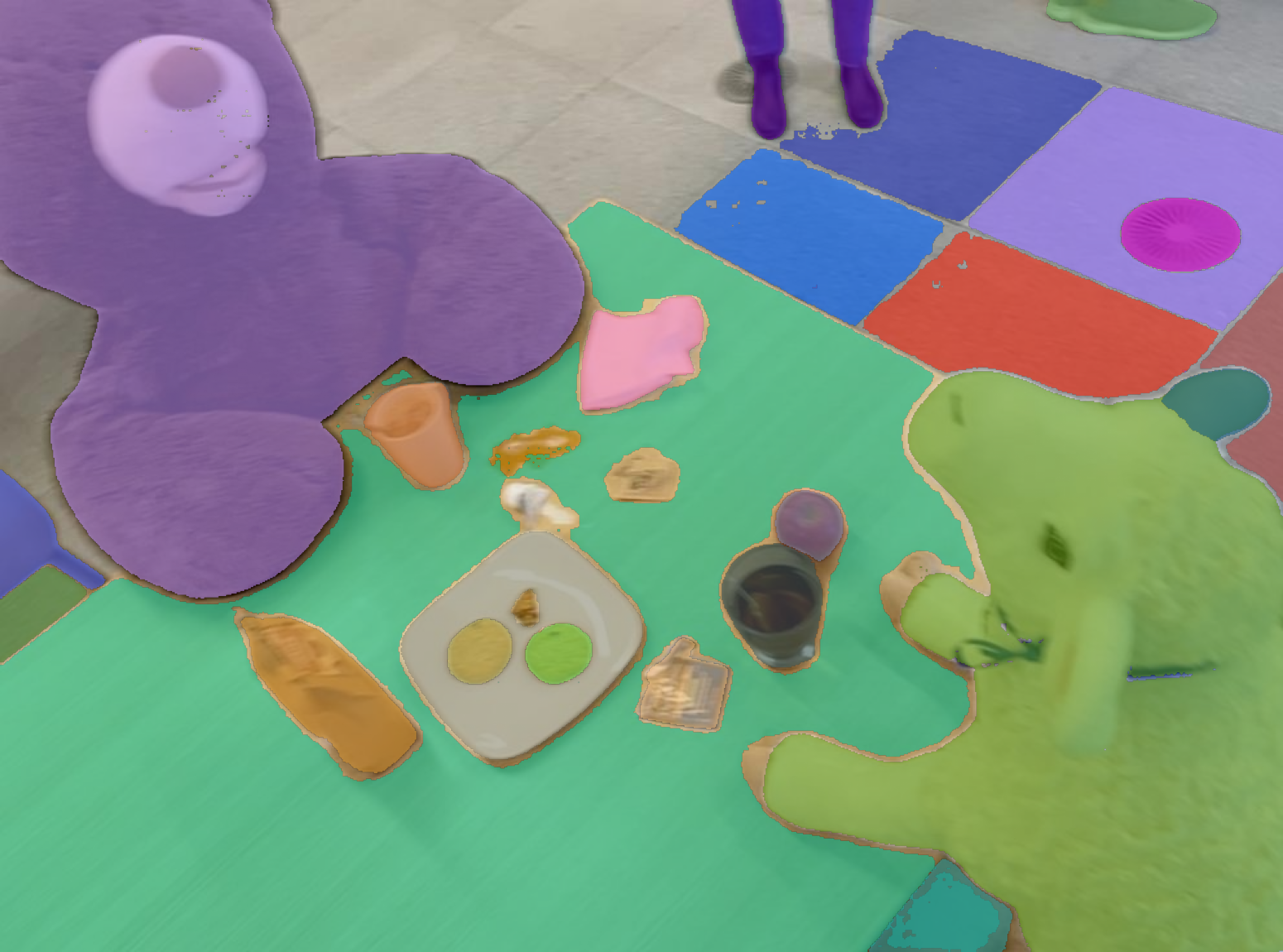} %
        \end{minipage}
        \begin{minipage}{.18\textwidth}
            \includegraphics[width=\textwidth]{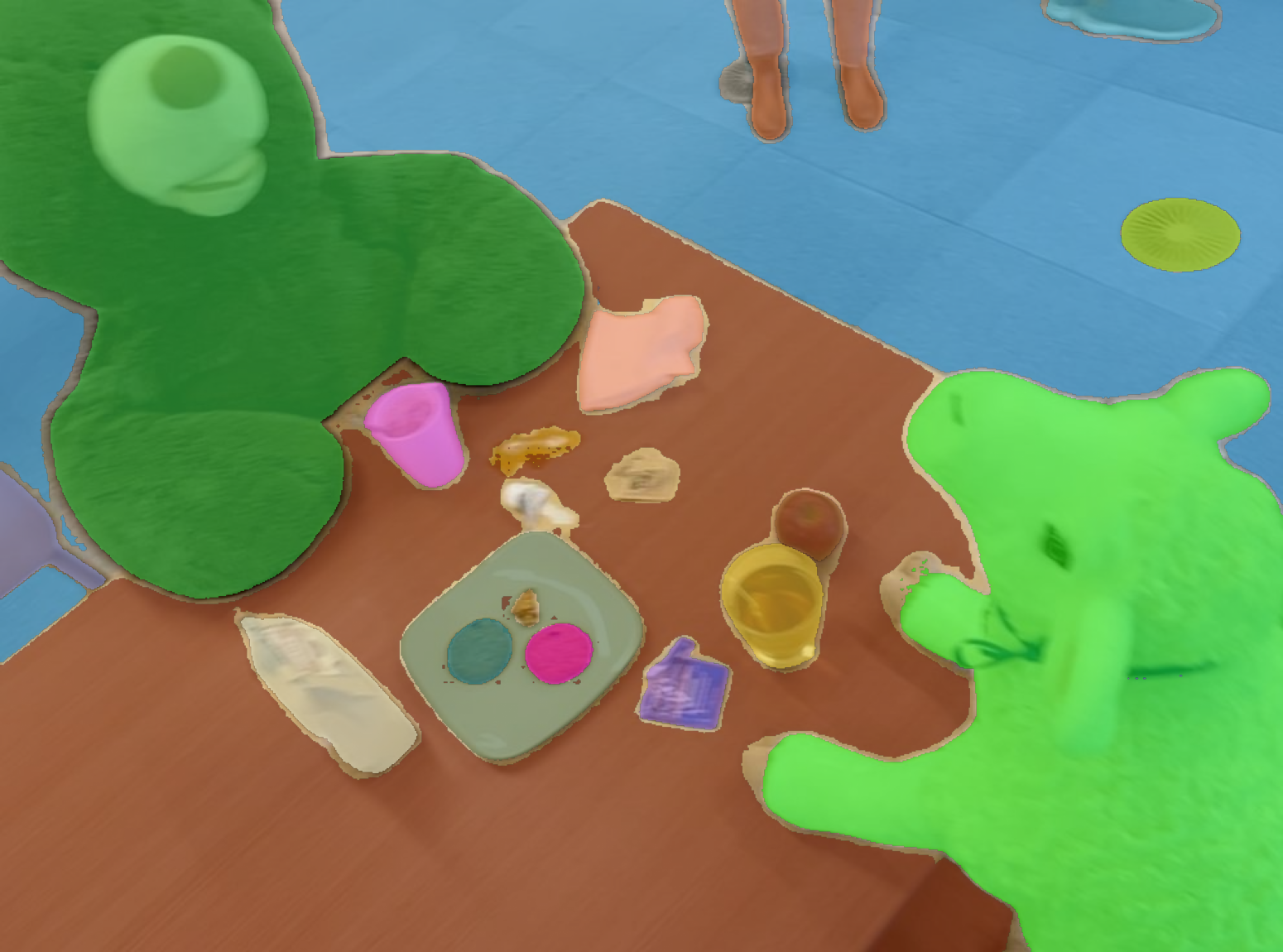} %
        \end{minipage}
    \end{minipage}\\[1mm]

    \begin{minipage}{\textwidth}
        \makebox[0pt][r]{%
            \raisebox{-0.3\height}{\rotatebox{90}{ScanNetv2/0062\_00}}%
            \hspace{2mm}%
        }%
        \centering
        \begin{minipage}{.18\textwidth}
            \includegraphics[width=\textwidth]{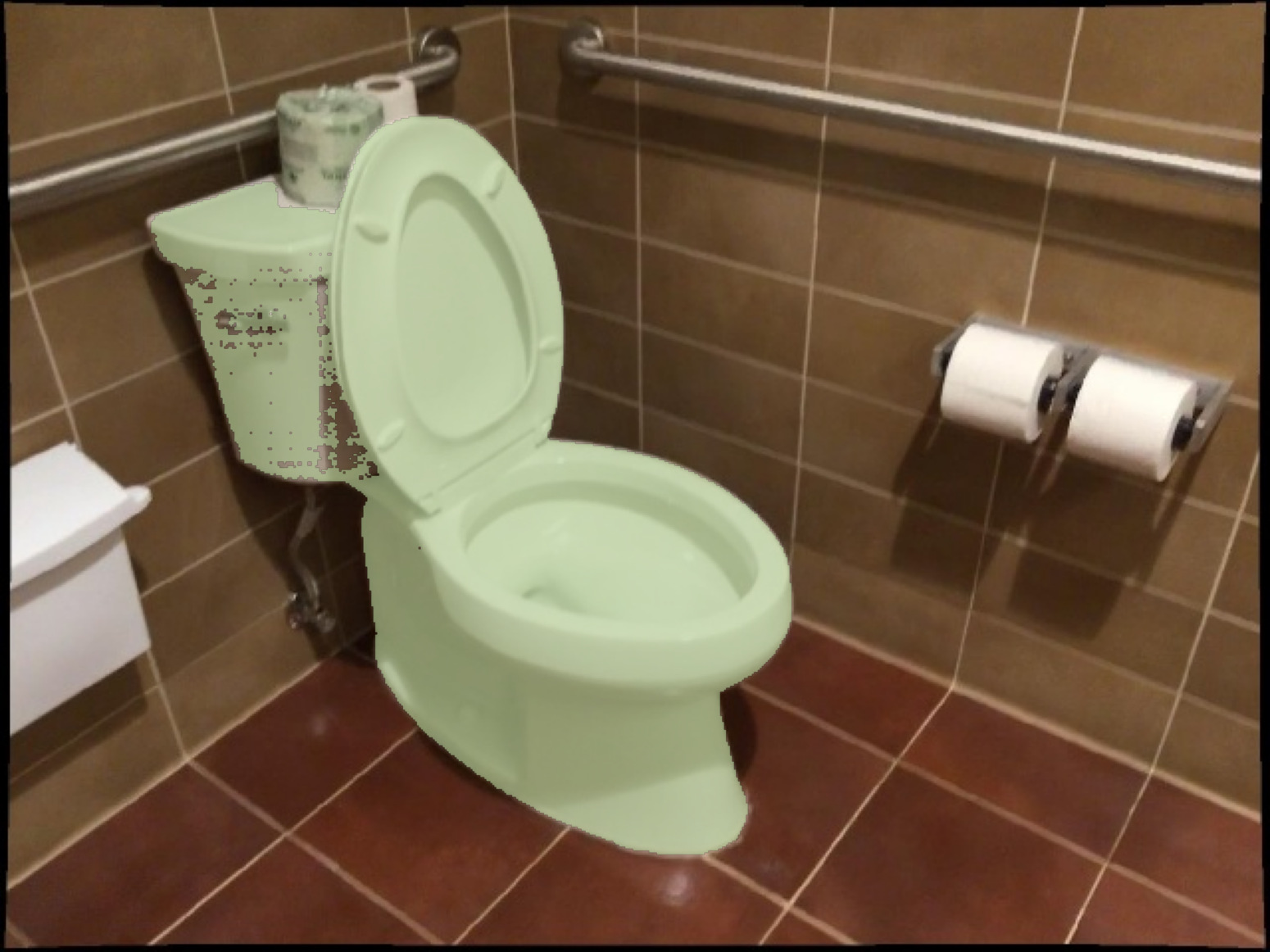} %
        \end{minipage}
        \begin{minipage}{.18\textwidth}
            \includegraphics[width=\textwidth]{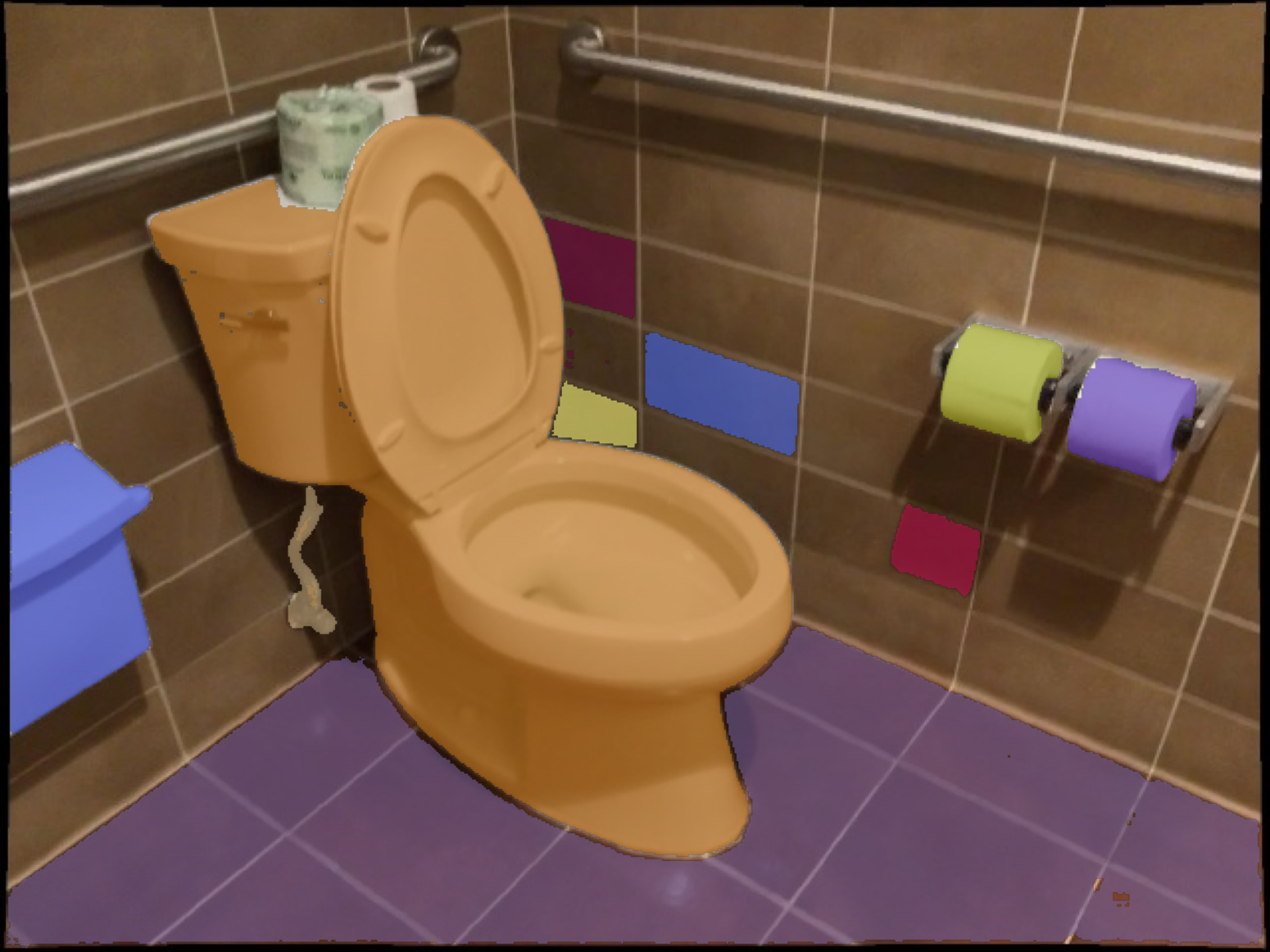} %
        \end{minipage}
        \begin{minipage}{.18\textwidth}
            \includegraphics[width=\textwidth]{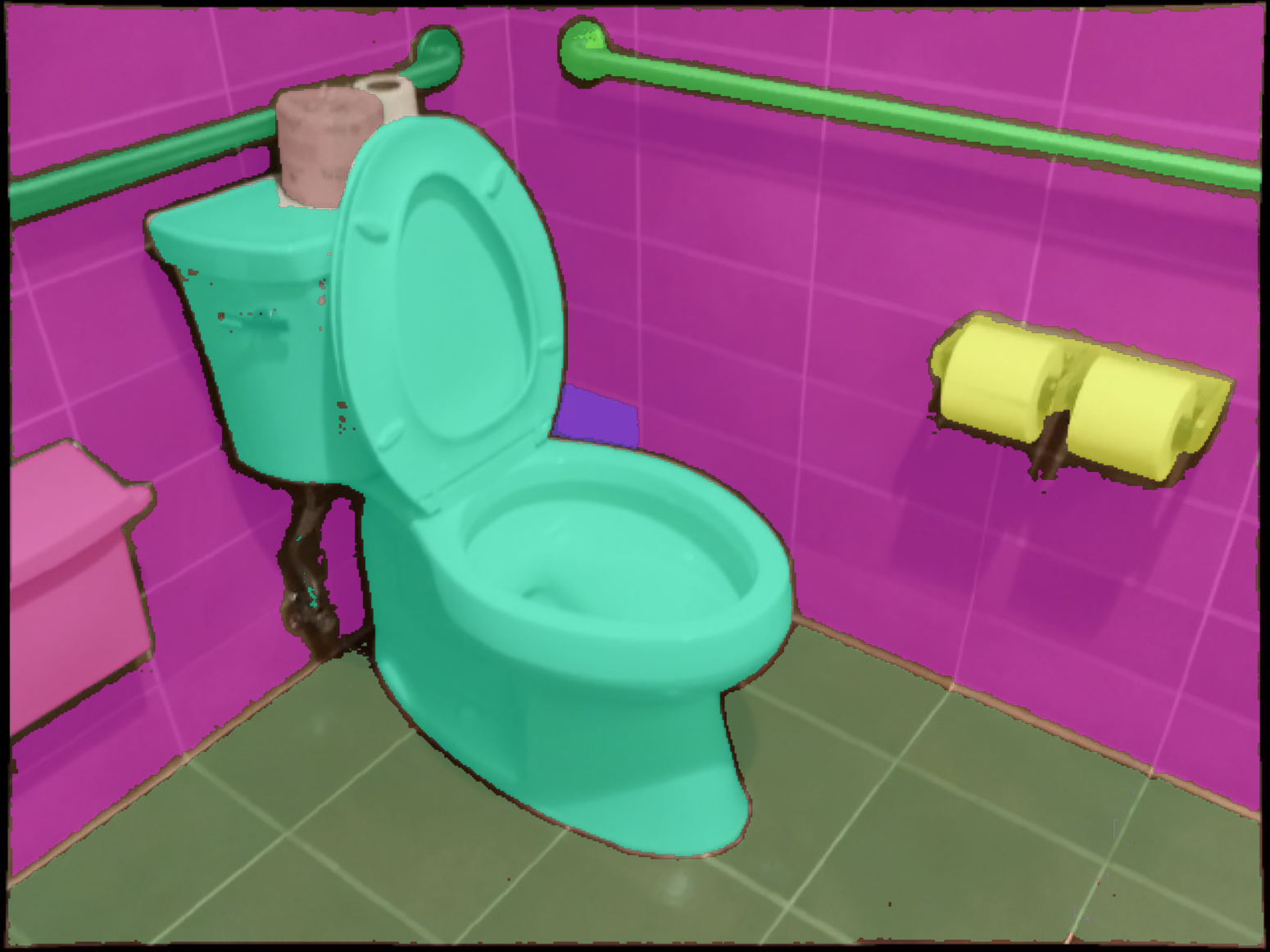} %
        \end{minipage}
        \begin{minipage}{.18\textwidth}
            \includegraphics[width=\textwidth]{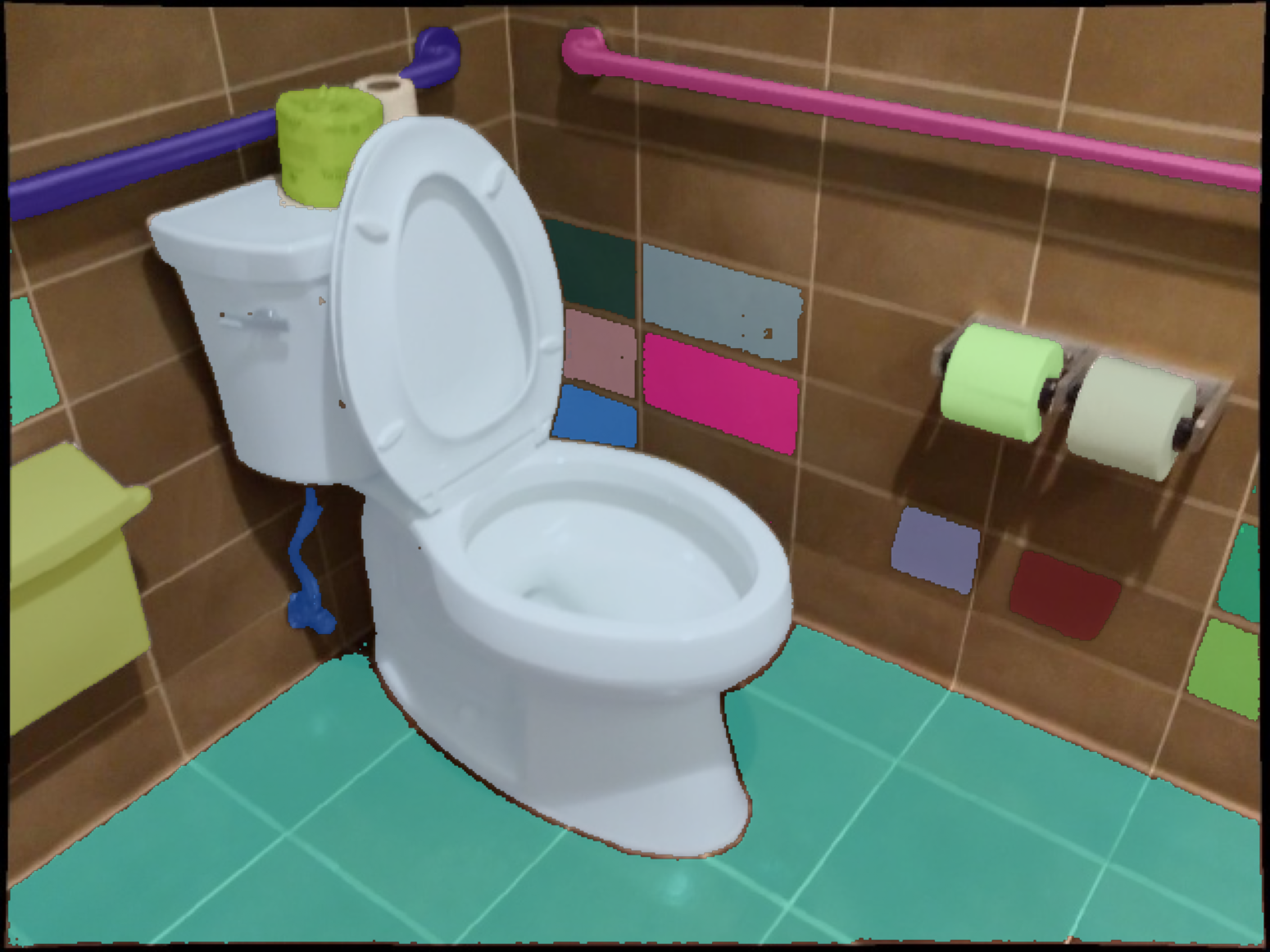} %
        \end{minipage}
        \begin{minipage}{.18\textwidth}
            \includegraphics[width=\textwidth]{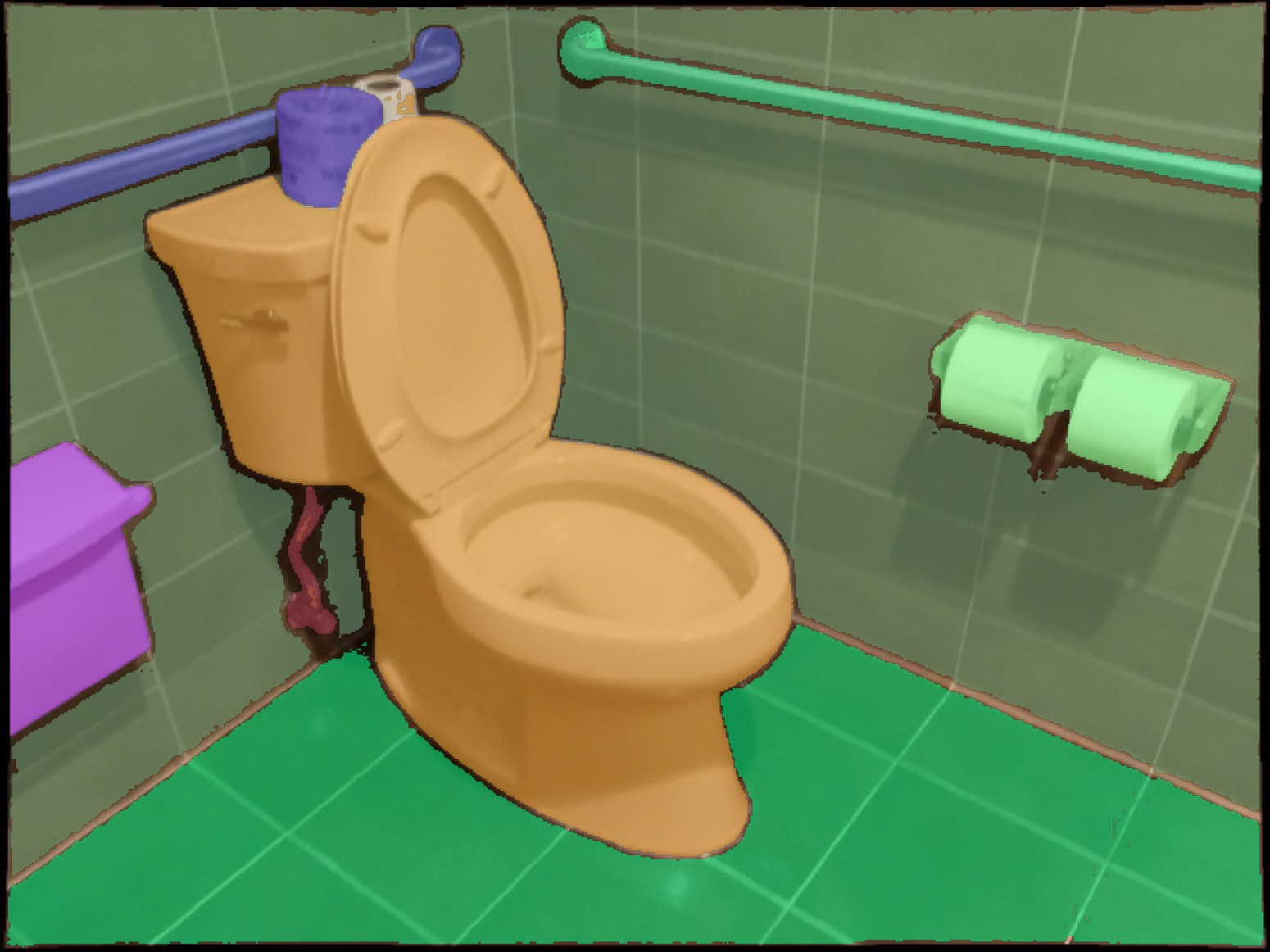} %
        \end{minipage}
    \end{minipage}\\[1mm]

    \begin{minipage}{\textwidth}
         \makebox[0pt][r]{%
            \raisebox{-0.3\height}{\rotatebox{90}{ScanNetv2/0347\_00}}%
            \hspace{2mm}%
        }%
        \centering
        \begin{minipage}{.18\textwidth}
            \includegraphics[width=\textwidth]{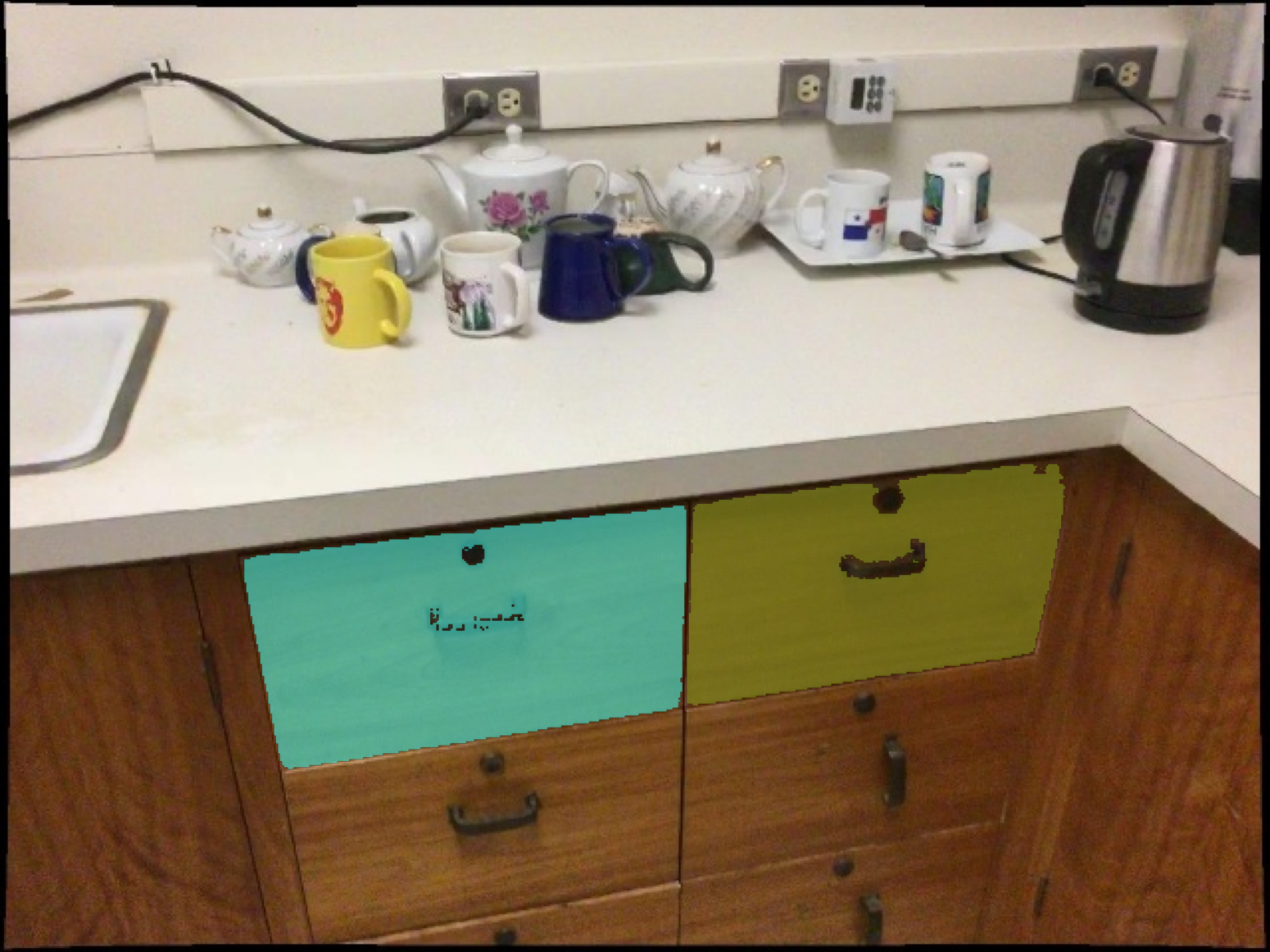} %
        \end{minipage}
        \begin{minipage}{.18\textwidth}
            \includegraphics[width=\textwidth]{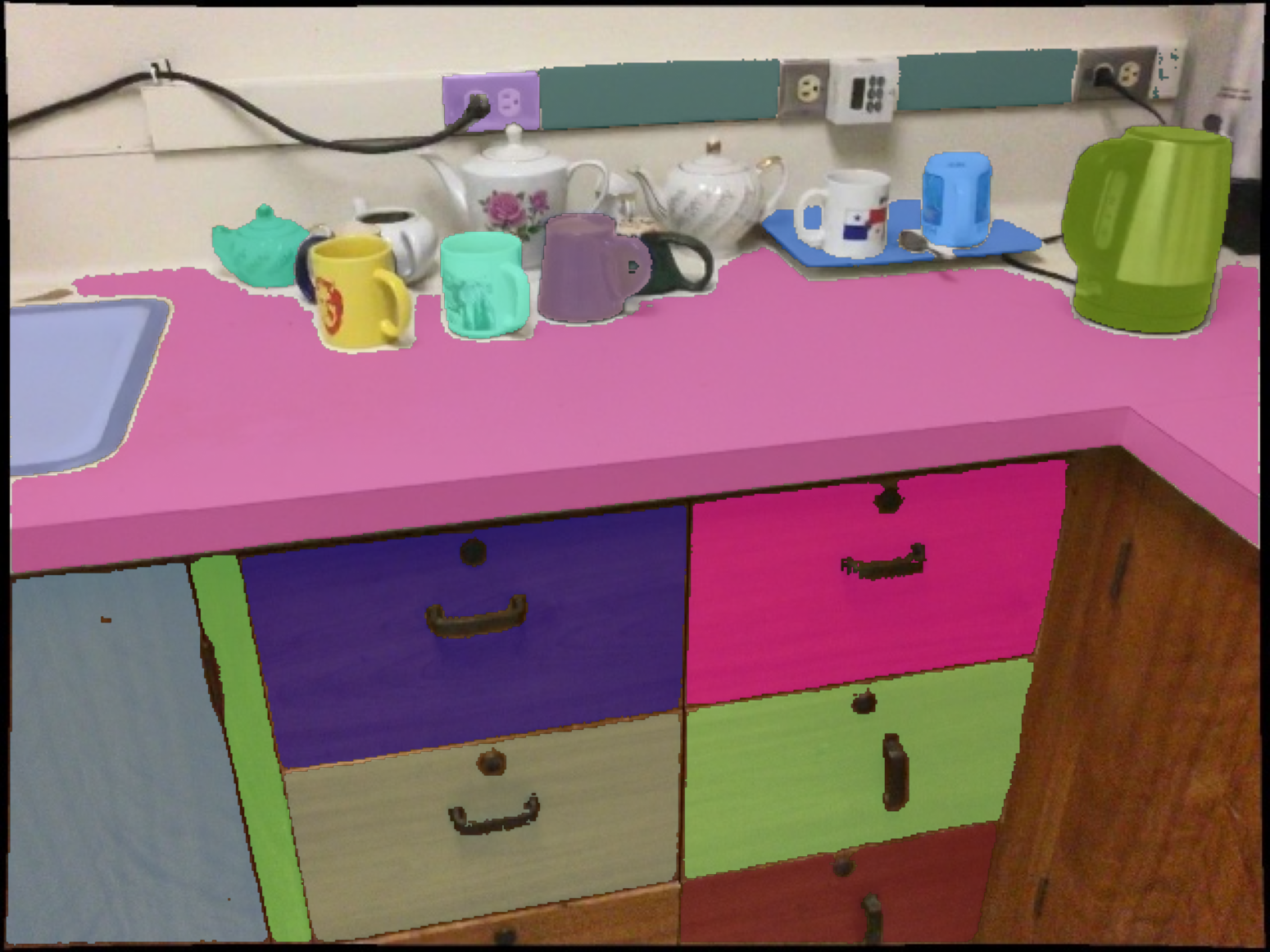} %
        \end{minipage}
        \begin{minipage}{.18\textwidth}
            \includegraphics[width=\textwidth]{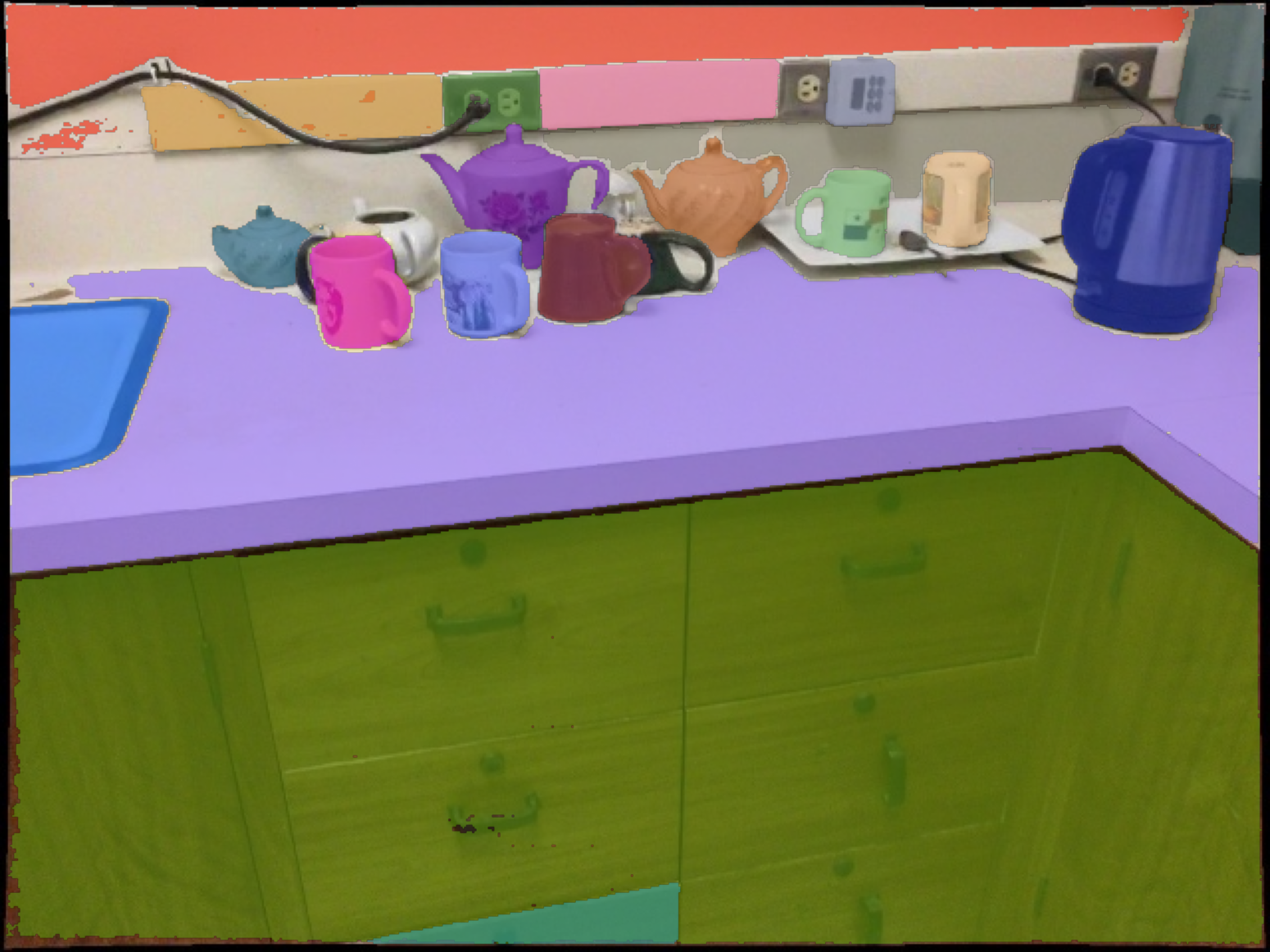} %
        \end{minipage}
        \begin{minipage}{.18\textwidth}
            \includegraphics[width=\textwidth]{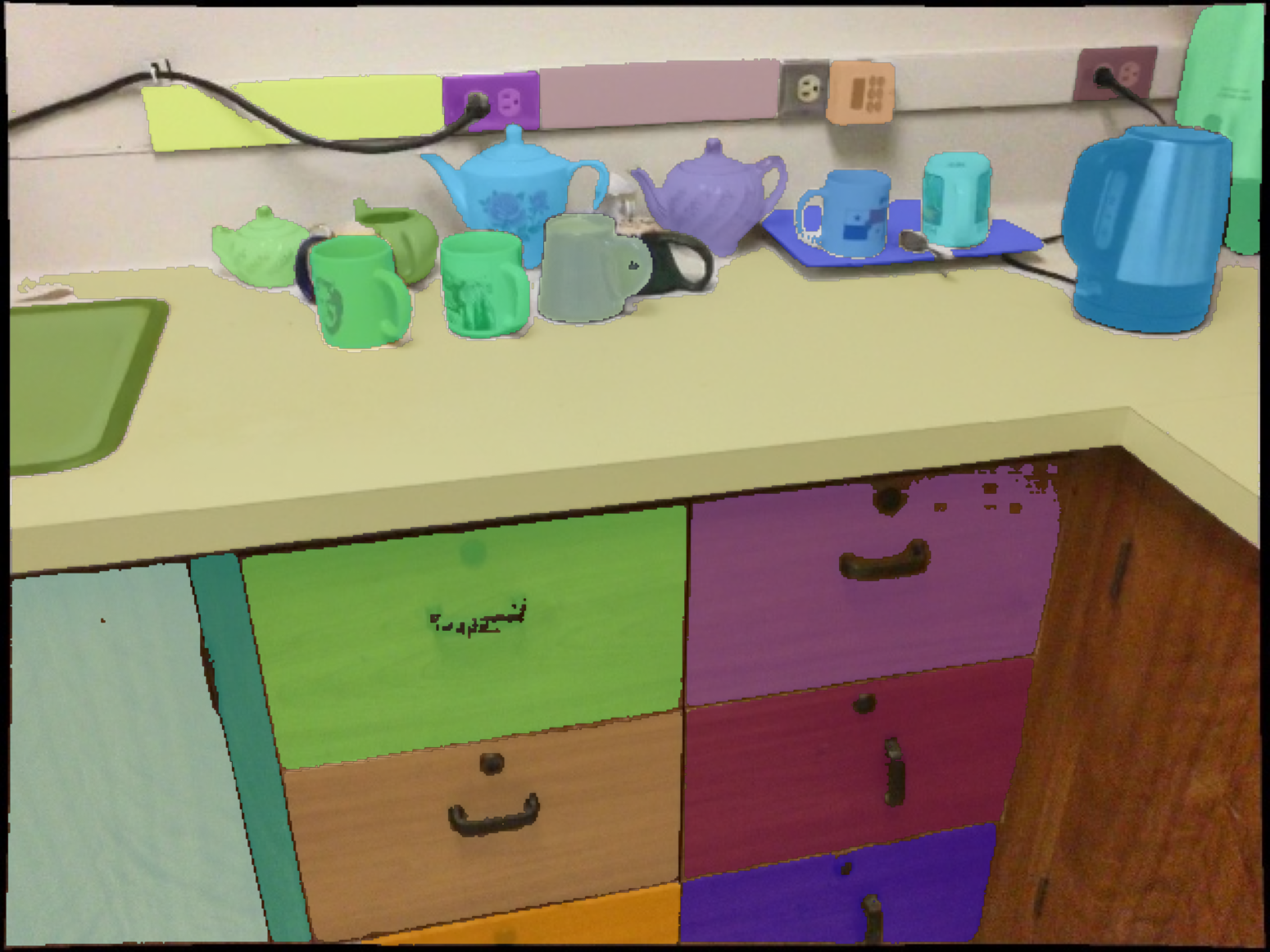} %
        \end{minipage}
        \begin{minipage}{.18\textwidth}
            \includegraphics[width=\textwidth]{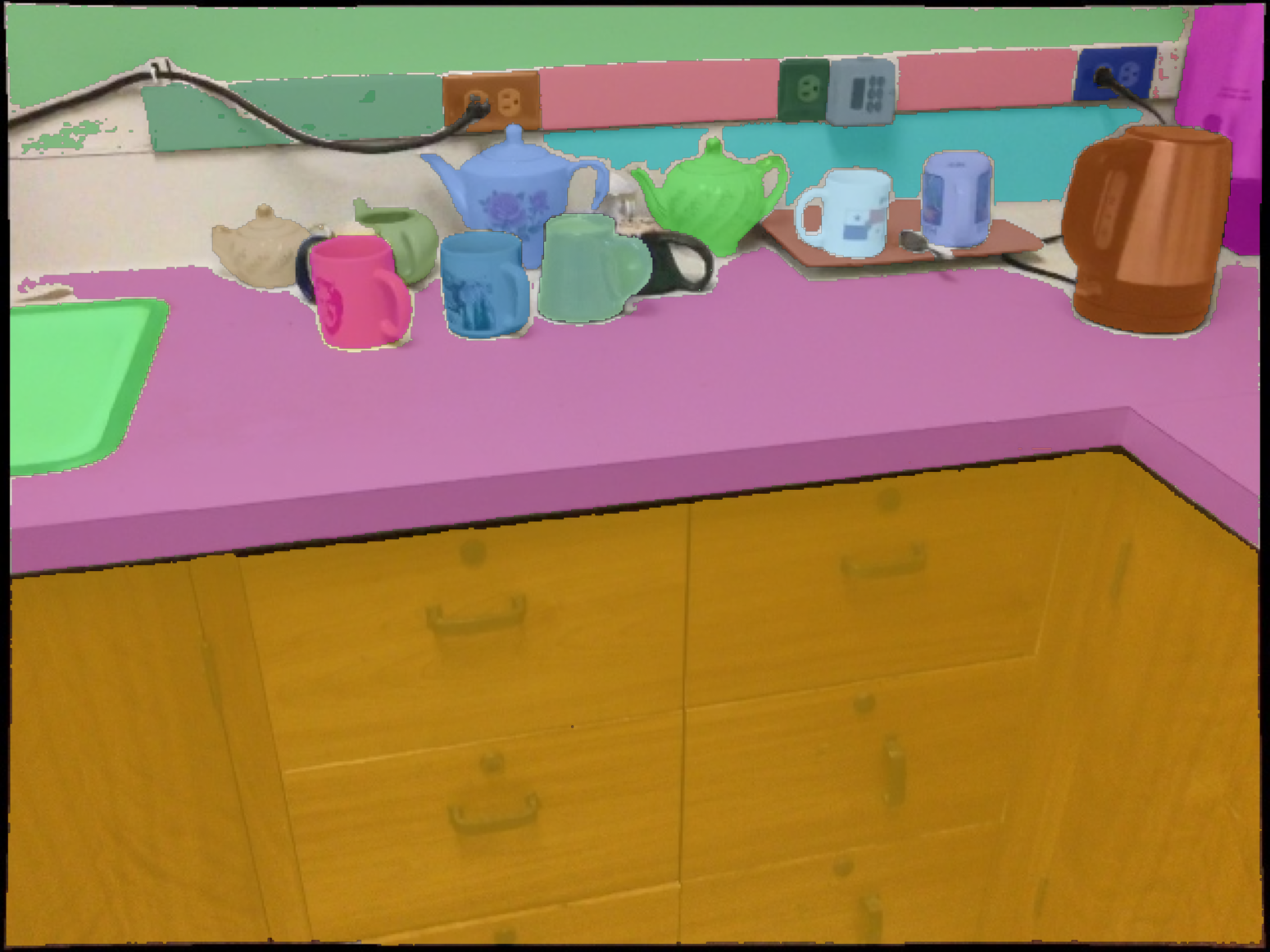} %
        \end{minipage}
    \end{minipage}

    \Description[Qualitative Results]{Mask generation}
    \caption{%
    Multi-resolution mask generation. Exploiting SAM2's multi-scale grid \textit{ping} point generation, we produce several masks that are subsequently merged to obtain the final result.
    }
    \label{fig:quali_grid}
\end{figure*}

The proposed pipeline has been implemented starting from the original Gaussian splatting \cite{kerbl3Dgaussians} codebase.
In the first stage (Sec.~\ref{subsec:mask_prop}), whenever depth information was not available in the dataset, the Murre approach \cite{guo2025murre} was employed for monocular depth estimation. 
This method was selected over other approaches \cite{depth_anything_v2} due to its scale consistency with respect to the SfM point cloud (obtained via COLMAP \cite{schonberger2016structure}) and consequently, with respect to Gaussian splatting.
Masks are instead computed with SAM2~\cite{sam2}, which ensures accurate mask boundaries (see Fig.~\ref{fig:quali_grid} for a visualization of different auto-segmentation settings).
Finally, semantic visual descriptors are extracted using CLIP \cite{clip}. %
During instance reconstruction, we run Gaussian splatting for $10k$ iterations in the LERF dataset, while the denser ScanNetv2 requires only $1k$ iterations. During instance merging, we run the splatting for $1k$ iterations after each merge. Weight $w_{mask}$ starts from a value of $0.05$ in the first composition and increases by $0.1$ after each (parallel) merging step until it reaches a maximum of $0.25$. We run our experiments on a single NVIDIA RTX 3090, with a maximum VRAM usage of $\sim\!\!10$GB (in the largest ScanNetv2 scene 0000\_00).

\begin{table}[t]
\centering \small
\caption{Per-scene instance segmentation accuracy on ScanNetv2 for our approach and InstanceGS~\cite{li2025instancegaussian}. Best in bold.
}
\label{tab:scannet_results}
\begin{tabular}{c|cc|cc}
\toprule
 & \multicolumn{2}{c|}{\textbf{Instance GS}} & \multicolumn{2}{c}{\textbf{Split\&Splat}} \\
\textbf{Scene} & mIoU & mAcc(25) & {mIoU} & mAcc(25) \\
\midrule
{0000\_00} & \textbf{51.71} & \textbf{85.07} & 43.97 & 76.12 \\
{0062\_00} & 50.76 & 85.33 & \textbf{63.46} & \textbf{100.00} \\
{0070\_00} & \textbf{48.86} & 82.61 & 48.70 & \textbf{86.96} \\
{0097\_00} & 58.03 & 82.61 & \textbf{64.98} & \textbf{100.00} \\
{0140\_00} & 54.32 & 91.49 & \textbf{59.76} & \textbf{95.74} \\
{0200\_00} & 45.07 & 68.42 & \textbf{59.98} & \textbf{84.21} \\
{0347\_00} & 57.70 & 89.29 & \textbf{69.81} & \textbf{96.43} \\
{0400\_00} & 48.58 & 73.91 & \textbf{57.93} & \textbf{82.61} \\
{0590\_00} & 47.57 & 78.33 & \textbf{48.58} & \textbf{86.67} \\
{0645\_00} & 40.42 & 67.44 & \textbf{46.70} & \textbf{75.58} \\
\midrule
\textbf{mean} & 50.30 & 80.45 & \textbf{56.39} & \textbf{88.43} \\
\bottomrule
\end{tabular}
\end{table}

\section{Results}
\label{sec:results}

The proposed approach produces a set of instance-level Gaussian splats, each containing a single object along with its semantic description. This opens the way to downstream tasks in the field of scene understanding, \ie, instance, panoptic, and open vocabulary segmentation. Additionally, instance-level modeling enables in-scene editing, \eg, adding, removing, or duplicating objects. 

We begin by presenting segmentation and reconstruction results on the ScanNetv2 dataset \cite{dai2017scannet} in Sec.~\ref{sec:results-segm}.
Then we consider a few downstream tasks that our approach is not directly designed for, \ie, open vocabulary segmentation (Sec.~\ref{sec:results-openvoc}) and scene editing (Sec.~\ref{sec:results-editing}), evaluated on the LERF dataset \cite{LERF2023}. 
Finally, Sec.~\ref{sec:ablation} presents an ablation study on the components.

\subsection{Instance Segmentation Results}
\label{sec:results-segm}
We followed the setup introduced by OpenGaussian \cite{wu2024opengaussian} and InstanceGS \cite{li2025instancegaussian} to evaluate 3D segmentation on the ScanNetv2 \cite{dai2017scannet} dataset using ground-truth (GT) instance-level annotations. Specifically, we report instance-level mIoU and mAcc (with a threshold of 25\%) on the 10 scenes selected in these works.
The mIoU is computed over all GT instances in a scene, while the mAcc($x$) measures the fraction of instances identified with at least $x\%$ IoU.
The ScanNetv2 benchmark provides a 3D labeled point cloud to be used as GT for metrics computation. Following the evaluation pipeline of the competitors, we compute our labeled point cloud by assigning new instance labels to the 3D points of the ground-truth point cloud ($P_{GT}$) using the output of the \textbf{Split\&Splat} pipeline. 
In detail, the Gaussian means are treated as points, and labels are assigned to each nearest-neighbor in $P_{GT}$.

In Tab.~\ref{tab:scannet_results}, we compare our method with the state-of-the-art Gaussian-based approach InstanceGS~\cite{li2025instancegaussian}, which relies on RGB images and SAM-generated masks as input.
Our approach achieves an average mIoU of $56.39\%$, improving by more than $6\%$ over the competitor.
Per-scene results confirm this trend, with \textbf{Split\&Splat} outperforming InstanceGS in 8 out of 10 scenes in terms of mIoU, and in 9 out of 10 when considering mAcc(25).

\begin{figure*}[t]
    \centering
    \scriptsize

    \begin{minipage}{0.9\textwidth}
        \centering
        \begin{minipage}{.22\textwidth}
            \centering      
            {Original mesh}
        \end{minipage}%
        \begin{minipage}{.22\textwidth}
            \centering
            {InstanceGS}
        \end{minipage}%
        \begin{minipage}{.22\textwidth}
            \centering
            \textbf{Split\&Splat}
        \end{minipage}%
        \begin{minipage}{.22\textwidth}
            \centering
            {GT}
        \end{minipage}%
    \end{minipage}\\[2mm]
    \begin{minipage}{0.9\textwidth}
        \makebox[0pt][r]{%
            \raisebox{-0.3\height}{\rotatebox{90}{ScanNetv2/{0062}}}%
            \hspace{2mm}%
        }%
        \centering
        \begin{minipage}{.22\textwidth}
            \includegraphics[width=\textwidth]{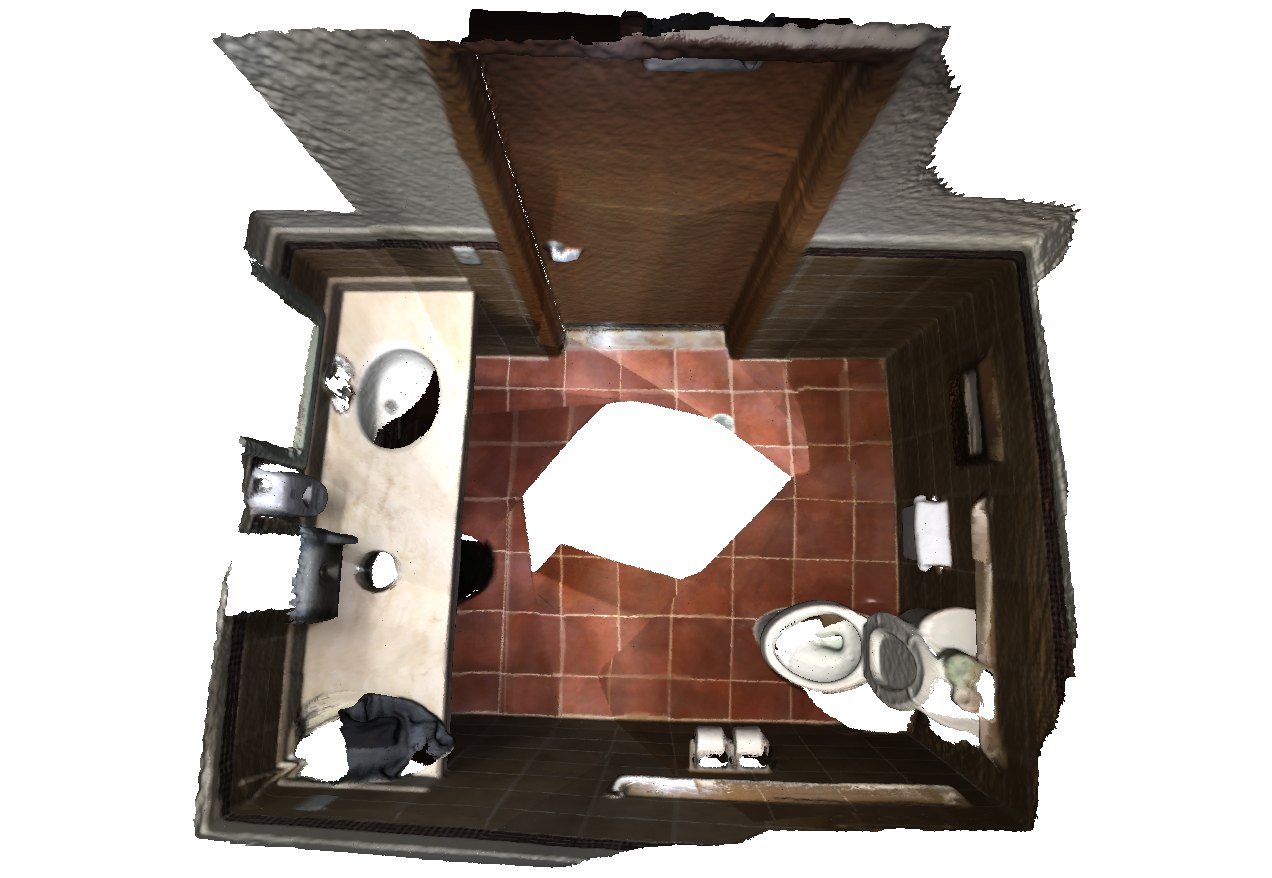} %
        \end{minipage}
        \begin{minipage}{.22\textwidth}
            \includegraphics[width=\textwidth]{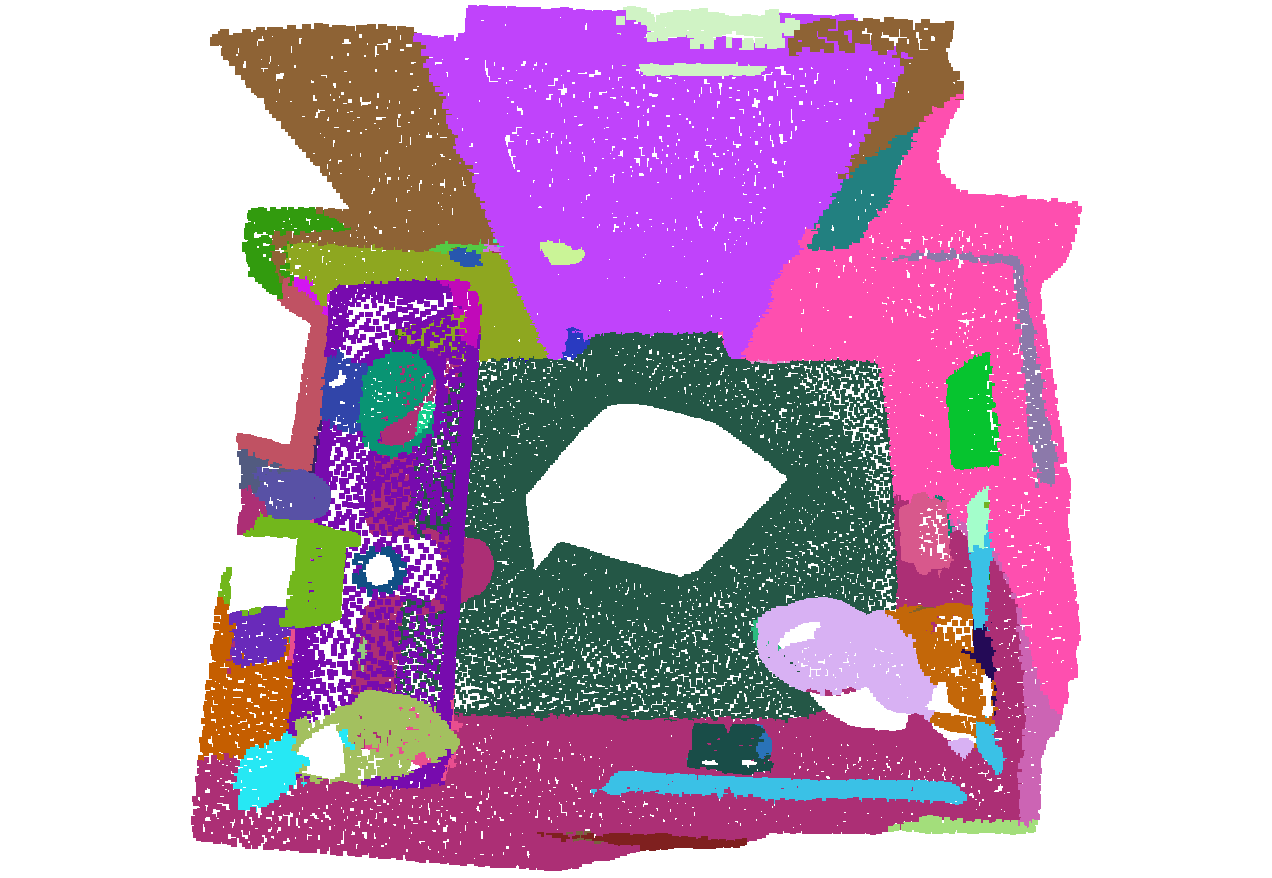} %
        \end{minipage}
        \begin{minipage}{.22\textwidth}
            \includegraphics[width=\textwidth]{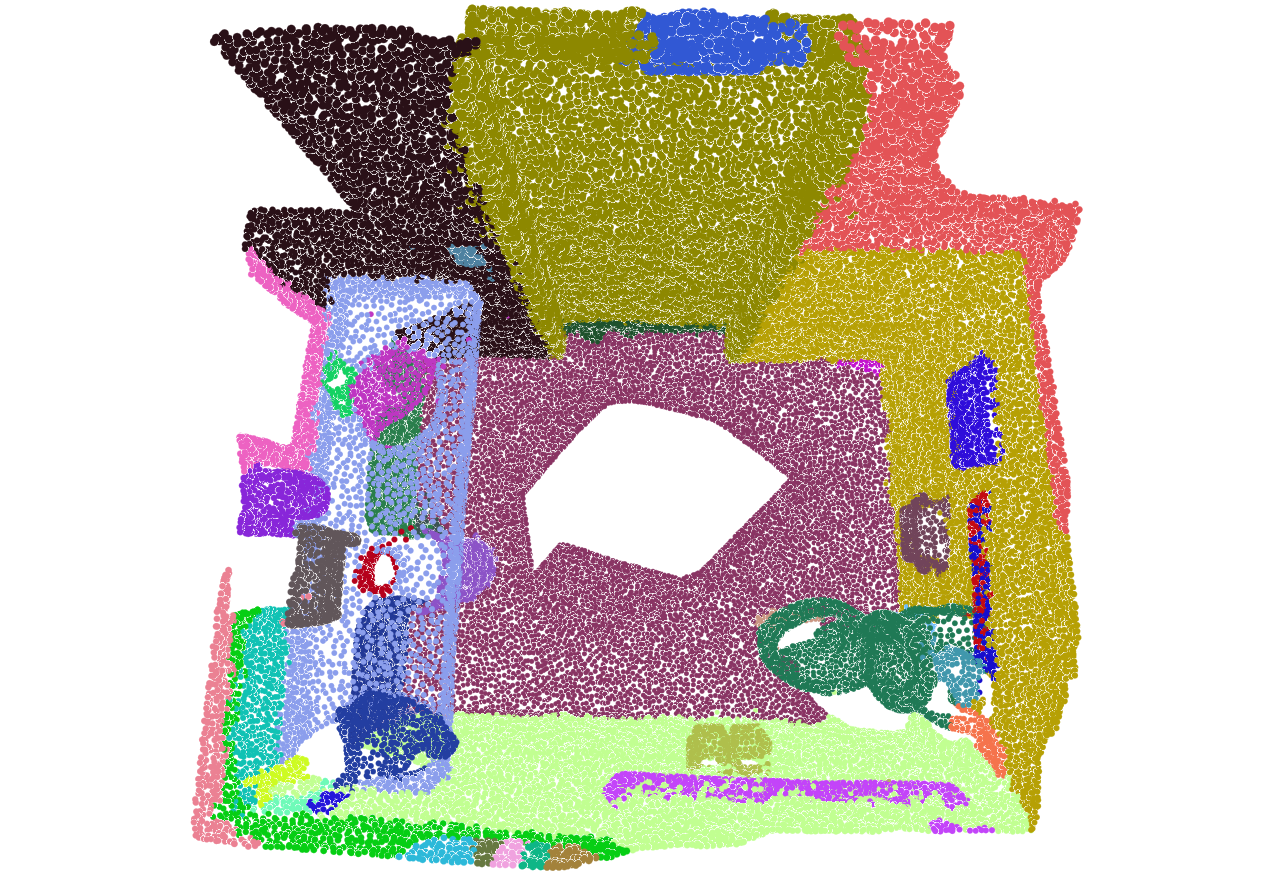} %
        \end{minipage}
        \begin{minipage}{.22\textwidth}
            \includegraphics[width=\textwidth]{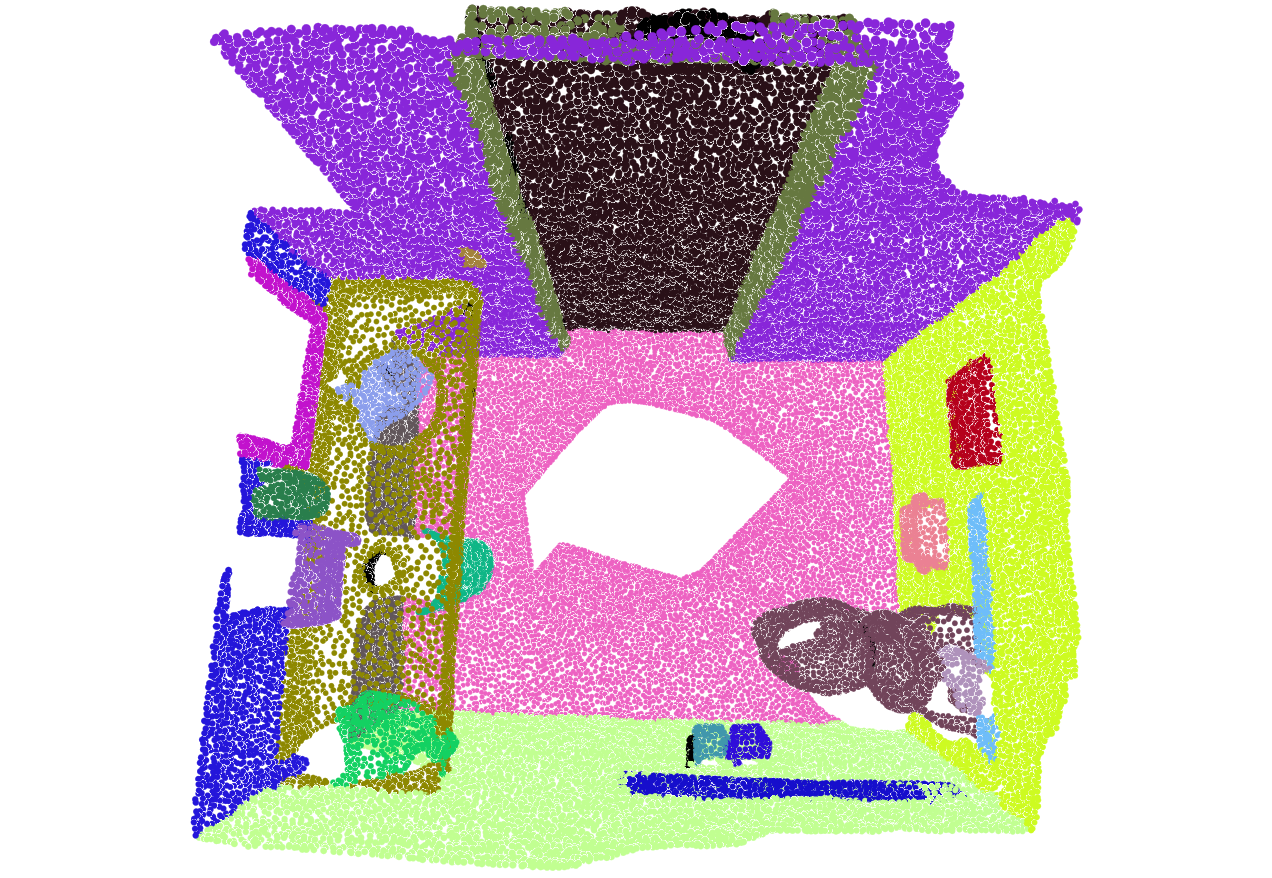} %
        \end{minipage}
    \end{minipage}\\[1mm]
    \begin{minipage}{0.9\textwidth}
         \makebox[0pt][r]{%
            \raisebox{-0.3\height}{\rotatebox{90}{ScanNetv2/{0097}}}%
            \hspace{2mm}%
        }%
        \centering
        \begin{minipage}{.22\textwidth}
            \includegraphics[width=\textwidth]{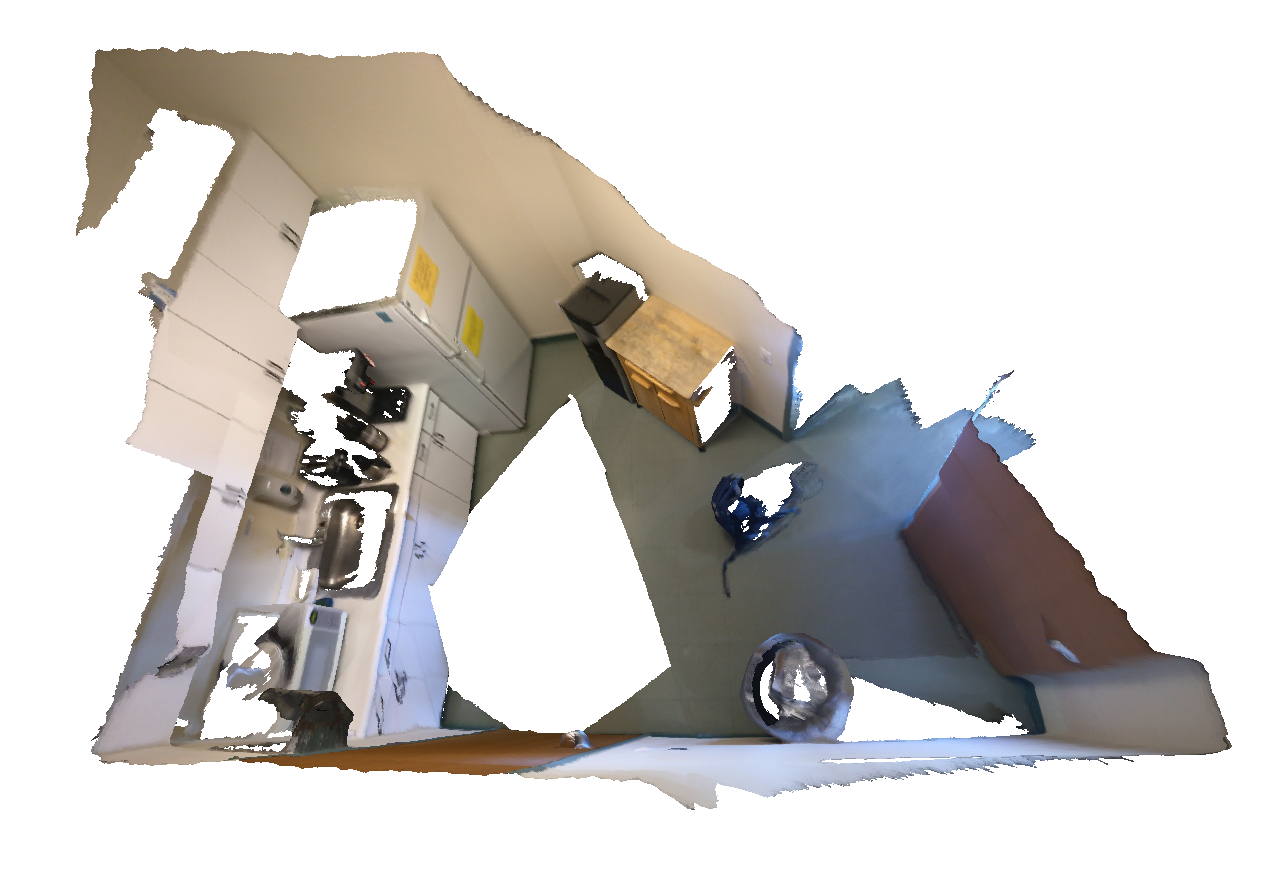} %
        \end{minipage}
        \begin{minipage}{.22\textwidth}
            \includegraphics[width=\textwidth]{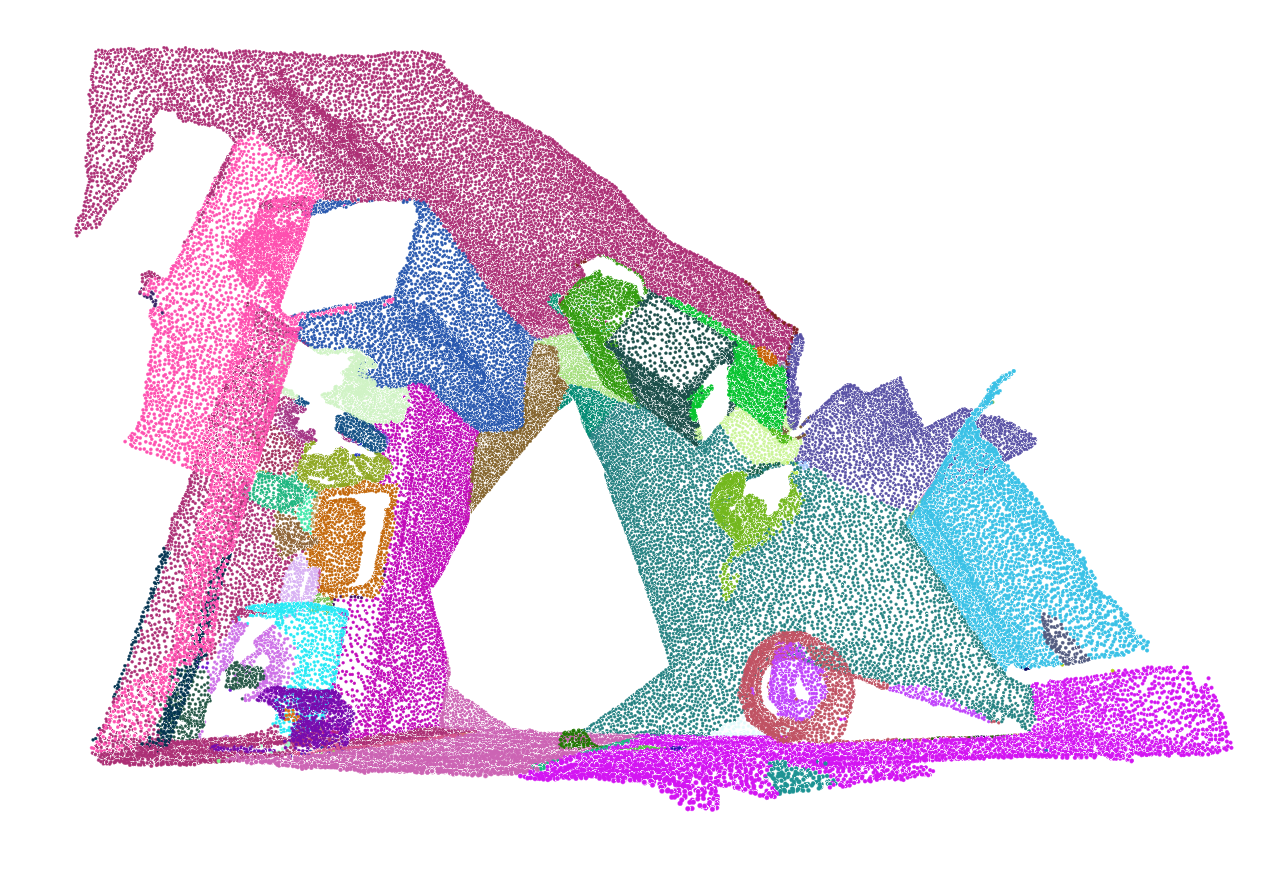} %
        \end{minipage}
        \begin{minipage}{.22\textwidth}
            \includegraphics[width=\textwidth]{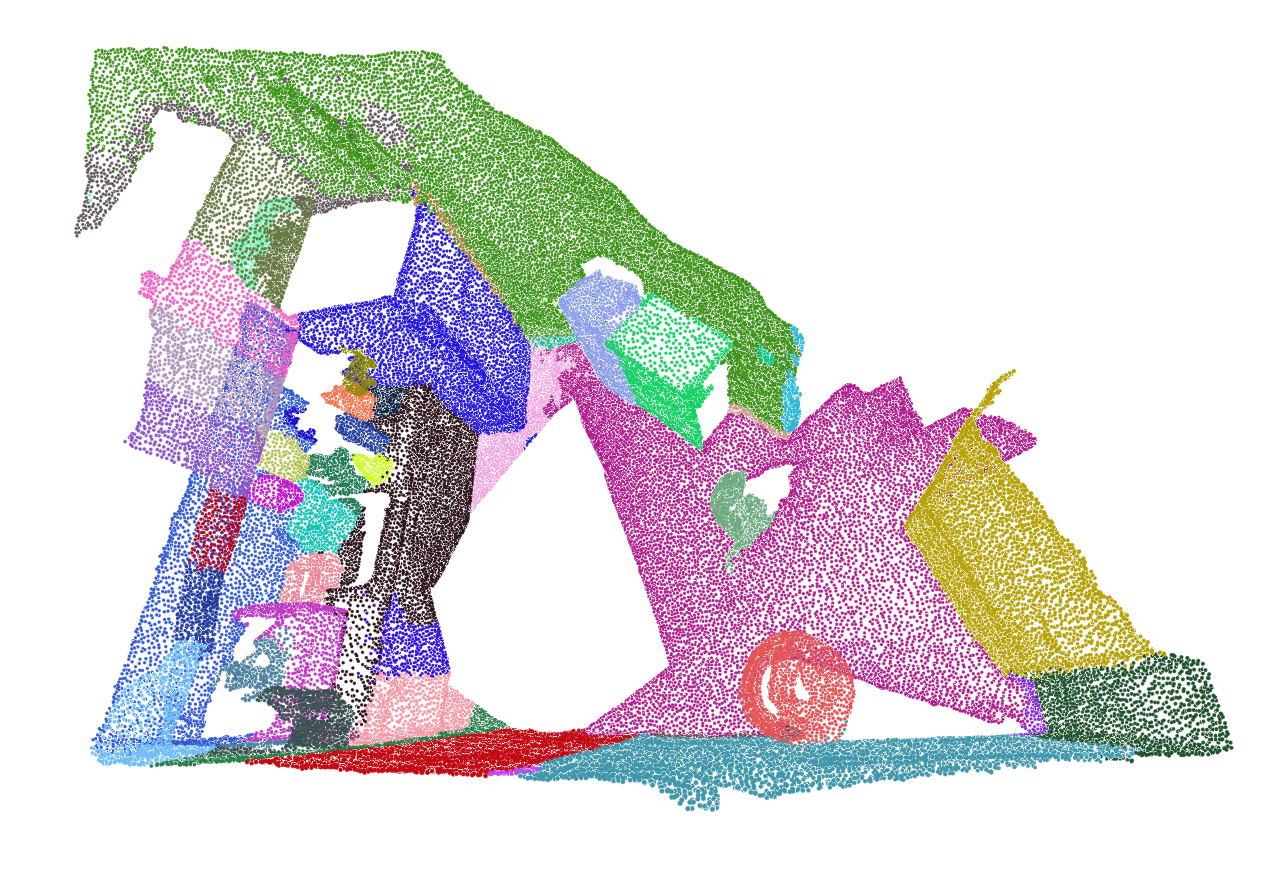} %
        \end{minipage}
        \begin{minipage}{.22\textwidth}
            \includegraphics[width=\textwidth]{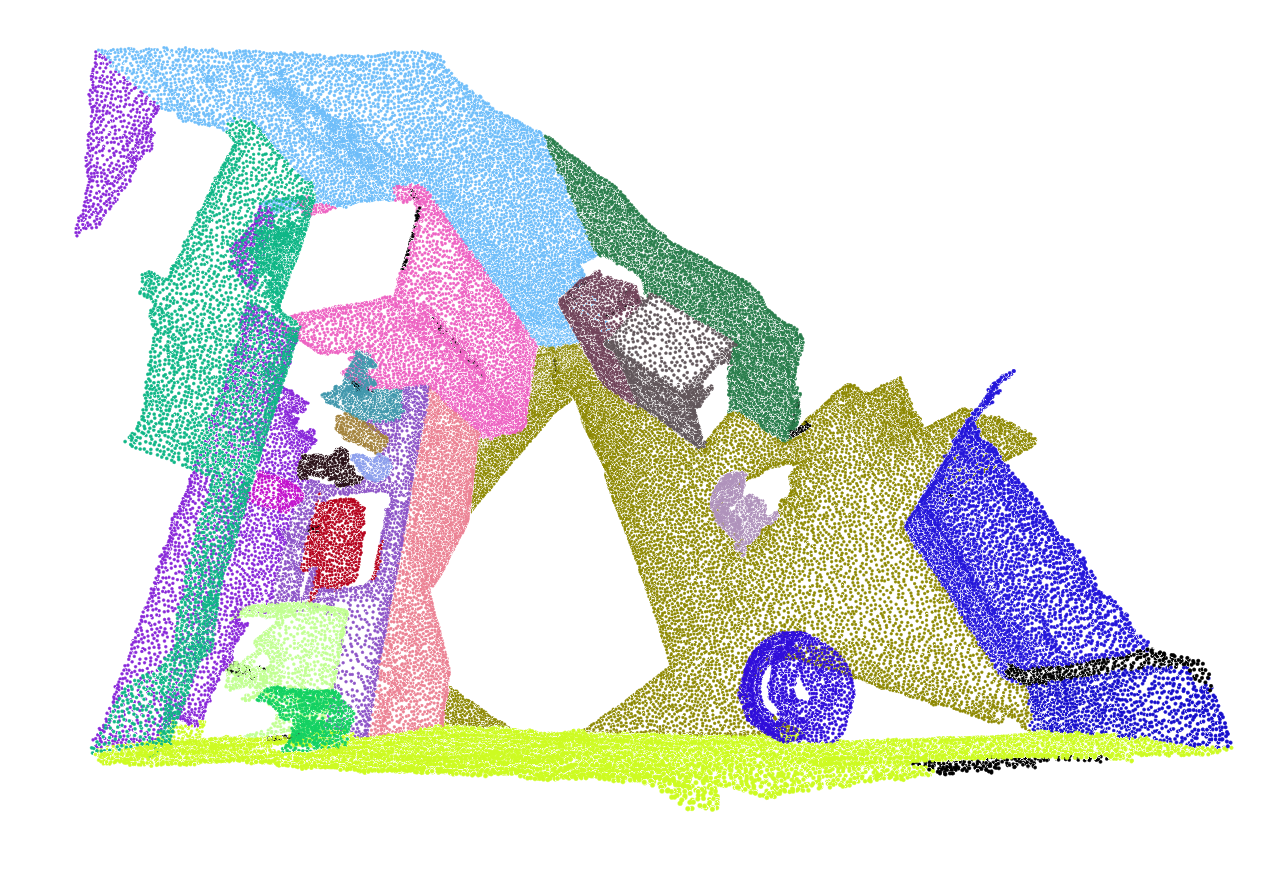} %
        \end{minipage}
       
    \end{minipage}\\[1mm]

    \begin{minipage}{0.9\textwidth}
        \makebox[0pt][r]{%
            \raisebox{-0.3\height}{\rotatebox{90}{ScanNetv2/{0347}}}%
            \hspace{2mm}%
        }%
        \centering
        \begin{minipage}{.22\textwidth}
            \includegraphics[width=\textwidth]{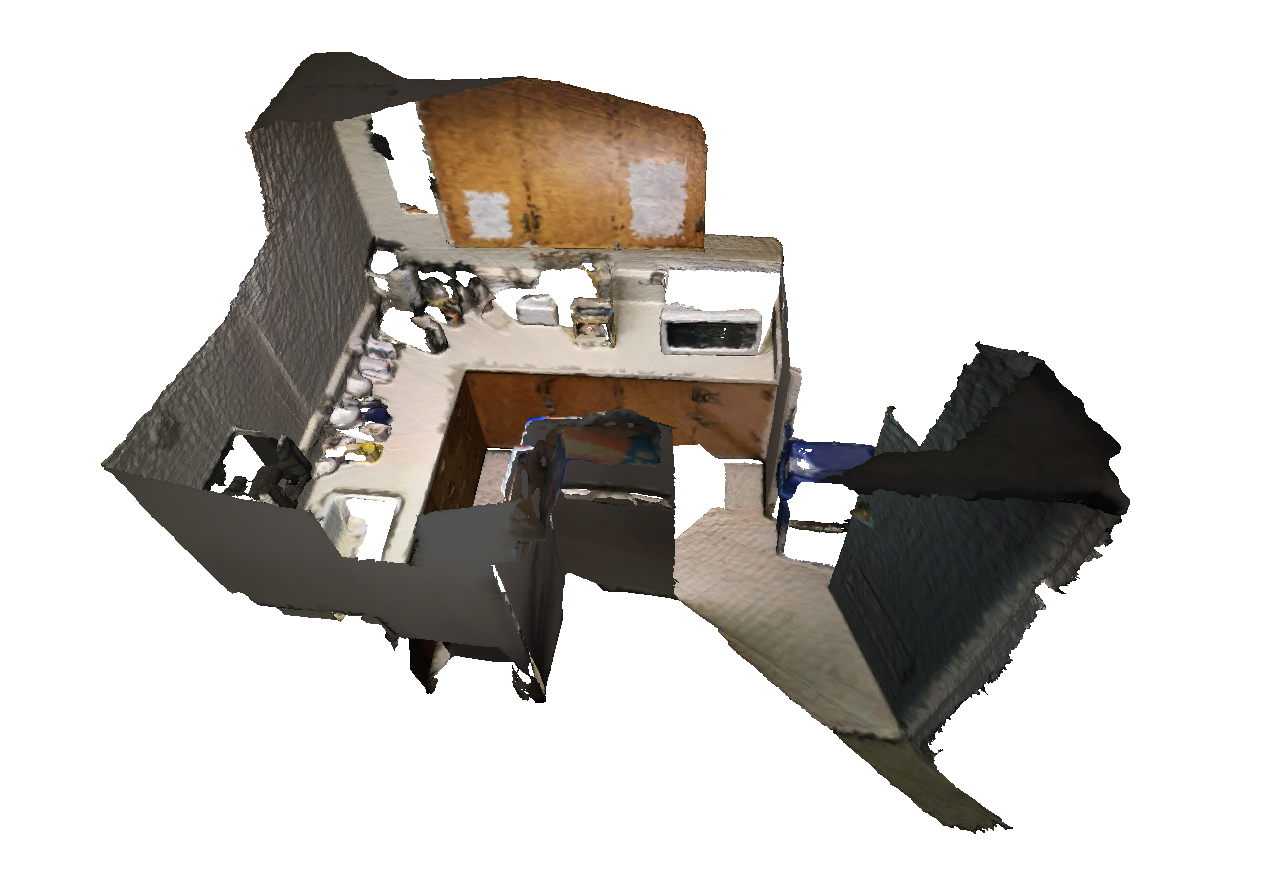} %
        \end{minipage}
        \begin{minipage}{.22\textwidth}
            \includegraphics[width=\textwidth]{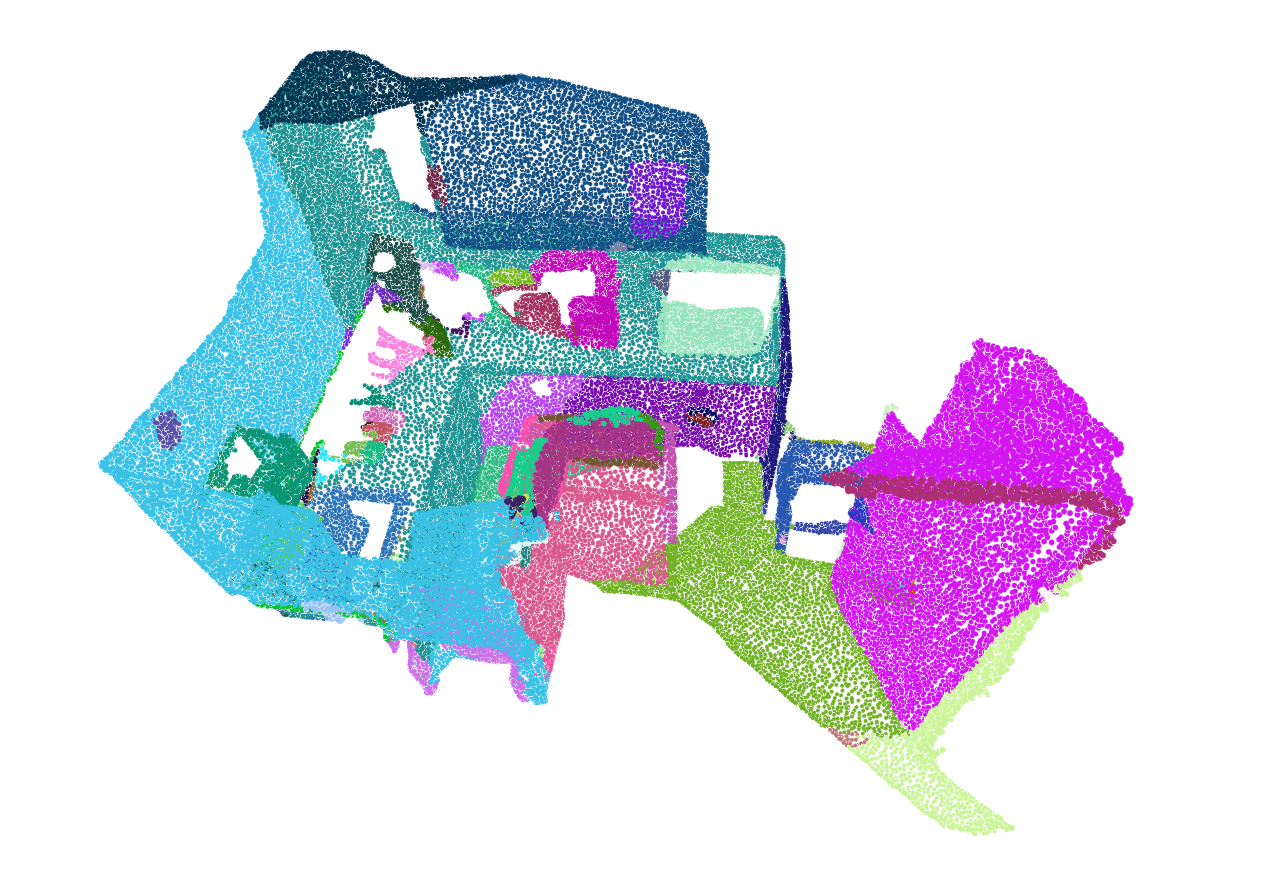} %
        \end{minipage}
        \begin{minipage}{.22\textwidth}
            \includegraphics[width=\textwidth]{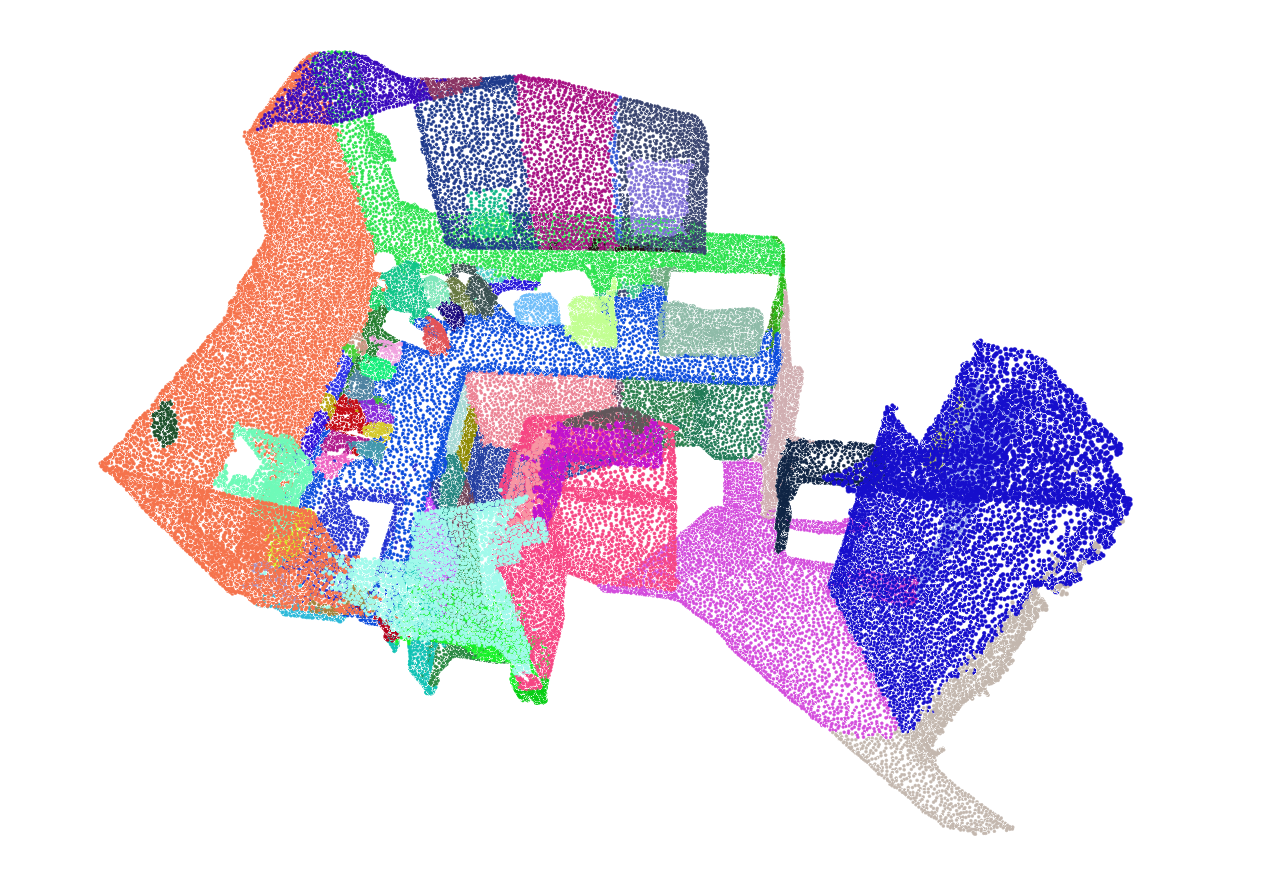} %
        \end{minipage}
        \begin{minipage}{.22\textwidth}
            \includegraphics[width=\textwidth]{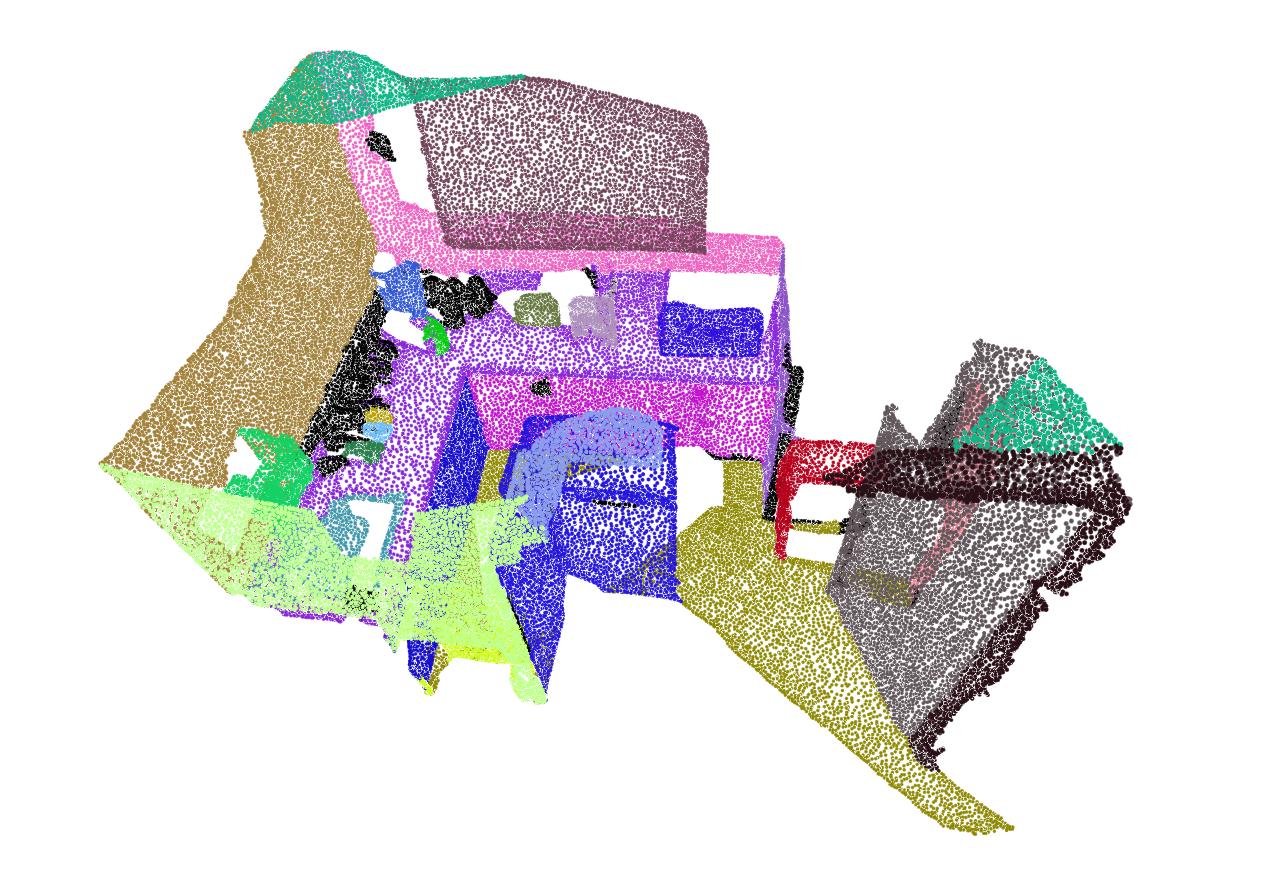} %
        \end{minipage}
    \end{minipage}\\[1mm]

    \begin{minipage}{0.9\textwidth}
        \makebox[0pt][r]{%
            \raisebox{-0.3\height}{\rotatebox{90}{ScanNetv2/{0400}}}%
            \hspace{2mm}%
        }%
        \centering
        \begin{minipage}{.22\textwidth}
            \includegraphics[trim=0cm 2cm 0cm 0cm, clip,width=\textwidth]{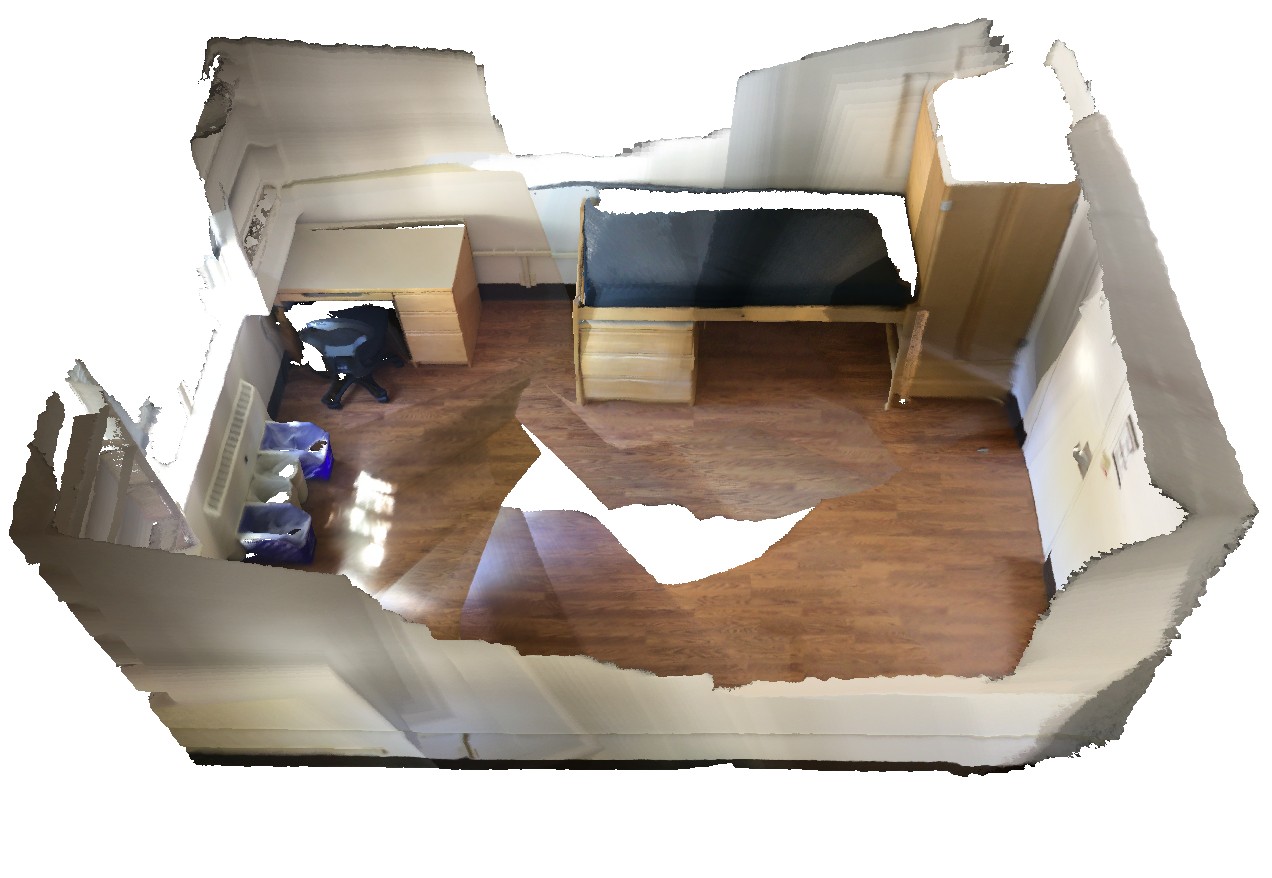} %
        \end{minipage}
        \begin{minipage}{.22\textwidth}
            \includegraphics[trim=0cm 2cm 0cm 0cm, clip,width=\textwidth]{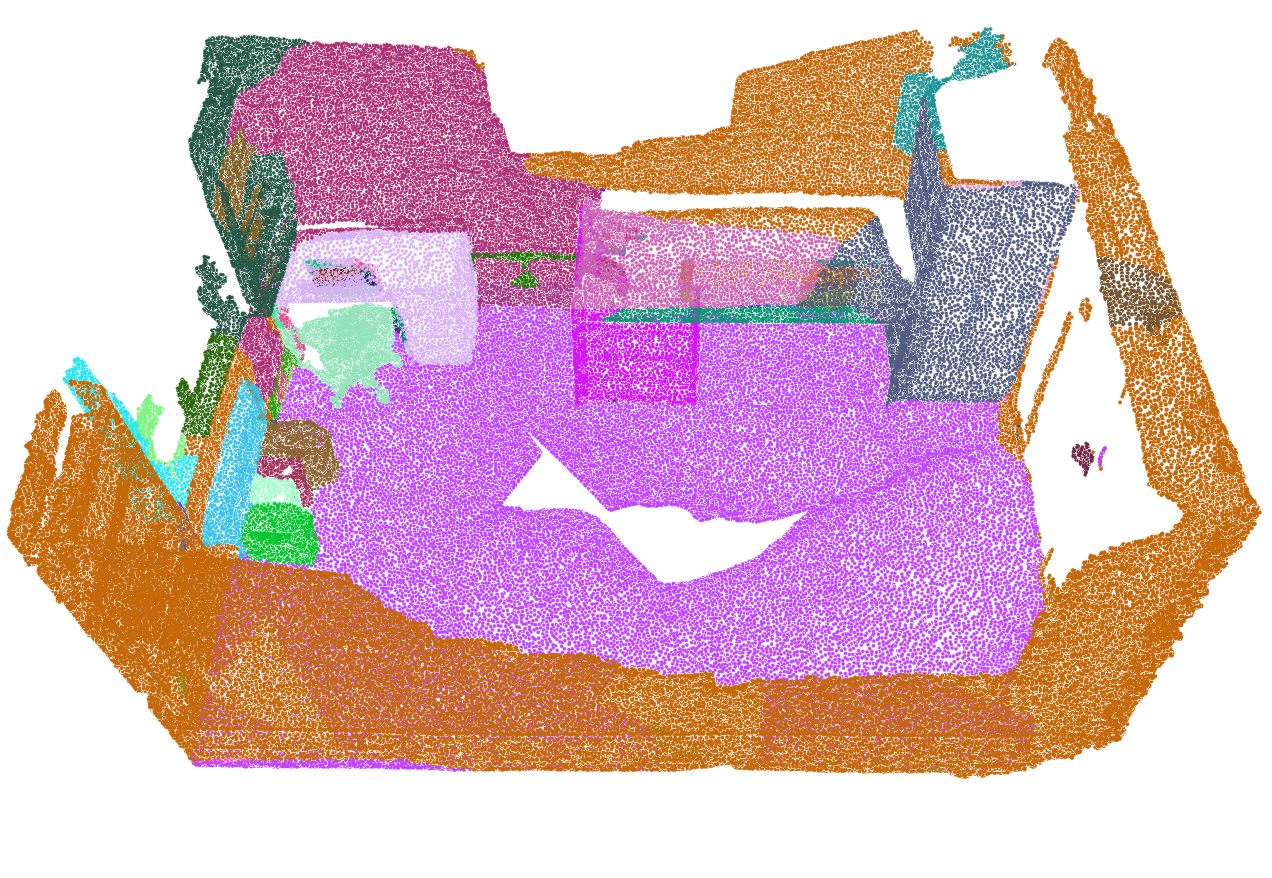} %
        \end{minipage}
        \begin{minipage}{.22\textwidth}
            \includegraphics[trim=0cm 2cm 0cm 0cm, clip,width=\textwidth]{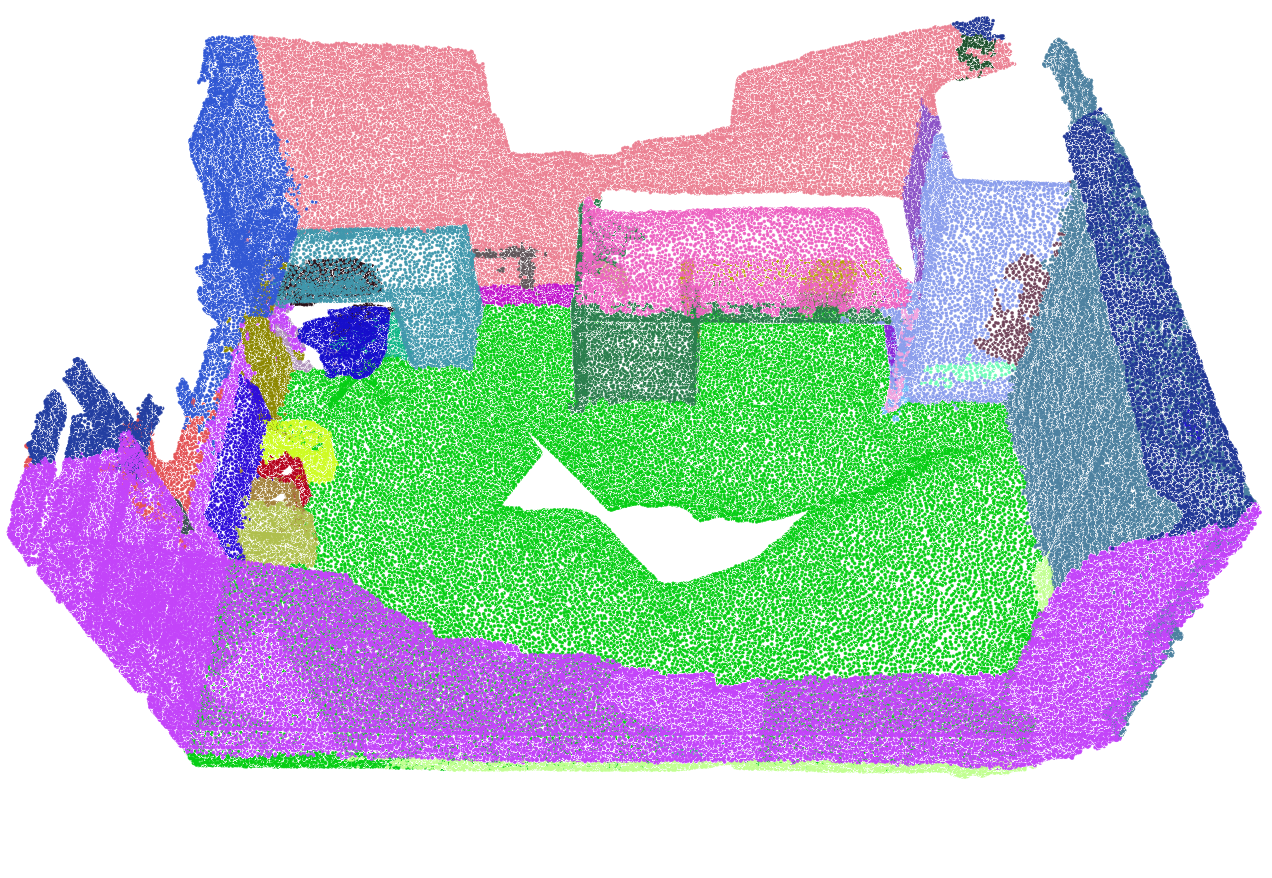} %
        \end{minipage}
        \begin{minipage}{.22\textwidth}
            \includegraphics[trim=0cm 2cm 0cm 0cm, clip,width=\textwidth]{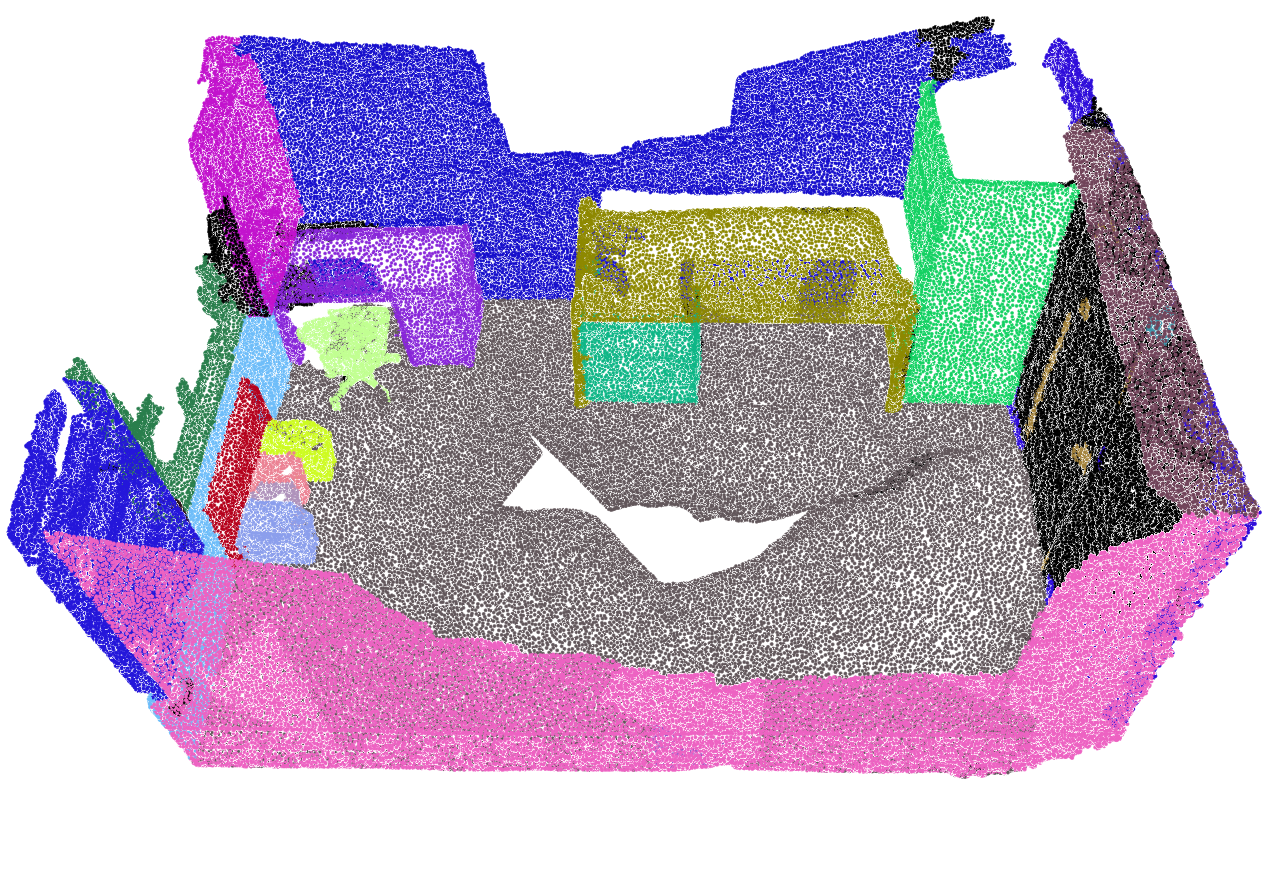} %
        \end{minipage}
    \end{minipage}\\[1mm]
    \Description[Qualitative Results]{Qualitative results on LERF/ScanNet datasets}
    \caption{Visualization comparison of category-agnostic 3D instance segmentation result. Split\&Splat outperforms InstanceGS, accurately distinguishing the different 3D objects. %
    }
    \label{fig:Instance_qualitatives}
\end{figure*}

Qualitative results presented in Fig.~\ref{fig:Instance_qualitatives} further support these findings.
Compared to InstanceGS, \textbf{Split\&Splat} produces sharper and more geometrically consistent object boundaries, especially in regions affected by occlusions or clutter.
This visual improvement is a direct consequence of explicitly modeling each object instance and enforcing multi-view consistency during the reconstruction process.
Regarding limitations, Tab.~\ref{tab:scannet_results} also shows that our pipeline degrades slightly in highly object-dense scenes.
In particular, scenes 0000\_00 and 0070\_00 contain $123$ and $90$ object instances, respectively, compared to an average of approximately $25$ in the other scenes.
In these cases, the increased instance density can lead to missed detections or label conflicts during propagation, which, in turn, affects the final segmentation accuracy.

\begin{table}[t]
\centering \small
\caption{Open vocabulary results on LERF~\cite{LERF2023}. Best in bold, second best underlined, third dashed underline.}
\label{tab:ov-results-mean}
\begin{tabular}{l|cc}
\toprule
\multirow{2}{*}{\textbf{Method}} & \multicolumn{2}{c}{\textbf{Average}}  \\
& mIoU & mAcc(25) \\
\midrule
LERF \cite{LERF2023} & 10.35 & 13.64 \\
LEGaussian \cite{shi2024language} & 16.21 & 23.82 \\
OpenGaussian~\cite{wu2024opengaussian} & 38.36 & 51.43 \\
SuperGSeg~\cite{liang2024supergseg} & 35.94 & 52.02  \\
Dr. Splat~\cite{jun2025dr} & 43.29 & 64.30  \\
InstanceGS~\cite{li2025instancegaussian} & 43.87 & 61.09  \\
CAGS~\cite{sun2025cags}  & \dashuline{50.79} & 69.62  \\
VoteSplat~\cite{jiang2025votesplat} & 50.10 & 67.38 \\
Occam's LGS~\cite{occams_lgs_2024} & 47.22 & \underline{74.84} \\
VALA \cite{vala} & \textbf{58.02} & \textbf{82.85}  \\
\midrule
\textbf{Split\&Splat} & \underline{55.68} & \dashuline{73.05}  \\
\bottomrule 
\end{tabular}%
\end{table}

\subsection{Open Vocabulary Segmentation Results}
\label{sec:results-openvoc}

As mentioned, while our approach does not natively support the open vocabulary segmentation task, it can be adapted by assigning a VLM descriptor to each of the identified instances in a scene.
More specifically, we mask the original views $\mathcal{I}_l$ using $\mathcal{M}^{gs}$ and crop them around the instance; the remaining part of the images is either blurred or filled in with one color, as explained in Sec.~\ref{sec:ov_desc}. This allows us to compute a VLM description for each image and average them to obtain a semantically descriptive representation for each instance that is not influenced by its surroundings. 
Note that, as discussed in Sec.~\ref{sec:related}, recent works that integrate semantic descriptors into GS~\cite{langsplatv2_2025,shi2024language} typically reduce descriptor dimensionality to limit memory usage, which may compromise semantic expressiveness. In contrast, our sparse formulation allows us to preserve their full dimensionality (in the Appendix, we propose an additional method to compute descriptors associated with Gaussians rather than with instances, to solve denser tasks). %
To perform the evaluation, given an open vocabulary query, we first find the best correlation among the existing 3D Gaussian instances and then extract their corresponding multi-view 2D segmentation masks from the rasterization pipeline. 
We then compute the correlation between each instance descriptor and the textual query using cosine similarity. The correspondences with a similarity lower than a threshold $\tau_{corr}$ are discarded (see Sec.~\ref{sec:ov_thres} for details). 
This object-centric representation enables more accurate and coherent mask extraction than InstanceGS (see Fig.~\ref{fig:OV_qualitative}), contributing to the improved performance reported in Tab.~\ref{tab:ov-results-mean}, where mIoU and mAcc(25) are computed on the LERF~\cite{LERF2023} dataset, following the protocol of state-of-the-art approaches such as LangSplat~\cite{langsplatv2_2025} and OpenGaussian~\cite{wu2024opengaussian} (a more detailed version, including per-scene results, is in the Appendix).
Remarkably, even if the approach has not been designed for this task, \textbf{Split\&Splat} ranks second in terms of mIoU and third in terms of mAcc(25) among a set of ten very recent works (while InstanceGS falls short by more than 10 points in both metrics), thus showing how it can be adapted to achieve competitive performance alongside state-of-the-art methods explicitly developed for the task.

\subsection{Editing Results}
\label{sec:results-editing}
\begin{figure*}[t]
\centering
\scriptsize
\hspace*{0.4cm}
\begin{minipage}[t]{0.48\textwidth}
\centering

\begin{minipage}{\textwidth}
\makebox[0pt][r]{%
\raisebox{-0.5\height}{\rotatebox{90}{\shortstack{\textbf{Figurines}\\``Miffy, Green apple,\\Handle, Old camera"}}}%
\hspace{1mm}}
\centering
\begin{minipage}{.44\textwidth}
\centering
\textbf{InstanceGS\vphantom{\textbf{p}}}\vspace{1mm}\\
\includegraphics[width=\linewidth]{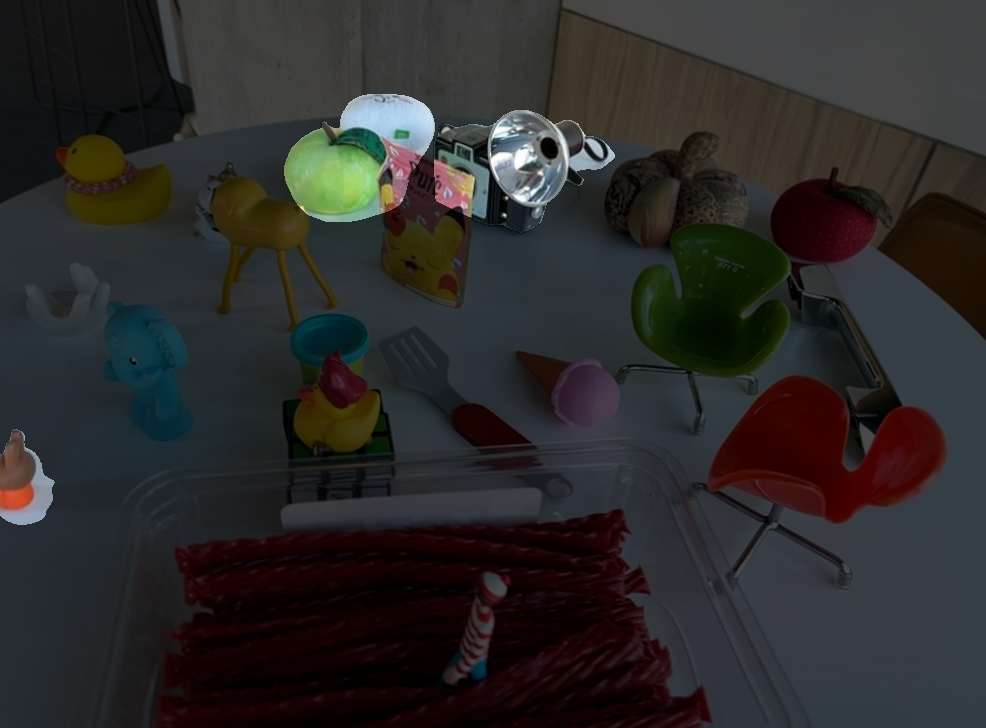}
\end{minipage}%
\hspace{0.01cm}
\begin{minipage}{.44\textwidth}
\centering
\textbf{Split\&Splat}\vspace{1mm}\\
\includegraphics[width=\linewidth]{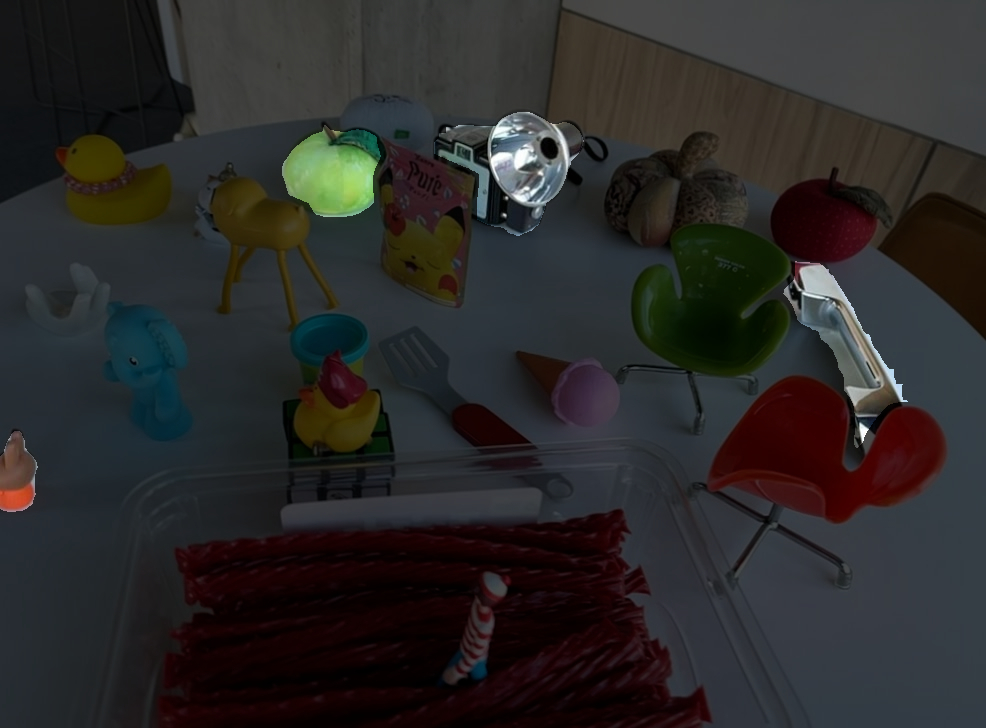}
\end{minipage}
\end{minipage}

\begin{minipage}{\textwidth}
\makebox[0pt][r]{%
\raisebox{-0.5\height}{\rotatebox{90}{\shortstack{\textbf{Ramen}\\``Weavy noodles, Chopstick,\\Sake cup, Kamaboko"}}}%
\hspace{1mm}}
\centering
\begin{minipage}{.44\textwidth}
\includegraphics[width=\linewidth]{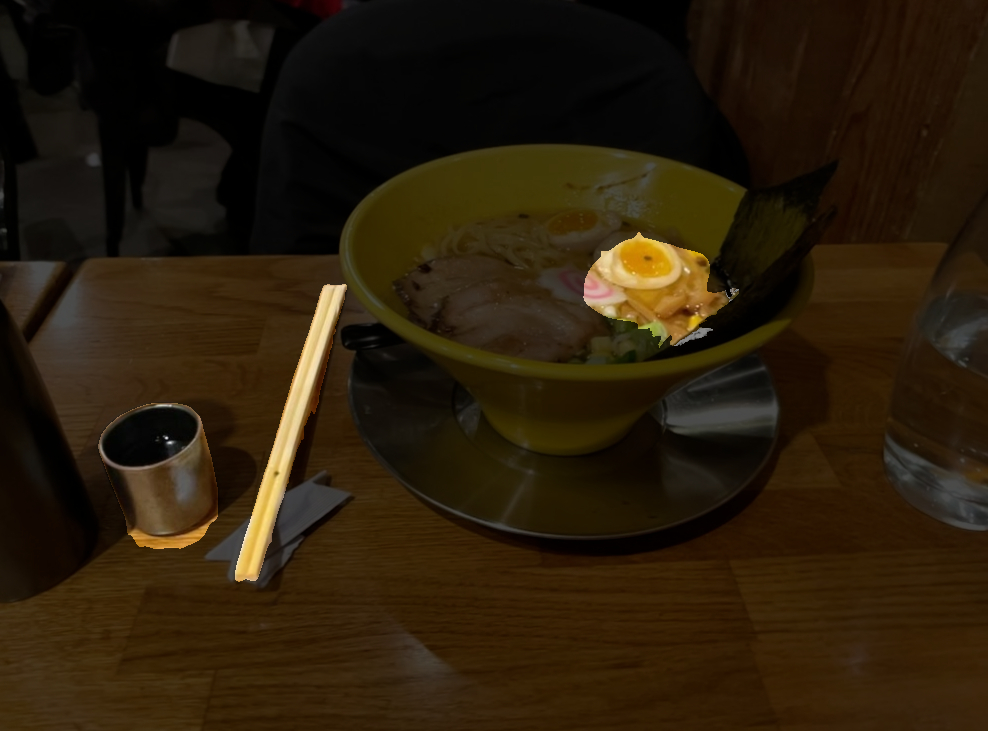}
\end{minipage}%
\hspace{0.01cm}
\begin{minipage}{.44\textwidth}
\includegraphics[width=\linewidth]{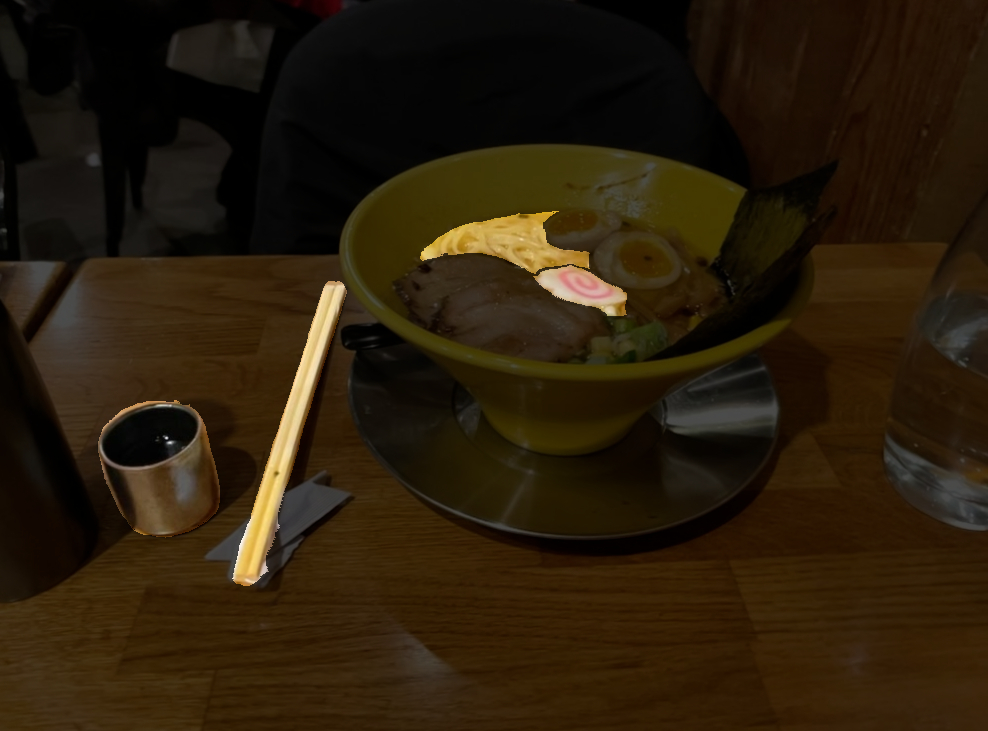}
\end{minipage}
\end{minipage}

\begin{minipage}{\textwidth}
\makebox[0pt][r]{%
\raisebox{-0.5\height}{\rotatebox{90}{\shortstack{\textbf{Teatime}\\``Stuffed bear, Napkin,\\Plate, Tea in glass"}}}%
\hspace{1mm}}
\centering
\begin{minipage}{.44\textwidth}
\includegraphics[width=\linewidth]{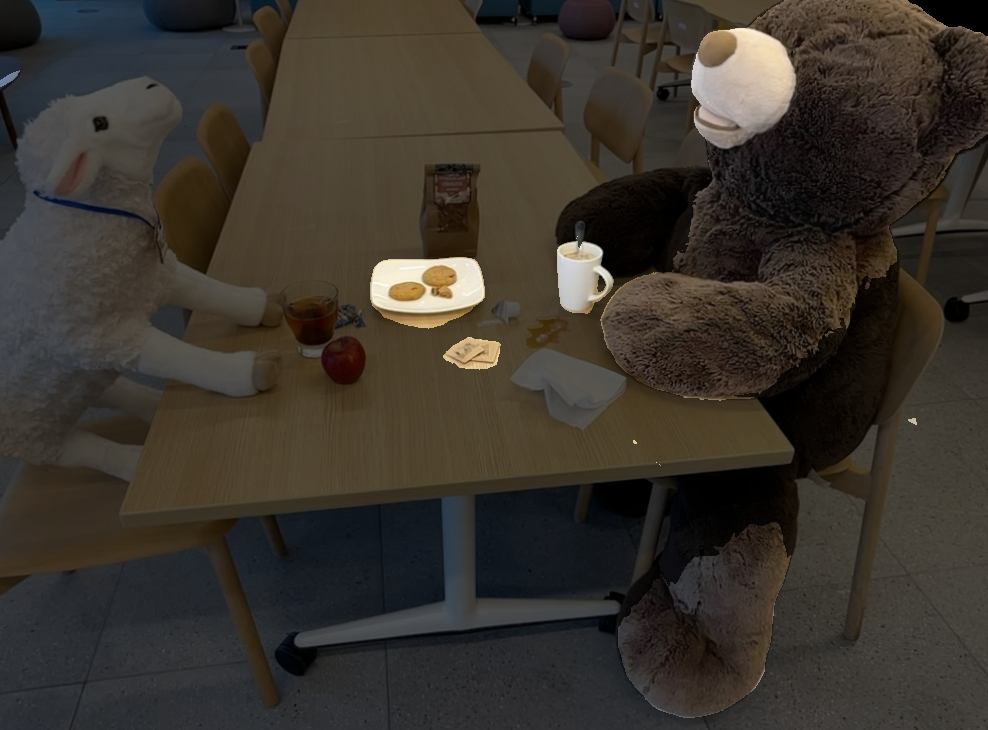}
\end{minipage}%
\hspace{0.01cm}
\begin{minipage}{.44\textwidth}
\includegraphics[width=\linewidth]{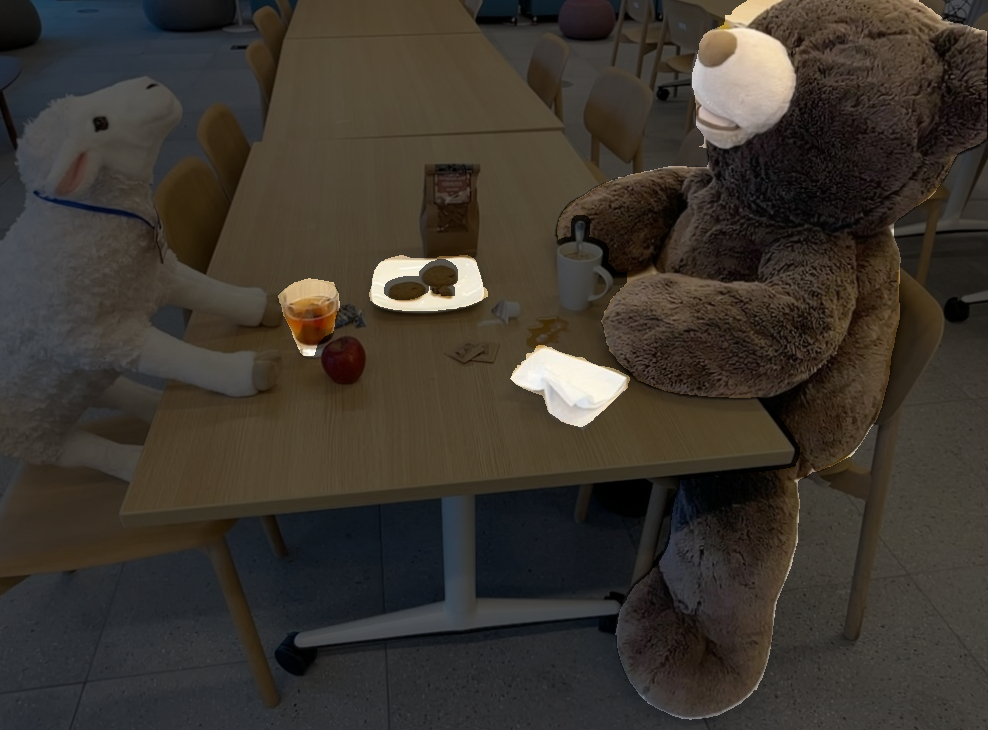}
\end{minipage}
\end{minipage}

\caption{Open-vocabulary query (retrieval) on LERF.}
\label{fig:OV_qualitative}
\end{minipage}%
\begin{minipage}[t]{0.48\textwidth}
\centering

\begin{minipage}{\textwidth}
\makebox[0pt][r]{%
\raisebox{-0.3\height}{\rotatebox{90}{\shortstack{\textbf{Removal}\\``jake'' removed}}}%
\hspace{1mm}}
\centering
\begin{minipage}{.44\textwidth}
\centering
\textit{Before}\vspace{1mm}\\
\includegraphics[width=\linewidth]{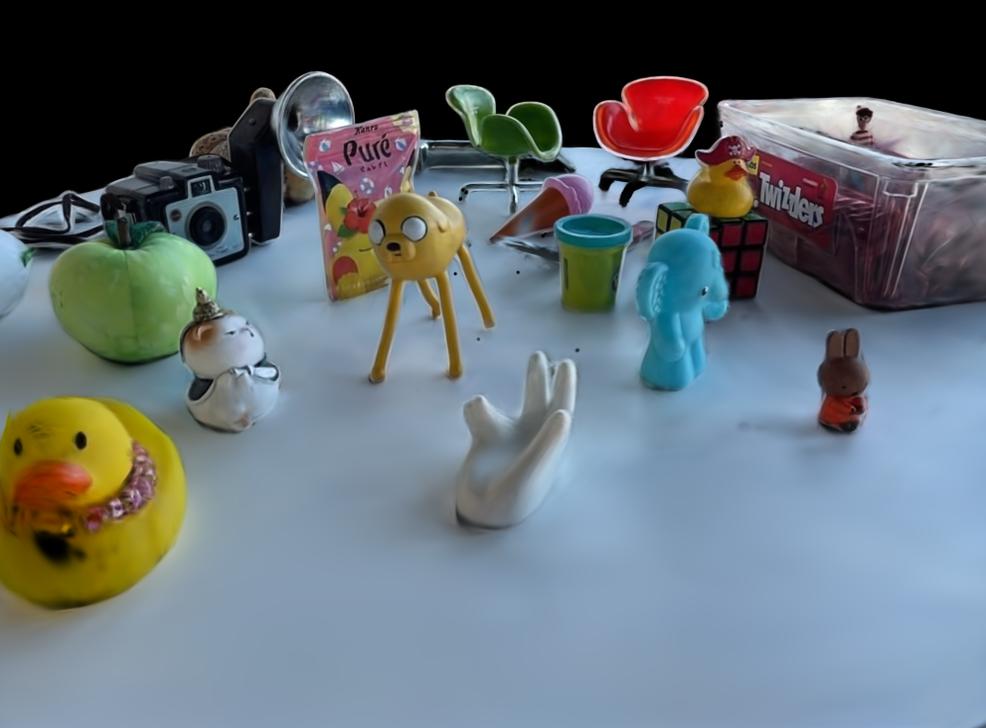}
\end{minipage}%
\hspace{0.01cm}
\begin{minipage}{.44\textwidth}
\centering
\textit{After}\vspace{1mm}\\
\includegraphics[width=\linewidth]{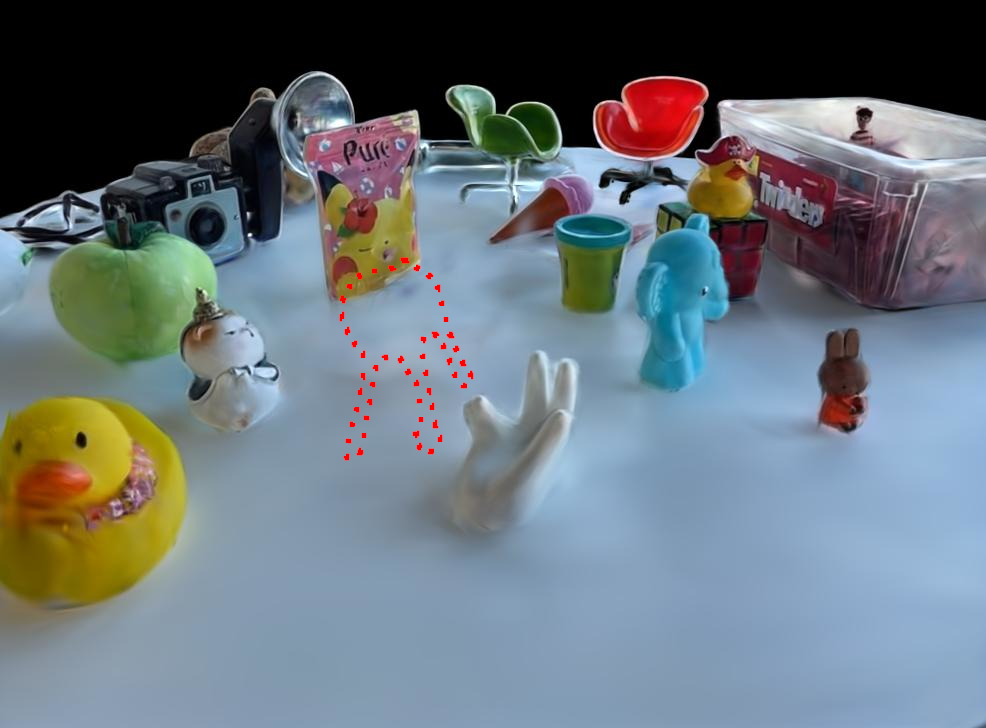}
\end{minipage}
\end{minipage}

\begin{minipage}{\textwidth}
\makebox[0pt][r]{%
\raisebox{-0.3\height}{\rotatebox{90}{\shortstack{\textbf{Duplication}\\``red apple''}}}%
\hspace{1mm}}
\centering
\begin{minipage}{.44\textwidth}
\includegraphics[width=\linewidth]{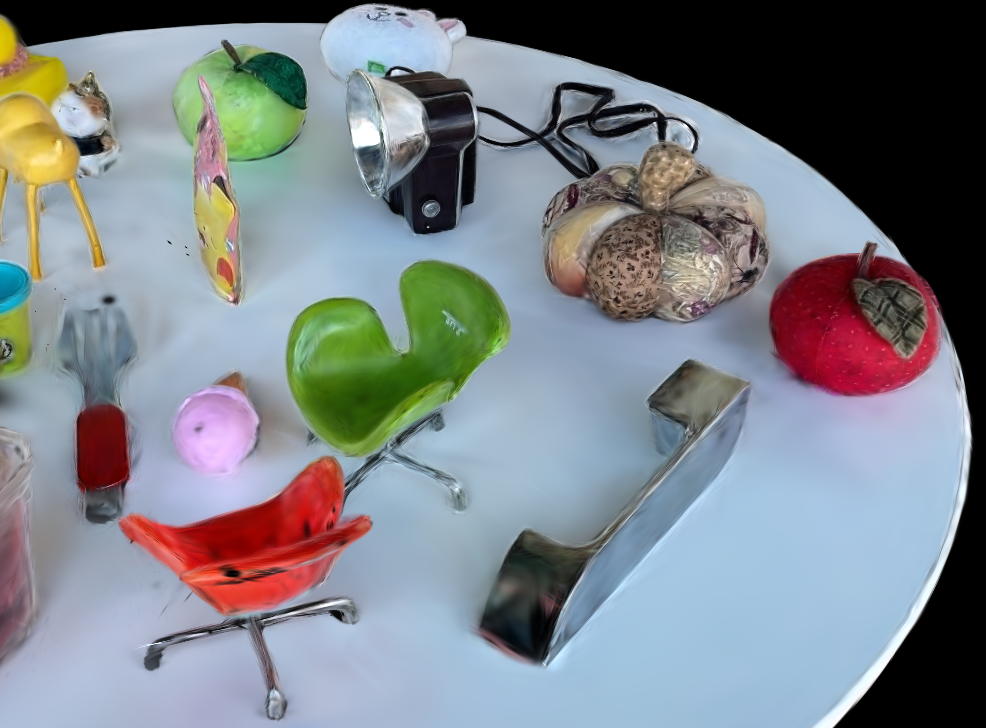}
\end{minipage}%
\hspace{0.01cm}
\begin{minipage}{.44\textwidth}
\includegraphics[width=\linewidth]{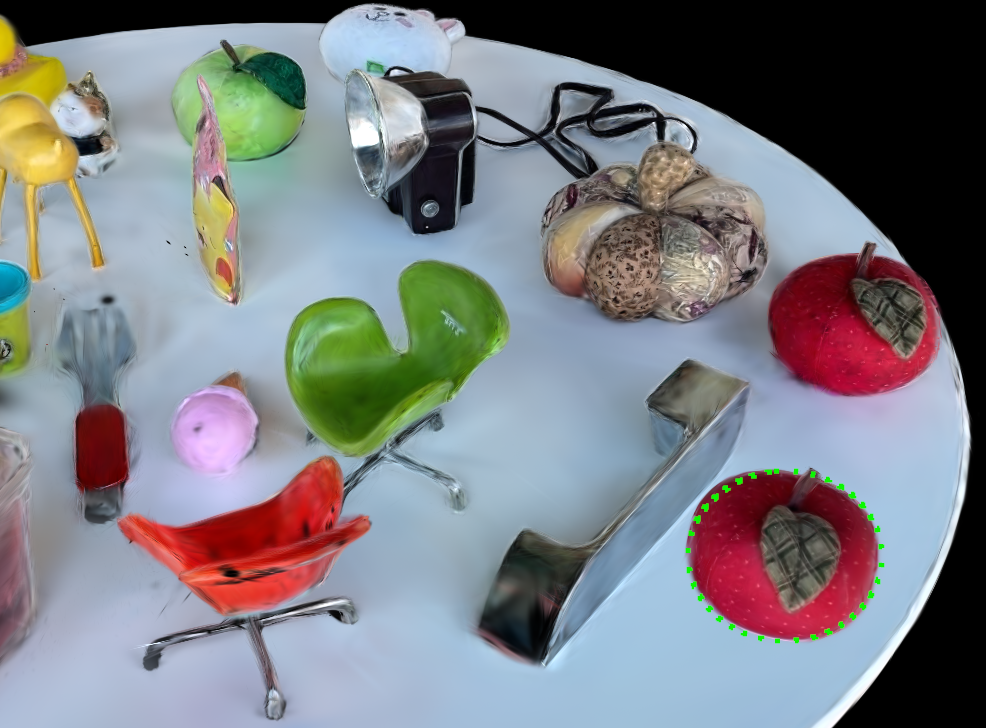}
\end{minipage}
\end{minipage}

\begin{minipage}{\textwidth}
\makebox[0pt][r]{%
\raisebox{-0.3\height}{\rotatebox{90}{\shortstack{\textbf{Recolor}\\``elephant"}}}%
\hspace{1mm}}
\centering
\begin{minipage}{.44\textwidth}
\includegraphics[width=\linewidth]{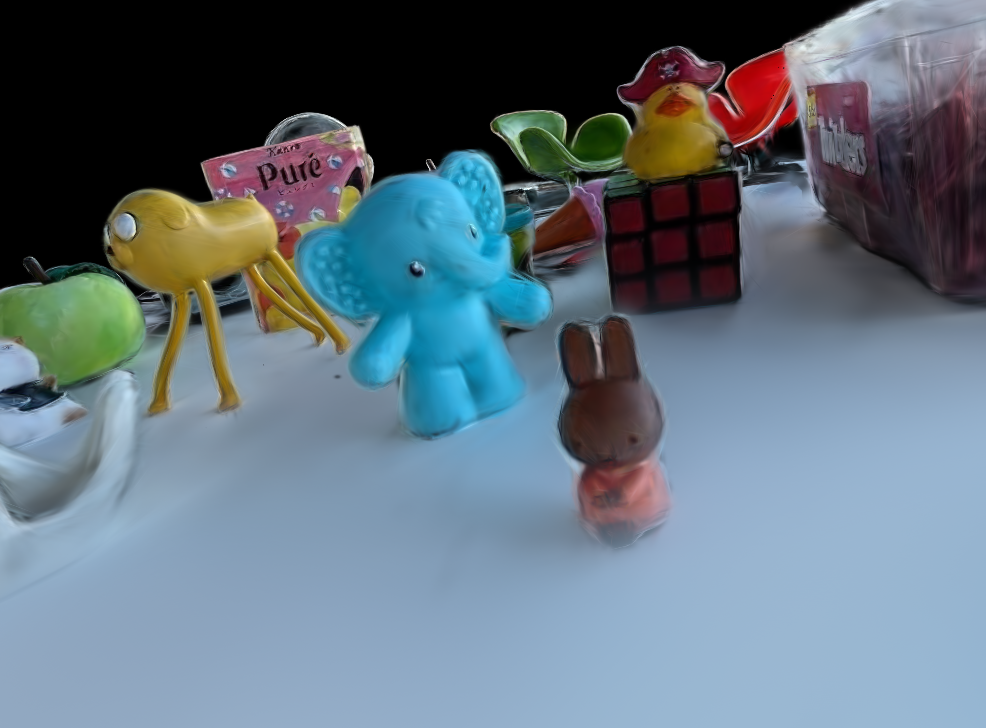}
\end{minipage}%
\hspace{0.01cm}
\begin{minipage}{.44\textwidth}
\includegraphics[width=\linewidth]{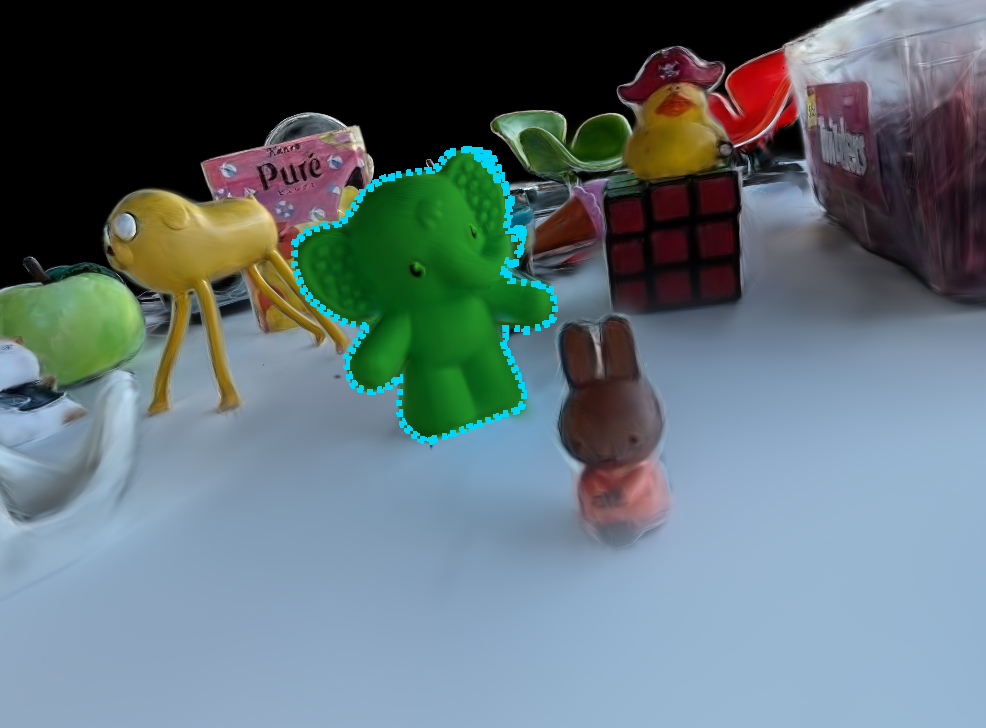}
\end{minipage}
\end{minipage}

\caption{Editing capabilities of Split\&Splat.}
\label{fig:edit_qualitative}
\end{minipage}
\Description[Editing Qualitatives]{Editing Qualitatives}
\end{figure*}

Since the output of \textbf{Split\&Splat} is an instance-labeled GS reconstruction, it naturally enables object-level scene editing operations, including \textit{object removal}, \textit{object duplication}, \textit{object movement}, and \textit{object recoloring}.
To validate this capability, we apply a variety of editing operations to the LERF \textit{figurines} scene and obtain visually compelling results.
Representative examples generated by combining our approach with external editing tools such as SuperSplat~\cite{supersplat} and Splatviz~\cite{barthel2024gaussian} are shown in Fig.~\ref{fig:edit_qualitative}.

\subsection{Ablation Study}
\label{sec:ablation}
We assess the contribution of each component of the proposed approach through an ablation study conducted on scene {0062\_00} of ScanNetv2, as reported in Tab.~\ref{tab:ablation_3d_0062}. 
Following InstanceGS, we applied image subsampling ($1:20$). This strategy results in only a $0.15\%$ performance drop while significantly reducing the computational cost. This confirms that the proposed approach remains reliable, even on datasets with a large number of acquisitions. 
In addition, we show that mask refinement is fundamental: without it, the mIoU drops to $48.69\%$ (a decrease of $\sim\!15\%$).

\begin{table}[t]
\centering \small
\addtolength{\tabcolsep}{-0.25em}
\caption{Ablation study for 3D instance segmentation on ScanNetv2 scene {0062\_00}.}%
\label{tab:ablation_3d_0062}
\begin{tabular}{l | ccc}
\toprule
\textbf{Ablation} & \textbf{mIoU} & \textbf{mAcc(25)} & \textbf{mAcc(50)} \\
\midrule
baseline & 63.46 & 100.00 & 70.83 \\
w/o subsampling & 64,00 & 100.00 & 75.00 \\
w/o mask refinement & 48.69 & 50.00 & 75.00 \\
\midrule
hb masks w/o propagation & 51.84 & 87.50 & 54.17 \\
hb masks w/ propagation & 58.50 & 95.83 & 66.67 \\
\bottomrule
\end{tabular}
\end{table}

\begin{table}[t]
\centering \small
\caption{Propagation performance metrics for scene {0062\_00} at different depth thresholds $\tau_{depth}$.}
\label{tab:depth_metrics}
\begin{tabular}{ l | ccc}
\toprule
& \multicolumn{3}{c}{\textbf{depth threshold $\tau_{depth}$}}\\
  & 0.1m & 0.02m & 0.001m \\
\midrule
\textbf{mIoU}     & 62.33 & 63.98 & 51.98 \\
\textbf{mAcc(25)} & 66.67 & 100.00 & 54.17 \\
\textbf{mAcc(50)} & 95.83 & 70.83 & 70.83 \\
\bottomrule
\end{tabular}
\end{table}

\begin{table}[t]
\centering \small
\addtolength{\tabcolsep}{-0.25em}
\caption{Open vocabulary segmentation accuracy with varying background masking color on LERF scenes.}
\label{tab:ov-ablation-results}
\begin{tabular}{l|cc|cc|cc}
\toprule
\multirow{2}{*}{\textbf{Dataset}} & \multicolumn{2}{c|}{\textbf{Mask-black}} & \multicolumn{2}{c|}{\textbf{Mask-blur}} & \multicolumn{2}{c}{\textbf{Mask-white}} \\
 & mIoU & mAcc(25) & mIoU & mAcc(25) & mIoU & mAcc(25) \\
\midrule
Figurines & 60.28 & 75.28 & 61.80 & 78.22 & 60.28 & 75.28 \\
Ramen & 58.29 & 74.27 & 58.89 & 75.95 & 51.53 & 67.38 \\
Teatime & 59.82 & 75.90 & 59.43 & 75.04 & 62.15 & 78.93 \\
Waldo Kitchen & 32.09 & 48.81 & 42.58 & 62.98 & 28.61 & 43.10 \\
\midrule
\textbf{Mean} & 52.62 & 68.57 & 55.68 & 73.05 & 50.64 & 66.17 \\
\bottomrule
\end{tabular}
\end{table}

\begin{table}[h!]
\centering \small
\addtolength{\tabcolsep}{-0.25em}
\caption{Performance metrics and average percentage of instances with label $l$ for different correlation thresholds $\tau_{corr}$.}
\label{tab:correlation_metrics}
\begin{tabular}{l | ccccccc}
\toprule
& \multicolumn{7}{c}{\textbf{correlation threshold $\tau_{corr}$}}\\
& 0.01 & 0.02 & 0.025 & 0.05 & 0.075 & 0.1 & 0.2 \\
\midrule
mIoU & 50.68 & 55.30 & 56.55 & 57.35 & 57.20 & 56.23 & 56.64 \\
mAcc(25) & 65.45 & 72.31 & 74.30 & 76.00 & 76.03 & 74.39 & 74.87 \\
mAcc(50) & 61.21 & 66.83 & 68.72 & 69.18 & 69.20 & 67.56 & 68.04 \\
\midrule
\textbf{avg \% lab. $l$} & 7.69\% & 12.24\% & 15.67\% & 39.24\% & 67.53\% & 86.23\% & 100.00\% \\
\bottomrule
\end{tabular}
\end{table}

\subsubsection{Exploiting hint-based mask generation}
To evaluate our mask generation and propagation pipeline, we compare its performance against a \textit{hint-based} (hb) baseline.
Following \cite{sam2}, a small number of views are provided with hand-crafted inputs in the form of arbitrary point prompts, \ie, a set of 2D coordinates used to approximate the location of object instances in a subset of views. These prompts are then propagated to the remaining views by treating them as consecutive video frames.
Results  in the lower part of Tab.~\ref{tab:ablation_3d_0062}, highlights how the lack of 3D consistency in the resulting masks propagates to the final reconstruction and leads to a performance drop of $11.6\%$.
Moreover, since our propagation pipeline treats these masks as uncorrelated, it partially compensates for segmentation errors, yielding a mIoU improvement of $6.34\%$ when applied on top of hint-based masks.

\subsubsection{Surface depth threshold} \label{sec:depth_th}
During the propagation step, $\tau_{depth}$ defines the depth threshold used to determine whether a point is considered part of the surface, which will be projected back onto the 2D mask.
Higher values retain more points, whereas lower values discard more, resulting in a more conservative policy with fewer but more reliable surface points.
Tab.~\ref{tab:depth_metrics} compares the labeled point cloud $P_{labeled}$ (\ie, the output of the propagation step) with the ground truth for different values of $\tau_{depth}$.
The results show that an intermediate value yields the best performance, avoiding the under-segmentation caused by excessively high thresholds and the over-segmentation introduced by overly low values.

\subsubsection{Open vocabulary descriptors ablation}\label{sec:ov_desc}
Since our descriptors are extracted from instance-centric crops defined by the binary masks produced by \textbf{Split\&Splat}, we investigate three masking strategies for computing CLIP \cite{clip} descriptors: background \textit{blurring}, \textit{white} background, and \textit{black} background.
As reported in Tab.~\ref{tab:ov-ablation-results}, background blurring yields the best overall performance, suggesting that preserving some surrounding context provides useful information about the instance.
In contrast, \textit{white} and \textit{black} backgrounds lead to different outcomes depending on the instance color. In particular, lighter objects tend to produce less informative descriptors when placed on a \textit{white} background, whereas a \textit{black} background can yield more discriminative descriptors.

\subsubsection{Open vocabulary correlation threshold}\label{sec:ov_thres}
Given the instance-centric nature of our approach, we use a threshold $\tau_{corr}$ to associate textual descriptors $(f_{text})$ with object-level descriptors ($f_{obj})$. This association is computed using the cosine distance $d_{cos} \! = \! d(f_{text},f_{obj})$.
For each object, we compute a correlation score $ \delta=|d_{cos}^* - d_{cos}|$ where $ d_{cos}^*$ denotes the minimum cosine distance for a given query.
A match between an instance and a textual query is established when $\delta\! < \! \tau_{corr}$.
As shown in Tab.~\ref{tab:correlation_metrics}, increasing this threshold improves overall segmentation performance.
However, a larger threshold effectively connects each textual query with multiple instances, resulting in test-time behavior that maximizes the mIoU independently for each ground-truth mask.
Thus, we adopt a more conservative, albeit suboptimal strategy, setting $\tau_{corr} = 0.02$.

\section{Conclusion} \label{sec:conclusions}
In this work, we introduced \textbf{Split\&Splat}, a novel 3D Gaussian splatting pipeline for panoptic scene understanding.
By explicitly modeling each object instance prior to reconstructing the full 3D scene, our method preserves instance-level separation throughout the reconstruction process, resulting in geometrically and semantically consistent representations.
Extensive quantitative and qualitative evaluations on ScanNetv2 demonstrate the effectiveness of our approach, with \textbf{Split\&Splat} consistently outperforming state-of-the-art competitors.
Thanks to its modular design, the proposed framework is highly flexible and naturally supports a variety of downstream tasks, including object retrieval and object-level scene editing (\eg, removal, duplication, and recoloring).
Experiments on the LERF benchmark further confirm the generalizability of our method, showing that \textbf{Split\&Splat} achieves competitive performance even in open-vocabulary settings.
Despite these encouraging results, there is still room for improvement, particularly in instance re-labeling and in the identification of object parts. Addressing these challenges will further strengthen the applicability of our framework and constitute an important direction for future work.

\clearpage
\newpage

\clearpage
\newpage

\appendix
\section{Appendix}

In this appendix, we report clarifications and additional findings that was not possible to fit in the main document. 

\subsection{Mask Propagation Algorithm} \label{subsec:semanticref}

Alg.~\ref{algo:labelAss} shows a procedural formulation of the mask propagation and label unification strategy described in Sec.~3.1.3 of the main document, and is reported here for clarity and reproducibility.

\begin{algorithm}[ht]
\small
\caption{Label via Mask--Point Intersection}
\label{algo:labelAss}
\DontPrintSemicolon
\KwIn{$P_{\mathrm{labeled}}$; mask set $\mathcal{M}_k$; warped mask $M^{*}_{k,j}$; bias $\lambda_{\text{init}}$}
\KwOut{label $l$ for each $M_{k,j} \in \mathcal{M}_k$ and updated $P_{\mathrm{labeled}}$}
$\mathcal{L} \gets \mathcal{J}_1$ \tcp{Initialize the set of global instance indices}
\For{$k \gets 2,\dots,K$}{
    \For{$j \in \mathcal{J}_k$}{
      \tcp{Collect labels overlapping with the virtual mask}
      $\mathcal{J}_{k-1 \cap k} \gets \{\, j \mid M^{*}_{k,j}(x,y)=1,\ M_{k,i}(x,y) = 1 \forall i,x,y \}$  \;
      
      \eIf{$\mathcal{J}_{k-1 \cap k} = \varnothing$}{
         $l \gets j$ \tcp*{new instance detected}
         $\mathcal{L} \gets \mathcal{L} \cup \{j\}$
      }{
         $l \gets {\arg\!\max}_{j \in \mathcal{J}_{k-1}} \left\{\sum\limits_{x,y} {M^{*}_{k,j}\cdot M_{k,i}}\right\}_{i \in \mathcal{J}_{k}}\;$ \hspace*{-6em}\tcp*{max overlap}
      }
    
    \tcp{Update label weights}
    \ForEach{$p \in P_{j,k}$}{
        \If{$w_p$ is uninitialized}{
            $w_p[i] \gets 1 + \lambda_{\text{init}}$\tcp*{first observation}
        }
        \Else{
            $w_p[j] \gets w_p[j] + 1$
        }
        $l_p \gets \arg\max\limits_{j \in \mathcal{J}} w_p[j]$
    }
}
}
\end{algorithm}

\subsection{Dense Semantic Embeddings} \label{subsec:semanticref}
In the main document, we assigned a single CLIP-based semantic descriptor to each object instance.
However, some scene recognition tasks, such as part segmentation and object-level editing, may require a denser embedding of semantic information embedded directly in the Gaussians.
To achieve this, we propose enriching object representations by assigning a semantic descriptor $f_{g_i}$ to each Gaussian $g_i \! \in\! \mathcal{G}_{l}$ of instance $l$.
To this end, we extract a new set of $N_{synth} = 72$ (corresponding to $6$ different elevations and $12$ azimuthal angles) multi-view images $\mathcal{I}_l$ around instance $l$ by sampling camera positions on a semi-spherical trajectory around the object and rendering it from these viewpoints. 
The trajectory is centered at the instance centroid $c_a = \frac{1}{|\mathcal{G}_l|}\sum_{g \in \mathcal{G}_l}, \;g \in \mathbb{R}^3$ and has a radius $r_a = 2\,d_{B_l}$, where $d_{B_l}$ is the diagonal of its bounding box $B_l$. This ensures sufficient coverage of the instance while providing viewpoint diversity.

From these masked views (where the background is blurred), we compute high-resolution 2D semantic descriptors using a feature extraction network $E_F(\cdot)$ (\eg, DINOv2), producing descriptor maps $f \in \mathbb{R}^{H \times W \times n}$. Each descriptor is then projected back onto the Gaussian splatting.

For ease of understanding, we align the following mathematical notation with that of \cite{kerbl3Dgaussians}. For a pixel location $\mathrm{q}$ in view $k$, we assign its descriptor to the Gaussian that contributes most to the alpha blending of the corresponding pixel rasterization:
\begin{equation}
g_i = \arg\!\max_i \left[
\alpha_i \cdot \exp\left(
-\tfrac{1}{2} \pi_{w,h}^\top
\Sigma_{i,2D}^{-1}
\pi_{w,h}
\right)
\right].
\end{equation}
where $\pi_{w,h} \coloneq \mathrm{w}-\Pi(h)$, and $\Sigma_{i,2D}^{-1}$ represents the covariance matrix in the standard Gaussian splatting formulation~\cite{kerbl3Dgaussians}.
For each Gaussian, the final descriptor $f_{g_i}$ is obtained by averaging contributions across all the $N_v$ views in which it is visible:
\begin{equation}
f_{g_i} = \frac{1}{N_v} \sum_{n=1}^{N_v} f_{g_i}^{n}, \; N_v < N.
\end{equation}

\begin{figure}[t]
  \centering
  \begin{subfigure}{0.23\textwidth}
    \centering
    \includegraphics[trim=8cm 8cm 8cm 8cm,clip,width=\linewidth]{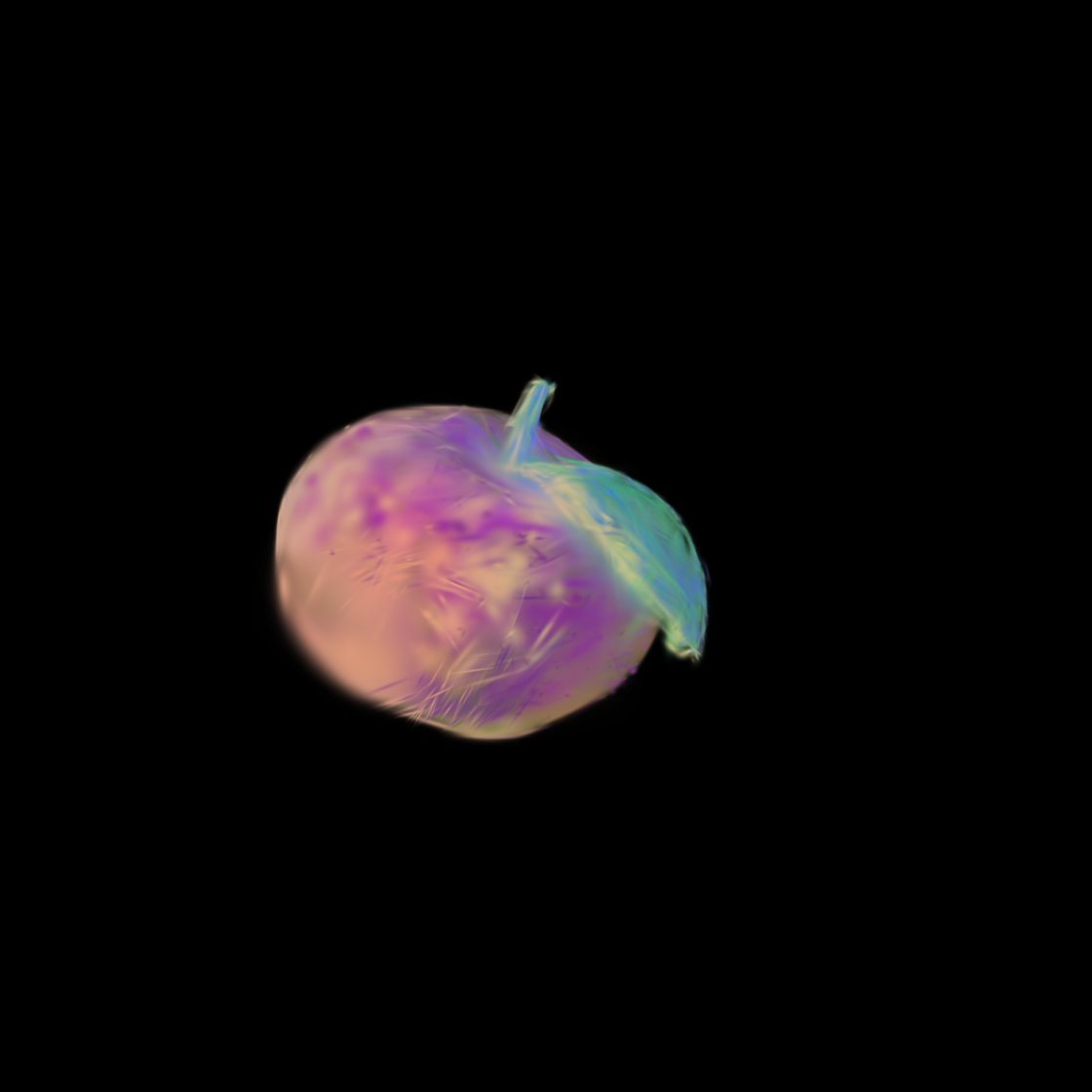}
  \end{subfigure}
  \begin{subfigure}{0.23\textwidth}
    \centering
    \includegraphics[trim=8cm 8cm 8cm 8cm,clip,width=\linewidth]{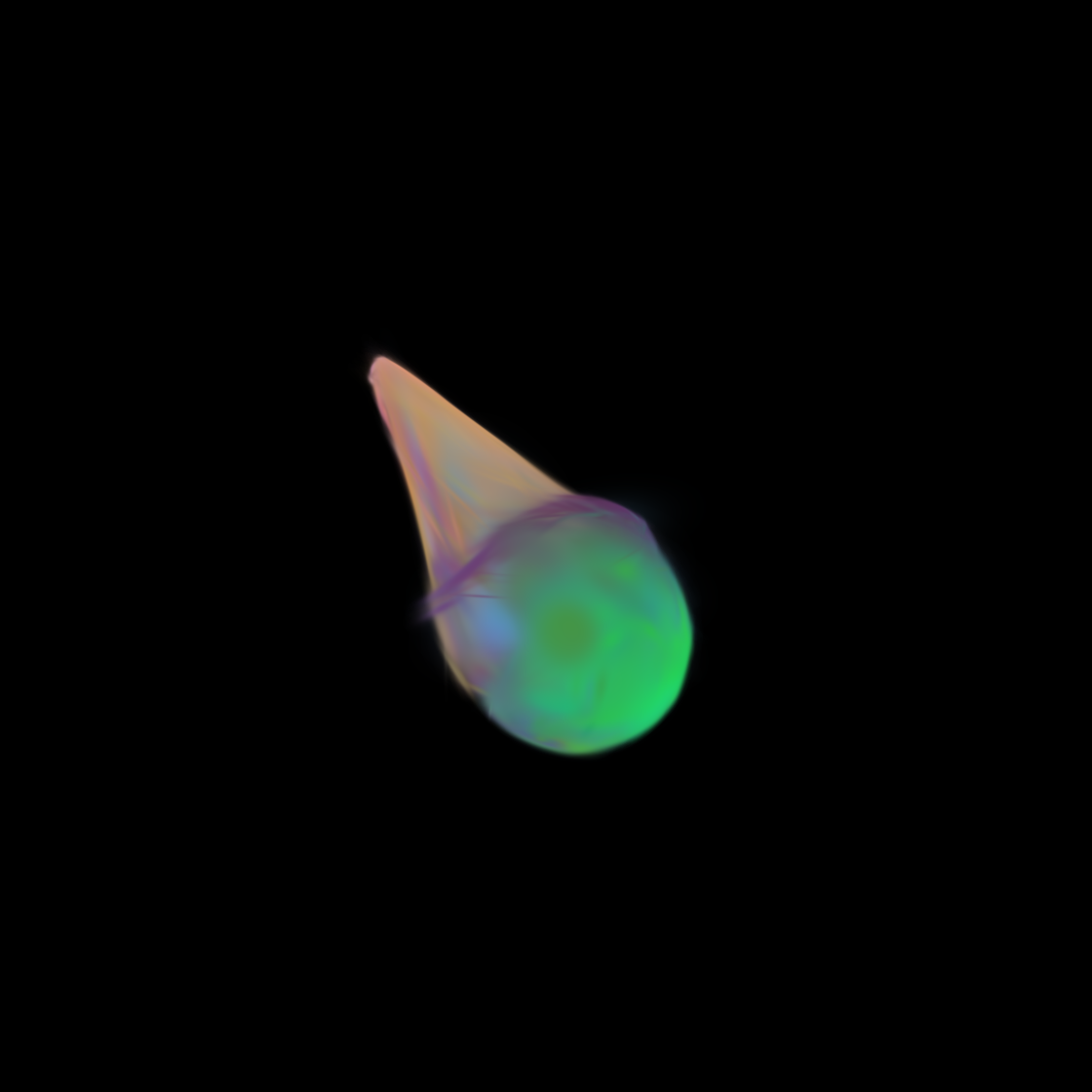}
  \end{subfigure}
  \caption{PCA on projected descriptors "red apple" (left) and "ice cream" (right).}
  \label{fig:app_descPCA}
\end{figure}

This produces a final 3D Gaussian reconstruction enriched with per-Gaussian semantic descriptors. %
Fig.~\ref{fig:app_descPCA} shows a PCA projection of the learned object descriptors for two textual queries (``red apple" and ``ice cream"). The learned features capture meaningful semantic information, showing differences between object parts in the two instances.

\begin{table*}[t]
\centering \small
\caption{Open vocabulary results on Lerf dataset \cite{LERF2023}, competitors data taken from the work of \cite{vala}. Best in bold, second best underlined, third dashed underline.} 
\label{tab:ov-results}
\begin{tabular}{l|cc|cccccccc}
\toprule
\multirow{2}{*}{\textbf{Method}} & \multicolumn{2}{c|}{\textbf{Average}} & \multicolumn{2}{c}{\textbf{Figurines}} & \multicolumn{2}{c}{\textbf{Ramen}} &  \multicolumn{2}{c}{\textbf{Teatime}} & \multicolumn{2}{c}{\textbf{Waldo Kitchen}} \\
& mIoU & mAcc & mIoU & mAcc & mIoU & mAcc & mIoU & mAcc & mIoU & mAcc \\
\midrule
LERF \cite{LERF2023} 
                                        & 10.35 & 13.64 & 7.27 & 10.71 & 10.05 & 9.86 & 14.38 & 20.34 & 9.71 & 9.09 \\
LEGaussian \cite{shi2024language} 
                                        & 16.21 & 23.82 & 17.99 & 23.21 & 15.79 & 26.76 & 19.27 & 27.12 & 11.78 & 18.18 \\
OpenGaussian~\cite{wu2024opengaussian}
                                        & 38.36 & 51.43 & 39.29 & 55.36 & 31.01 & 42.25 & 60.44 & 76.27 & 22.70 & 31.82 \\
SuperGSeg~\cite{liang2024supergseg}
                                        & 35.94 & 52.02 & 43.68 & 60.71 & 18.07 & 23.94 & 55.31 & 77.97 & 26.71 & 45.45 \\
Dr.Splat~\cite{jun2025dr}
                                        & 43.29 & 64.30 & 54.42 & 80.36 & 24.33 & 35.21 & 57.35 & 77.97 & 37.05 & \dashuline{63.64} \\
InstanceGS~\cite{li2025instancegaussian}
                                        & 43.87 & 61.09 & 54.87 & 73.21 & 25.03 & 38.03 & 54.13 & 69.49 & 41.47 & \dashuline{63.64} \\

CAGS~\cite{sun2025cags}           
                                        & \dashuline{50.79} & 69.62 & \dashuline{60.85} & \dashuline{82.14} & 36.29 & 46.48 & \underline{68.40} & \dashuline{86.44} & 37.62 & \dashuline{63.64} \\
VoteSplat~\cite{jiang2025votesplat}
                                        & 50.10 & 67.38 & \textbf{68.62} & \underline{85.71} & \dashuline{39.24} & \dashuline{61.97} & \dashuline{66.71} & \underline{88.14} & 25.84 & 33.68 \\
Occam's LGS~\cite{occams_lgs_2024}       
                                        & 47.22 & \underline{74.84} & 52.90 & 78.57 & 32.01 & 54.92 & 61.02 & \textbf{93.22} & \underline{42.95} & \underline{72.72} \\
VALA \cite{vala}                      
                                        & \textbf{58.02} & \textbf{82.85} & 60.38 & \textbf{89.29} & \underline{45.41} & \underline{67.61} & \textbf{70.61} & \underline{88.14} & \textbf{55.71} & \textbf{86.36} \\

\midrule
\textbf{Split\&Splat}   
                                        & \underline{55.68} & \dashuline{73.05} &  \underline{61.80} & 78.22 & \textbf{58.89} & \textbf{75.95} & 59.43 & 75.04 & \dashuline{42.58} & 62.98 \\
\bottomrule 
\end{tabular}%
\end{table*}

\subsection{Additional Results} %

Tab.~\ref{tab:ov-results} extends Tab.~2 of the main document by  reporting open-vocabulary segmentation results across the four main scenes of LERF, comparing  \textbf{Split\&Splat} to other methods.
Although our method is not specifically designed for open-vocabulary segmentation, \textbf{Split\&Splat} achieves competitive performance, ranking second in average mIoU ($55.68$) among all compared approaches and being outperformed only by the recent VALA method, while significantly outperforming strong baselines such as InstanceGS, SuperGSeg, and OpenGaussian.
Performance varies across scenes: the most challenging case is Waldo Kitchen, where the number of object instances ($155$) is substantially higher than in the other scenes (on average $\simeq 25$). This highlights a current limitation of our approach when handling scenes with an extremely large number of instances.
Finally, note that the reported scores underestimate the actual visual quality of our results. \textbf{Split\&Splat} performs object-level segmentation, whereas the ground truth often splits a single object into multiple parts. As a result, a visually correct object mask may be penalized by the evaluation protocol, leading to lower quantitative scores.

Tab.~\ref{tab:per-instance} reports per-instance IoU values on the ScanNetv2 scene 0062\_00. The results show a large variability across instances, with IoU values ranging from $26.9$ to $97.8$, and a mean IoU of $63.5$.
This spread reflects the varying difficulty of different objects, especially in the presence of clutter, occlusions, and fine-grained structures. High scores (\eg, instances 7, 10, and 11) correspond to large and well-isolated objects, while lower values are typically associated with small or heavily occluded instances.

\begin{table}[t]
\centering \small
\addtolength{\tabcolsep}{-1em}
\caption{Per-instance IoU on scene {0062\_00}.}
\label{tab:per-instance}
\begin{tabular}{p{1.6cm}cccc}
\toprule
\textbf{Instance ID} & 0 & 1 & 2 & 3 \\
\textbf{Name} & wall & wall & wall & wall \\
\textbf{IoU\%} & 40.4 & 53.3 & 74.1 & 56.7 \\
\midrule
\textbf{Instance ID} & 4 & 5 & 6 & 7 \\
\textbf{Name} & wall & mirror & counter & \parbox{1.4cm}{\centering\linespread{.8}\selectfont trash can} \\
\textbf{IoU\%} & 26.9 & 83.4 & 81.3 & 93.6 \\
\midrule
\textbf{Instance ID} & 8 & 9 & 10 & 11 \\
\textbf{Name} &
\parbox{1.4cm}{\centering\linespread{.8}\selectfont trash can} &
\parbox{2.2cm}{\centering\linespread{.8}\selectfont toilet seat\\cover dispenser} &
\parbox{2.0cm}{\centering\linespread{.8}\selectfont paper towel dispenser} &
floor \\
\textbf{IoU\%} & 75.4 & 88.2 & 97.8 & 90.0 \\
\midrule
\textbf{Instance ID} & 12 & 13 & 14 & 15 \\
\textbf{Name} & toilet &
\parbox{1.6cm}{\centering\linespread{.8}\selectfont toilet paper} &
\parbox{1.6cm}{\centering\linespread{.8}\selectfont toilet paper} &
\parbox{1.6cm}{\centering\linespread{.8}\selectfont light switch} \\
\textbf{IoU\%} & 77.9 & 41.9 & 33.8 & 47.4 \\
\midrule
\textbf{Instance ID} & 16 & 17 & 18 & 19 \\
\textbf{Name} & jacket & rail & rail &
\parbox{1.8cm}{\centering\linespread{.8}\selectfont soap dispenser} \\
\textbf{IoU\%} & 81.5 & 57.2 & 37.3 & 67.0 \\
\midrule
\textbf{Instance ID} & 20 & 21 & 22 & 23 \\
\textbf{Name} & door & sink &
\parbox{1.6cm}{\centering\linespread{.8}\selectfont toilet paper} &
doorframe \\
\textbf{IoU\%} & 52.3 & 72.1 & 61.9 & 31.6 \\
\midrule
\multicolumn{3}{c}{\textbf{mean IoU\%}} & \multicolumn{2}{c}{\textbf{63.5}} \\
\bottomrule
\end{tabular}
\end{table}

\clearpage
\bibliographystyle{ACM-Reference-Format}
\bibliography{main}

\end{document}